\documentclass[preprint,12pt]{aastex}
\usepackage{lscape}

\newcommand{\ee}[1]{\mbox{${} \times 10^{#1}$}}

\newcommand{\am}{\mbox{\arcmin}}
\newcommand{\as}{\mbox{\arcsec}}

\newcommand{\um}{$\mu$m}

\newcommand{\lsun}{\mbox{L$_\odot$}}
\newcommand{\msun}{\mbox{M$_\odot$}}
\newcommand{\rsun}{\mbox{R$_\odot$}}

\newcommand{\lbol}{\mbox{$L_{bol}$}} 
\newcommand{\lir}{\mbox{$L_{IR}$}} 
\newcommand{\lint}{\mbox{$L_{int}$}} 
\newcommand{\lext}{\mbox{$L_{ext}$}} 
\newcommand{\tbol}{\mbox{$T_{bol}$}} 

\newcommand{\lbolsmm}{$\lbol/$\lsmm}

\newcommand{\lsmm}{\mbox{$L_{smm}$}} 

\newcommand{\nthp}{\mbox{{\rm N$_2$H}$^+$}}

\newcommand{\andre}{Andr\'{e}}
\newcommand{\alcala}{Alcal\'{a}}

\slugcomment{Accepted for publication in ApJS}
\begin{document}

\title {Identifying the Low Luminosity Population of Embedded Protostars in the c2d Observations of Clouds and Cores}

\author{Michael M. Dunham\altaffilmark{1,2}, Antonio Crapsi\altaffilmark{3,4}, Neal J. Evans II\altaffilmark{1}, Tyler L. Bourke\altaffilmark{5}, Tracy L. Huard\altaffilmark{5}, Philip C. Myers\altaffilmark{5}, and Jens Kauffmann\altaffilmark{6,5}}

\altaffiltext{1}{Department of Astronomy, The University of Texas at Austin, 1 
University Station, C1400, Austin, Texas 78712--0259}

\altaffiltext{2}{E-mail: mdunham@astro.as.utexas.edu}

\altaffiltext{3}{Sterrewacht Leiden, Leiden University, P.O. Box 9513, 2300 RA Leiden, the Netherlands}

\altaffiltext{4}{Observatorio Astron\'{o}mico Nacional (IGN), Alfonso XII, 3, E-28014 Madrid, Spain}

\altaffiltext{5}{Harvard-Smithsonian Center for Astrophysics, 60 Garden Street, Cambridge, MA 02138}

\altaffiltext{6}{Initiative in Innovative Computing at Harvard, 60 Oxford Street, Cambridge, MA 02138}

\begin{abstract}
We present the results of a search for all embedded protostars with internal luminosities $\le$ 1.0 \lsun\ in the full sample of nearby, low-mass star-forming regions surveyed by the \emph{Spitzer Space Telescope} Legacy Project ``From Molecular Cores to Planet Forming Disks'' (c2d).  The internal luminosity of a source, \lint, is the luminosity of the central source and excludes luminosity arising from external heating.  On average, the \emph{Spitzer} c2d data are sensitive to embedded protostars with \lint\ $\geq 4 \times 10^{-3}$ $(d/140 \, \rm{pc})^2$ \lsun, a factor of 25 better than the sensitivity of the \emph{Infrared Astronomical Satellite (IRAS)} to such objects.  We present a set of selection criteria used to identify candidates from the \emph{Spitzer} data and examine complementary data to decide whether each candidate is truly an embedded protostar.  We find a tight correlation between the 70 \um\ flux and internal luminosity of a protostar, an empirical result based on both observations and detailed two-dimensional radiative transfer models of protostars.  We identify 50 embedded protostars with \lint\ $\le$ 1.0 \lsun; 15 have \lint\ $\le$ 0.1 \lsun.  The intrinsic distribution of source luminosities increases to lower luminosities.  While we find sources down to the above sensitivity limit, indicating that the distribution may extend to luminosities lower than probed by these observations, we are able to rule out a continued rise in the distribution below \lint\ $= 0.1$ \lsun.  Between $75-85$\% of cores classified as starless prior to being observed by \emph{Spitzer} remain starless to our luminosity sensitivity; the remaining $15-25$\% harbor low-luminosity, embedded protostars.  We compile complete Spectral Energy Distributions for all 50 objects and calculate standard evolutionary signatures (\lbol, \tbol, and \lbolsmm), and argue that these objects are inconsistent with the simplest picture of star formation wherein mass accretes from the core onto the protostar at a constant rate.
\end{abstract}



\section{Introduction}\label{intro}

Recently, the \emph{Spitzer Space Telescope} Legacy Project ``From Molecular Cores to Planet Forming Disks'' (c2d; Evans et al. 2003) completed an extensive $3.6-160$ \um\ imaging survey of nearby, low-mass star-forming regions.  One of the results to come out of this survey is the discovery of very low luminosity objects (VeLLOs; Young et al. 2004).  If the internal luminosity of a source, \lint, is the total luminosity of the central protostar and circumstellar disk (if present), a VeLLO is defined to be an object embedded within a dense core with $\lint \leq 0.1$ \lsun\ (Di Francesco et al. 2007).  The bolometric luminosity of an embedded protostar, an observable quantity that can be calculated by integrating over the full Spectral Energy Distribution (SED), is composed of both internal and external luminosity ($\lbol\ = \lint\ + \lext$).  The external luminosity is usually that arising from heating of the circumstellar envelope by the Interstellar Radiation Field (ISRF), and will add, on average, a few tenths of a solar luminosity to \lbol\ (e.g., Evans et al. 2001).  Thus, the distinction between \lbol\ and \lint\ is most relevant for embedded protostars with $\lint \la 1.0$ \lsun, where the external luminosity can be a significant fraction of the observed \lbol.  For VeLLOs, the external luminosity can dominate the observed \lbol.  Radiative transfer modeling of the SEDs of embedded protostars, including both the emission from the envelope at submillimeter and millimeter wavelengths and the emission from the central source itself at infrared wavelengths, is required to decouple internal and external luminosities (e.g., Shirley et al. 2002; Young et al. 2004; Dunham et al. 2006).

Several VeLLOs have been discovered in cores that were previously classified as starless prior to being observed by \emph{Spitzer}.  In fact, the very first starless core observed by c2d, L1014, was found to harbor a VeLLO with \lint\ $\sim$ 0.09 \lsun\ (Young et al. 2004).  This discovery reinforces that the known sample of embedded protostars is not complete.  This sample has been assembled primarily by two methods:  (1) searching for \emph{IRAS} sources that are associated with dense cores and have colors consistent with those expected for embedded protostars (e.g., Myers et al. 1987), and (2) identifying molecular outflows and radio point sources associated with dense cores indicating the presence of protostars too deeply embedded to detect with \emph{IRAS} (e.g., \andre\ et al. 1993).  Myers et al. (1987) found that the \emph{IRAS} data could detect protostars with \lint\ $\ga 0.1$ $(d/140 \, \rm{pc})^2$ \lsun, where $d$ is the distance to the protostar, although this does not include the younger, more deeply embedded protostars that were only identified on a case-by-case basis by the second method.  The regions surveyed by c2d with \emph{Spitzer} are located at distances ranging from $125-500$ pc.  Even in the closest of these regions VeLLOs are likely to fall below the \emph{IRAS} sensitivity limit.  In the more distant regions, no protostars with \lint\ $\la$ 1 \lsun\ would be detected.  While some of these protostars might have been identified on a case-by-case basis as described above, the full sample of embedded protostars with \lint\ $\le$ 1 \lsun\ is clearly incomplete.

Constructing a complete sample of embedded protostars with \lint\ $\le$ 1 \lsun\ is important for studies of low-mass star formation.  Despite substantial progress in recent decades, the details of the physical processes regulating mass accretion from the envelope to the protostar remain poorly understood (see McKee \& Ostriker [2007] for a recent review).  Several authors have attempted to constrain evolutionary models of the formation of low-mass stars by determining the observational signatures of these models and comparing them to the properties of known protostars (e.g., Myers et al. 1998; Young \& Evans 2005).  A result common to all such studies is a substantial population of protostars with luminosities below model predictions.  An idea proposed to explain this discrepancy is that the mass accretion is episodic in nature and the protostars with the lowest luminosities are those observed in quiescent accretion states (e.g., Kenyon \& Hartmann 1995; Young \& Evans 2005; Enoch 2007; Enoch et al. 2008a, in preparation).  Theoretical studies have provided several mechanisms by which such a process may occur, such as material piling up in a circumstellar disk until gravitational instabilities drive angular momentum outward and mass inward in short-lived bursts (Vorobyov \& Basu 2005, 2006).  Alternatively, quasi-periodic magnetically driven outflows in the envelope can cause mass accretion onto the protostar to occur in magnetically controlled bursts (Tassis \& Mouschovias 2005).  Indeed, the evidence for non-steady mass accretion in young protostellar systems still in the embedded phase is steadily growing (e.g., Hartmann \& Kenyon 1985; Dunham et al. 2006; Acosta-Pulido et al. 2007; K\'{o}sp\'{a}l et al. 2007; Fedele et al. 2007).  However, as the sample of embedded, low luminosity protostars is incomplete, the true nature of the discrepancy between evolutionary models and observations of protostars is unknown.  Future work devoted to assessing the validity of various models by comparing their predictions to the properties of known protostars depends on the existence of a sample that is as complete and unbiased as possible.

The VeLLOs are a particularly interesting subset of embedded, low-luminosity protostars; in essence, they are an extreme case of the problem discussed above.  To date, only three VeLLOs have been studied in detail:  L1014-IRS (\lint\ $\sim$ 0.09 \lsun; Young et al. 2004), L1521F-IRS (\lint\ $\sim$ 0.06 \lsun; Bourke et al. 2006), and IRAM 04191-IRS (\lint\ $\sim$ 0.08 \lsun; Dunham et al. 2006).  Despite the fact that all three have similar internal luminosities, they differ greatly in envelope and outflow properties, as discussed by Bourke et al. (2006).  IRAM 04191 drives a large, bipolar molecular outflow, features bright molecular line and dust continuum emission, and shows evidence for infall, depletion, and deuteration (\andre\ et al. 1999; Belloche et al. 2002).  L1521F is also bright in molecular line and dust continuum emission and also shows evidence for infall, depletion, and deuteration (Crapsi et al. 2004), but the envelope is not as centrally condensed as IRAM 04191 and the presence of an outflow is uncertain (Crapsi et al. 2004; Bourke et al. 2006).  L1014 does not show significant evidence for infall, depletion, or deuteration (Crapsi et al. 2005a), but it does drive a compact, weak molecular outflow detected only in interferometer observations (Bourke et al. 2005; Crapsi et al. 2005a).  A systematic search for all VeLLOs in the regions surveyed by c2d will allow their properties to be examined in detail both on a case-by-case basis and as a class of objects.  Identifying the complete sample of VeLLOs will also allow us to determine how many cores classified as starless prior to being observed by \emph{Spitzer} truly are starless, a question with important implications for estimates of the lifetime of starless cores (e.g., Kirk et al. 2005).

In this paper, we present the results of a search for all embedded protostars with \lint\ $\leq$ 1.0 \lsun\ in the full c2d imaging dataset.  Depending on the distance to each individual source, some will already have been detected by \emph{IRAS}, while others will be new sources.  We consider this work to be complementary to several related studies:  A search by Kirk et al. (2007) for embedded protostars in 22 cores classified as starless prior to being observed by \emph{Spitzer}; a search by J\o rgensen et al. (2007; 2008, in preparation) for all embedded protostars in Perseus and Ophiuchus, regardless of luminosity, conducted by combining \emph{Spitzer} and SCUBA 850 \um\ dust continuum emission data; and a search by Enoch (2007) and Enoch et al. (2008a, in preparation) for all embedded objects in the Perseus, Ophiuchus, and Serpens molecular clouds, regardless of luminosity, conducted by combining \emph{Spitzer} and Bolocam 1.1 mm dust continuum emission data.  

The key difference between the work presented here and the searches for embedded protostars listed above is that we do not start by identifying dense cores from their millimeter dust continuum emission and then look for associated \emph{Spitzer} sources embedded within them.  Instead, we develop a set of criteria to identify candidate embedded, low-luminosity protostars based on the $3.6-70$ \um\ \emph{Spitzer} data.  This way, we are able to identify all of the candidates in the full c2d dataset, regardless of the availability and quality of millimeter wavelength observations for each region.  Only after we identify all candidates based on \emph{Spitzer} data alone do we turn to other observations to distinguish the objects of interest from various contaminants masquerading in our sample.  Our method identifies candidates for further examination once large-scale surveys of nearby star-forming regions are completed with SCUBA-2 (Ward-Thompson et al. 2007) and Herschel, and the method can easily be extended to search for embedded, low-luminosity protostars in the additional nearby, low-mass star-forming regions being surveyed by the \emph{Spitzer} Gould Belt Legacy Project (L. Allen et al. 2008, in preparation).

The organization of this paper is as follows:  In \S \ref{observations}, we provide a brief description of the c2d observations and data reduction, emphasizing those aspects relevant to this work.  The criteria for identifying candidate embedded, low-luminosity protostars from the $3.6-70$ \um\ \emph{Spitzer} data are discussed in \S \ref{id}, along with the possibilities for estimating source internal luminosities directly from observable quantities.  A general proof-of-concept demonstrating the validity of these criteria is given in \S \ref{proofofconcept}.  In \S \ref{confirmation}, we discuss the contamination expected in the list of candidates, both from background extra-galactic sources and from more evolved Young Stellar Objects (YSOs) no longer embedded within their dense cores.  We discuss the necessary requirements to prove that a candidate is truly an embedded protostar in \S \ref{prove}, we apply these requirements to our candidate list in \S \ref{groups}, and we discuss the difficulties in including regions lacking good quality 70 \um\ data in \S \ref{need70}.  We discuss several general results of this work in \S \ref{discussion}.  Finally, we present our conclusions in \S \ref{conclusions}.

\section{Observations and Data Reduction}\label{observations}

The \emph{Spitzer} c2d data have been published in several papers focusing on individual regions in the survey.  We summarize here the observation strategy and data reduction method used by c2d; greater detail on individual regions can be found in the references given below.

\subsection{Observations}

\subsubsection{Molecular Clouds}

The c2d project obtained complete $3.6-160$ \um\ \emph{Spitzer} maps of five nearby, large molecular clouds.  The clouds (and approximate areas) surveyed include 1 deg.$^2$ of Serpens (Harvey et al. 2006; 2007a; 2007b), 2.5 deg.$^2$ total of the Lupus I, Lupus III, and Lupus IV clouds\footnote{For simplicity, we refer to the ensemble of these three portions of the Lupus Molecular Cloud Complex as simply ``Lupus''.} (Chapman et al. 2007; Mer\'{i}n et al. 2008), 1 deg.$^2$ of Chamaeleon II (Young et al. 2005; Porras et al. 2006; \alcala\ et al. 2008), 7 deg.$^2$ of Ophiuchus (Padgett et al. 2007), and 4 deg.$^2$ of Perseus (J\o rgensen et al. 2006; Rebull et al. 2007).  These clouds were chosen to span a wide range of low-mass star-forming environments (Evans et al. 2003).  We adopt distances of $260 \pm 10$, $150 \pm 20$, $200 \pm 20$, $150 \pm 20$, $178 \pm 18$, $125 \pm 25$, and $250 \pm 50$ pc for Serpens, Lupus I, Lupus III, Lupus IV, Chamaeleon II, Ophiuchus, and Perseus, respectively (see discussion of distances in the references listed above).  The observations of Serpens, Lupus, Chamaeleon II, Ophiuchus, and Perseus were obtained in Program IDs (PIDs) 174, 175, 176, 177, and 178, respectively.

Observations at 3.6, 4.5, 5.8, and 8.0 \um\ were obtained with the Infrared Array Camera (IRAC; Fazio et al. 2004).  Each region was observed in two epochs to allow for the detection and removal of asteroids, and each region was observed twice per epoch with small dithers between the observations.  The integration time per pointing was selected to be 12 s, providing an effective integration time per pixel of 10.4 s.  This strategy results in four total observations per region (2 epochs of 2 dithers each), resulting in a total integration time of $\sim$ 42 s per pixel.  All observations were obtained in the ``High Dynamic Range'' mode, which provides an additional short (0.6 s) exposure to enable photometry of bright sources saturated in the longer exposures.

Observations at 24, 70, and 160 \um\ were obtained with the Multiband Imaging Photometer for \emph{Spitzer} (MIPS; Rieke et al. 2004).  The observations consist of two epochs of fast scan maps with a spacing between adjacent scan legs of 240\as. The second epoch was offset by 125\as\ from the first in the cross-scan direction to fill in the 70 \um\ sky coverage that would have otherwise been missed due to detector problems, and by 80\as\ in the scan direction to minimize missing 160 \um\ data.  These mapping parameters result in two epochs of observations at 24 \um\ and one epoch at 70 and 160 \um, with total integration times per pixel of 30, 15, and 3 s, respectively.  Small gaps of missing coverage remain in the 160 \um\ maps.

For technical reasons relating to the fact that the length of a MIPS scan map leg must be chosen from a fixed list, the regions observed by MIPS are often significantly larger than those observed by IRAC in order to ensure full coverage.  The approximate areas quoted above are for the IRAC+MIPS overlap regions; in general, the areas observed by MIPS are a factor of $2-3$ larger.

\subsubsection{Small Dense Cores}

In addition to the five large molecular clouds, c2d also obtained small maps of 82 regions containing 95 small, dense cores.  These cores are selected on the basis of being nearby (within 500 pc), relatively small (generally less than 5\am, which corresponds to $\sim$0.7 pc at a distance of 500 pc), relatively isolated, and showing evidence for dense gas and dust (Evans et al. 2003).  The list of all 82 regions is given in Evans et al. (2007).  A discussion of the overall results of the \emph{Spitzer} observations of these cores and estimates of the distance to each core are presented in T. Huard et al. (2008, in preparation).  Although ``relative isolation'' was one of the criteria used to select these cores, we note here that many are not truly isolated in the sense that they are often loosely associated (both in projection on the sky and in velocity) with larger complexes.  Approximately 70\% of these cores were classified as starless prior to the launch of \emph{Spitzer}, based primarily on showing no association with \emph{IRAS} sources (Evans et al. 2003 and references therein).

Observations at 3.6, 4.5, 5.8, and 8.0 \um\ were obtained with IRAC in PID 139 in a manner nearly identical to the clouds.  The main differences are that the maps are much smaller (often only a single $5 \times 5$\am\ frame or a small map $< 30$\am\ on a side), the number of epochs observed (one or two) depends on the location of each core relative to the ecliptic, and short-exposure HDR images were only obtained for cores expected to contain bright sources based on previous surveys.  Cores observed in only one epoch featured 4 dithers instead of 2 so that all cores featured a total integration time per pixel identical to the clouds.

Observations at 24 and 70 \um\ were obtained with MIPS in PID 139; 160 \um\ observations were not obtained.  Unlike the clouds, the MIPS observations of the core regions were obtained in the pointed photometry mode.  The exposure time per pointing was 3 seconds at 24 \um\ and either 3 or 10 seconds at 70 \um, and small raster maps were obtained to match the IRAC maps.  As a result, the core regions, unlike the clouds, do not feature significantly larger areas observed with MIPS than with IRAC.  The exact integration time per pixel varies from region to region due to the differences in field and map size, but is generally $30-60$ s at 24 \um\ and $50-100$ s at 70 \um.

\subsection{Data Reduction}

The IRAC and MIPS images were processed by the \emph{Spitzer} Science Center (SSC), using their standard pipeline, version S13, to produce Basic Calibrated Data (BCD) images.  These images were then improved by the c2d Legacy Project to correct for artifacts remaining in the BCD images.  A complete description of the improvements made, as well as the source extraction and band-merging process summarized below, is given in Evans et al. (2007).

After correcting for artifacts, mosaics were produced using the MOPEX software provided by the SSC.  Photometry at $3.6-24$ \um\ was obtained using c2dphot, a modified version of DOPHOT (Schechter et al. 1993) that utilizes a digitized rather than analytic point source profile to better match the real \emph{Spitzer} data, as well as incorporating several other changes (a complete description of which are given in Harvey et al. 2006).  Sources that were not detected in all 5 bands from $3.6-24$ \um\ were band-filled, a process whereby the c2dphot source extractor was forced to obtain fluxes at the source position (known from the band(s) in which it was detected) in the bands for which the source was not detected in the original source extraction (see Evans et al. (2007) for details).  Finally, photometry at $3.6-24$ \um\ were band-merged into a final source catalog.

Photometry at 70 \um\ was obtained using the SSC's MOPEX point-source fitting package on filtered BCD data for faint sources ($\la$ 2 Jy) and unfiltered BCD data for bright sources ($\ga$ 2 Jy).  Sources extracted at 70 \um\ were band-merged into the $3.6-24$ \um\ catalogs described above, although this process is imperfect due to the significantly worse resolution at 70 \um\ ($\sim$18\as) compared to 24 \um\ ($\sim$6\as) and $3.6-8.0$ \um\ ($\sim$2\as) (see Evans et al. 2007 and \S \ref{need70}).

All of the above reduction and source extraction is part of the standard c2d pipeline.  However, this pipeline does not include source extraction at 160 \um\ due to problems with diffuse emission and low spatial resolution.  Thus, photometry at 160 \um\ for the five large clouds was obtained using the SSC's MOPEX point-source fitting package.  The source extraction lists were then inspected and sources that appeared to be extended dust ``clumps'' with little or no central condensation were removed (T. Brooke, private communication).  As a redundancy check, we also visually inspected all of the 160 \um\ images at the positions of the candidates identified in this work and performed aperture photometry on those that have associated 160 \um\ sources.  The aperture sizes and sky annuli were chosen to allow us to use one of the standard aperture corrections for 160 \um\ point sources determined by the SSC.  In general, we find very good agreement between the PSF and aperture photometry.  The ratio of PSF to aperture flux for sources extracted by both methods has a mean and median of 1.07 and 1.02, respectively, with a standard deviation of 0.2.  We thus add an overall 20\% systematic uncertainty into our photometry at 160 \um.  Both methods assume all 160 \um\ sources are point sources; fluxes of extended sources may thus be under-estimated.

\section{Identification of Candidate Embedded, Low-Luminosity Protostars}\label{id}

Generally speaking, we identify candidate embedded, low-luminosity protostars from the \emph{Spitzer} observations by selecting sources with rising fluxes from shorter to longer wavelengths and infrared luminosities \lir, calculated over the 2MASS and \emph{Spitzer} bands ($1.25-70$ \um), indicating \lint\ $\le 1$ \lsun\ (see below).  Throughout this paper we denote flux as $F_{\nu}$ ($F_{\nu} = \nu S_{\nu}$).  When we refer to the shape of the Spectral Energy Distribution (SED) of an object, it is always evaluated in terms of flux ($F_{\nu}$) rather than flux density ($S_{\nu}$).

At this stage, we emphasize completeness over reliability; many objects identified as candidates are likely either more evolved Young Stellar Objects (YSOs) no longer embedded in their dense cores, or background galaxies.  Separating these from the true embedded protostars is addressed in \S \ref{confirmation}.  We have assembled a set of seven criteria that must be met for a source to be identified as a candidate.  We list these criteria below, followed by an explanation of each one individually in \S \ref{criteria}.  We provide a detailed discussion of the motivation for the third criterion in \S \ref{importance70}, and investigate the results of relaxing the first criterion in \S \ref{relaxone}.  Finally, we discuss the need to visually inspect each candidate in \S \ref{visual} and present a general proof-of-concept in \S \ref{proofofconcept}.

The seven criteria that must be met are:
\begin{enumerate}
\item Detected at 24 \um\ with a signal-to-noise ratio (SNR) $\geq$ 3.
\item Located at a position observed by a minimum of two \emph{Spitzer} photometric bands between 3.6 and 70 \um.
\item Detected at 70 \um, unless confused or not observed at this wavelength.
\item Rising (or flat) SED between the longest IRAC wavelength at which the source is detected and 24 \um.
\item Rising (or flat) SED between 24 and 70 \um, unless confused or not observed at 70 \um.
\item Observed luminosity calculated using all detections ranging from $1.25-70$ \um\ of \lir\ $\leq 0.5$ \lsun.
\item Not classified as a candidate galaxy in the classification method of Harvey et al. (2007b) unless $Log(P_{gal}) \leq -1.25$, where $P_{gal}$ is the un-normalized probability of being a galaxy.
\end{enumerate}

\subsection{Selection Criteria}\label{criteria}

We impose the first criterion simply because 24 \um\ observations are convenient for determining whether or not a source features a rising SED in the wavelength region covered by our \emph{Spitzer} observations ($3.6-70$ \um).  First of all, 24 \um\ is located approximately halfway between 3.6 and 70 \um\ in log space, making it useful for determining the general shape of the SED over the wavelengths observed by \emph{Spitzer}.  Additionally, the diffraction limited resolution of \emph{Spitzer}, $\theta = 1.2 \frac{\lambda}{D}$, is a factor of 2.92 better at 24 \um\ than 70 \um.  Thus, sources clustered close together (at least in projection on the sky) can often be resolved into individual sources at 24 \um\ even when they are confused at 70 \um, providing reliable flux information at a wavelength beyond the $3.6-8.0$ \um\ region covered by IRAC and better enabling us to evaluate the shape of the SED.  We discuss the effects of relaxing this criterion in \S \ref{relaxone}.

The second criterion ensures that we have flux information available at a minimum of one other \emph{Spitzer} wavelength besides 24 \um.  The MIPS observations of the five large clouds were obtained in the scan map observing mode and often cover significantly larger areas than the IRAC observations of the same regions for technical reasons.  Furthermore, there are minor differences between the exact areas observed at 24 and 70 \um, resulting in small regions that are observed at 24 \um\ but not at any other wavelength.  Since it is impossible to determine the shape of the SED of any source observed at only one wavelength, all sources in these ``MIPS-24 only'' regions are removed from consideration.  This issue is not as significant in the regions with dense cores, where the MIPS data were obtained in the pointed observations mode and often cover very similar areas to the IRAC observations.

The third criterion is imposed because 70 \um\ is a crucial wavelength for determining \lint\ of embedded, low-luminosity protostars.  We will return to this point in greater detail below.

The fourth and fifth criteria select sources with rising or flat SEDs from shorter to longer wavelengths.  Simple, 1-D models of protostars still embedded within their cores show that they should essentially always feature rising SEDs both between the IRAC wavelength regime ($3.6-8.0$ \um) and 24 \um\ and between 24 and 70 \um\ (e.g., Young et al. 2005).  Two-dimensional models that include the effects of outflow cavities show that the exact shape of the $3.6-70$ \um\ SED depends on the inclination of the source (e.g., Whitney et al. 2003a; Whitney et al. 2003b; Robitaille et al. 2006).  In particular, embedded protostars observed at nearly pole-on inclinations exhibit much flatter SEDs over these wavelengths since more of the protostellar emission can escape through the outflow cavities instead of being reprocessed to longer-wavelength emission by the envelope.  While this flattening of the SED can be very significant, even at the most extreme inclinations the models predict essentially flat rather than falling SEDs at these wavelengths.  We return to this point in \S \ref{proof_enoch}.

We thus require each source to have an increasing or constant flux between the longest IRAC wavelength at which the source is detected and 24 \um, and between 24 and 70 \um.  If a source is not detected at any IRAC wavelength, it is considered to pass the fourth criterion.  We take the photometric uncertainties into account; a source may actually have a decreasing flux between the two wavelengths considered as long as it is consistent with being at least flat given the uncertainties.  Not all sources have 70 \um\ detections; we will return to these sources at the end of this section, but for now, we keep any source that meets the first rising criterion but has no flux information at 70 \um.  We do not place any rising requirements on the IRAC fluxes themselves since the effects of geometry (Whitney et al. 2003a; Whitney et al. 2003b), solid-state ice features (Boogert et al. 2004), and shocked emission from outflows (e.g., Noriega-Crespo et al. 2004) greatly complicate the exact shape of the $3.6-8.0$ \um\ SED.

The sixth criterion separates out the low-luminosity sources, those of interest to this study, from the more luminous sources.  As described in \S \ref{intro}, the internal luminosity of an embedded, low-luminosity protostar is not an observable quantity and is usually derived from radiative transfer modeling.  Here we investigate whether or not we can estimate \lint\ without the need to construct detailed models of each source.  We have identified 11 embedded protostars from the literature that have been observed with \emph{Spitzer} by c2d and have accurate internal luminosity determinations from radiative transfer models.  We list these 11 objects in Table \ref{lintlspitzertab}, along with \lir\ calculated from the observations and \lbol\ and \lint\ derived from the published models.

\begin{figure}[t]
\plotone{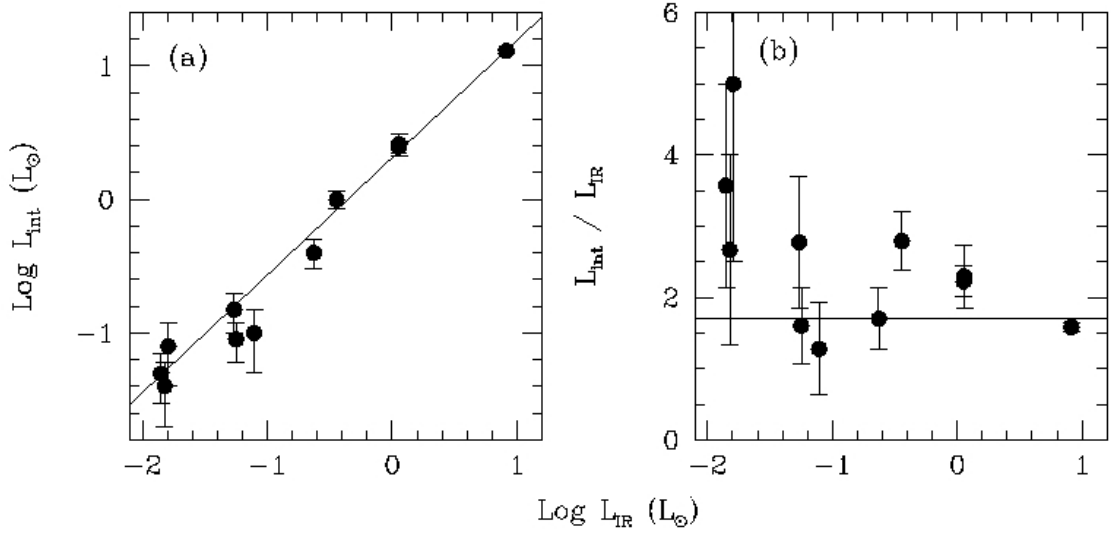}
\caption{\label{fig_lintlspitzer}(a): Log \lint\ vs. Log \lir\ for the objects listed in Table \ref{lintlspitzertab}.  The solid line shows the results of a linear least-squares fit in log-log space; it has a slope of 0.88 and a y-intercept of 0.32.  (b): \lint$/$\lir\ vs. Log \lir\ for the same objects.  The solid line shows the average ratio of 1.7, weighted by the uncertainties in \lint.}
\end{figure}

Figure \ref{fig_lintlspitzer}a plots \lint\ vs. \lir\ in log-log space for these 11 objects; a linear correlation is clearly observed.  The solid line shows the results of a linear least-squares fit; it has a slope of 0.88 and a y-intercept of 0.32.  However, any linear correlation in a luminosity-luminosity plot is suspect since it may simply arise from plotting the square of the distance to the object versus itself.  Thus, Figure \ref{fig_lintlspitzer}b plots the ratio of \lint$/$\lir, a quantity that is independent of distance, vs. the log of \lir.  An approximately constant ratio is observed over several orders of magnitude of \lir, as would be expected if a true linear correlation exists between these quantities.  The solid line shows the average ratio of 1.7, weighted by the uncertainties in \lint.  Inverting this quantity, we find that the luminosity calculated between $1.25-70$ \um\ gives a result that is approximately half the true value of the internal luminosity.  Thus, to select sources with \lint\ $\la 1.0$ \lsun, we impose the requirement that \lir\ $\leq 0.5$ \lsun.  We emphasize that this relationship between \lint\ and \lir\ is only an approximation; Figure \ref{fig_lintlspitzer}b clearly shows there is some variation from source to source.  In fact, the lowest-luminosity sources may have a larger ratio of \lint$/$\lir\ than more luminous sources, although the uncertainties are large.  However, even if the ratio of \lint$/$\lir\ does increase for lower-luminosity sources, they will have \lir\ far below the cut of 0.5 \lsun\ and will thus be included in our sample.

The seventh criterion is related to the fact that background galaxies are difficult to separate from YSOs when analyzing \emph{Spitzer} data of star-forming regions.  A number of authors have recently shown that such galaxies often exhibit similar mid-infrared colors and SEDs to YSOs (e.g., Harvey et al. 2006; J\o rgensen et al. 2006).  In an effort to accurately separate out background galaxies from YSOs, Harvey et al. (2007b) developed a classification method that assigns to each object that can not be fitted by an extincted stellar photosphere an un-normalized ``probability'' of being a galaxy, $P_{gal}$.  The probabilities are assigned based on the position of each source in various color-color and color-magnitude diagrams, along with information about the nature of the source (point-like vs. extended) given by the c2d source extraction software (see \S \ref{observations} and references therein).  Based on its value of $P_{gal}$, Harvey et al. classified each source as a \emph{candidate} YSO or \emph{candidate} Galaxy.

\begin{figure}[t]
\plotone{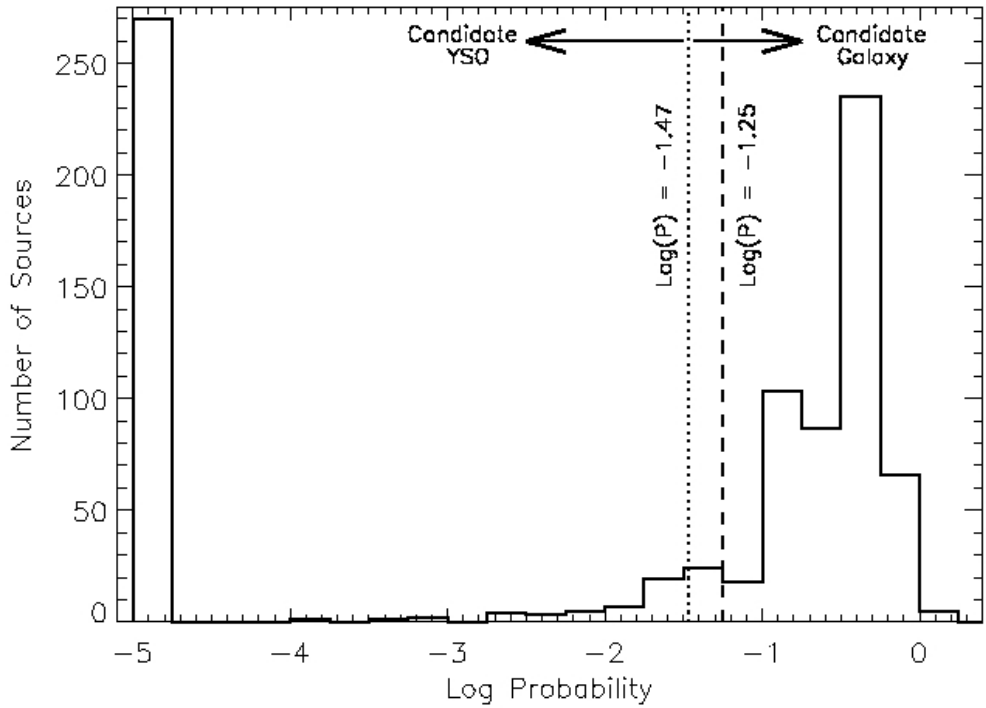}
\caption{\label{fig_probhist}Histogram of the un-normalized ``probability'' of being a galaxy, $P_{gal}$, using the method of Harvey et al. (2007b) for the 851 sources assigned such a probability in the ensemble of 82 regions with dense cores observed by c2d.  The dotted line shows $Log(P_{gal})=-1.47$ while the dashed line shows $Log(P_{gal})=-1.25$.}
\end{figure}

Figure \ref{fig_probhist} shows a histogram of $P_{gal}$ for all 851 sources assigned a probability in the ensemble of 82 regions with dense cores observed by c2d.  The dotted line shows $Log(P_{gal})=-1.47$ while the dashed line shows $Log(P_{gal})=-1.25$.  This figure is similar to Figure 4 of Harvey et al. (2007b), which plots the distribution of $P_{gal}$ for all sources assigned a probability in the c2d observations of the Serpens Molecular Cloud.  As noted by Harvey et al., there are two distinct peaks in probability space:  one at $Log(P_{gal})=-5.00$ and one at $Log(P_{gal}) \sim -0.4$, with a tail connecting the two.  In a similar analysis on a \emph{Spitzer} dataset of extragalactic objects, Harvey et al. found only the peak at $Log(P_{gal}) \sim -0.4$ and not the other peak or the tail connecting the two.  Based on this, they set the boundary between candidate YSO and candidate galaxy at $Log(P_{gal})=-1.47$, such that objects in the peak at $Log(P_{gal})=-5.00$ and the tail between the two peaks are considered candidate YSOs while objects in the peak at $Log(P_{gal}) \sim -0.4$ are considered candidate Galaxies.  However, as is evident both from Figure \ref{fig_probhist} of this paper and Figure 4 of Harvey et. al, there is likely some overlap in probability space between the tail of objects considered candidate YSOs and the peak of objects considered candidate galaxies.  Some of the objects near the boundary of $Log(P_{gal})=-1.47$, but slightly into the regime of galaxies, may in fact still be YSOs.  Furthermore, there was no galaxy in the \emph{Spitzer} dataset of extragalactic objects considered by Harvey et al. with $Log(P_{gal}) \leq -1.25$.  Thus, for all the objects that pass the first six criteria, we reject all those classified as candidate galaxies based on this classification method except for those with $-1.47 \leq Log(P_{gal}) \leq -1.25$.  Many of these intermediate objects are likely to be galaxies, but we include them as candidates at this stage for the sake of completeness.  Separating out these false candidates is the subject of \S \ref{confirmation}.

\subsection{Importance of 70 \um\ Data}\label{importance70}

\begin{figure}[t]
\plotone{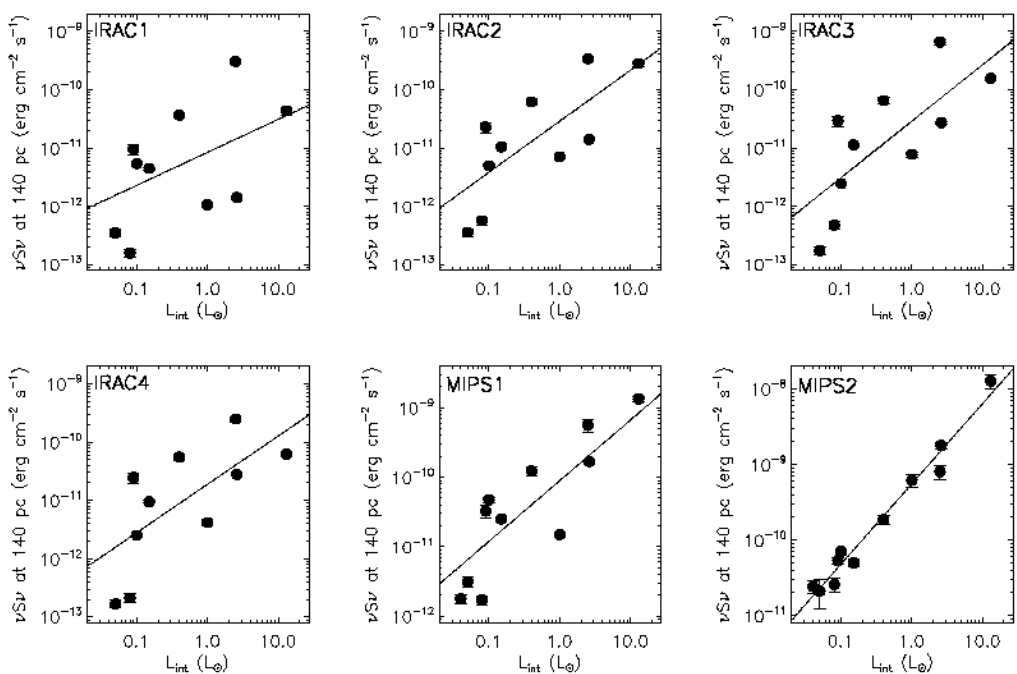}
\caption{\label{fig_fluxlintobs}$F_{\nu}$ (normalized to 140 pc) vs. \lint\ for the 11 embedded protostars listed in Table \ref{lintlspitzertab}.  From left-to-right, top-to-bottom:  IRAC1 (3.6 \um), IRAC2 (4.5 \um), IRAC3 (5.8 \um), IRAC4 (8.0 \um), MIPS1 (24 \um), and MIPS2 (70 \um).  The lines represent the results of linear least-squares fits at each wavelength.  The parameters of the fits are given in Table \ref{tab_fluxlintfits}.}
\end{figure}

We now return to the issue of detection at 70 \um.  This is a crucial wavelength for determining \lint\ for embedded, low-luminosity protostars.  Radiative transfer models of these objects are strongly constrained by this wavelength since the flux at 70 \um, $F_{70}$, is mostly unaffected by the details of the source geometry and presence or absence of a circumstellar disk that significantly affect the $3.6-24$ \um\ SED.  $F_{70}$ is also quite unaffected by the amount of external heating from the ISRF that is important at wavelengths $\ga 100$ \um\ (e.g., Dunham et al. 2006).  To examine this more fully, we again use the sources listed in Table \ref{lintlspitzertab} observed with \emph{Spitzer} by c2d and possessing accurate determinations of \lint\ through radiative transfer models.  Figure \ref{fig_fluxlintobs} plots $F_{\nu}$, normalized to 140 pc, for all six wavelengths observed by \emph{Spitzer} vs. \lint\ for these 11 sources, along with the results of linear least-squares fits to the data in log-log space.  These fits show that a correlation exists between the flux at each wavelength and \lint; in general terms, more luminous sources emit more energy at all wavelengths.  However, the correlation is clearly strongest at 70 \um, weaker at 24 \um, and very poor at $3.6-8.0$ \um.  To quantify this, at each wavelength we calculate $\chi_r^2$, a reduced chi-squared for the fit:

\begin{equation}
\chi_r^2 = \frac{1}{n-p}\displaystyle\sum_{i=0}^n \frac{(y_i-mx_i-b)^2}{\sigma_i^2} \quad ,
\end{equation}

where $n$ is the number of datapoints ($n=11$), $p$ is the number of free parameters ($p=2$), $y_i$ and $x_i$ are the log of the x and y values for each datapoint, $\sigma_i$ is the size of the uncertainty in log space, and $m$ and $b$ are the slope and y-intercept, respectively, derived from the fit in log-log space.  The values for $m$ and $b$, along with their statistical uncertainties from the fit and the results of the $\chi_r^2$ calculations, are presented in Table \ref{tab_fluxlintfits}.  We find $\chi_r^2=3$ at 70 \um, $\chi_r^2=85$ at 24 \um, and $\chi_r^2 > 100$ for all four IRAC wavelengths, in agreement with the qualitative conclusions from above.  

These results suggest that a direct correlation exists between $F_{70}$ and \lint.  However, this analysis is based on only 11 data points, and furthermore, many of the radiative transfer model determinations of \lint\ listed in Table \ref{lintlspitzertab} are based on the same simple, 1-D modeling setup that, among other things, uses a very simple method of including a disk in a 1-D model (Butner et al. 1994).  We do not expect this to be significant, as previous work has concluded that even these simple, 1-D models accurately constrain \lint\ (e.g., Dunham et al. 2006), but to test the validity of these results, we attempt to reproduce them by running a grid of two-dimensional models that are more physically realistic than those on which many of the values of \lint\ in Table \ref{lintlspitzertab} are based.

We used the two-dimensional, axisymmetric, Monte Carlo dust radiative transfer code RADMC (Dullemond \& Dominik 2004) to compute the radiative transfer and calculate the emergent SEDs of embedded, low-luminosity protostars and look for correlations between \lint\ and $F_{\nu}$ in the \emph{Spitzer} bands.  The density structure assumed for the models consists of a protostellar disk and a rotationally flattened, infalling envelope with an outflow cavity, and is described in greater detail in Crapsi et al. (2008).  The details of the models are nearly identical to those of Crapsi et al.; we do not describe them here except to note that they have been tuned to their specific purpose within this work in the following ways:

\begin{itemize}
\item The protostar has a fixed temperature of 3000 K and a random luminosity in the range of $0.03 - 10$ \lsun.
\item The disk has a fixed size of 100 AU, variable mass between $10^{-3}$ and $10^{-5}$ \msun, and fixed flaring parameters (a scale height $H$ of 20 AU at the outer radius and then decreasing as $H \propto r^{9/7}$, corresponding to the self-irradiated passive disk of Chiang \& Goldreich [1997]).
\item The envelope has a fixed radius of 14,000 AU, a variable mass in the range of $1-10$ \msun, and a variable centrifugal radius in the range of $100-900$ AU.
\item The envelope is externally heated by the ISRF, for which we adopt that of Black (1994), modified at the ultraviolet wavelengths to reproduce the ISRF of Draine (1978).  This Black-Draine ISRF is then attenuated by dust with properties given by Draine \& Lee (1984) to simulate being embedded in a parent cloud and multiplied by a scale factor to account for the fact that the ISRF is not uniform everywhere, where the number of magnitudes of visual extinction and scale factor are chosen randomly.
\end{itemize}

The observed SED was obtained by raytracing the density, temperature, and scattering structures calculated by RADMC at five different inclinations: 20, 35, 50, 65, and $80\mbox{$^{\circ}$}$ from the axis of symmetry.  Fluxes were integrated inside apertures comparable to the resolution of \emph{Spitzer}:  2\as\ for wavelengths shorter than 10 \um, 6\as\ for wavelengths in the range of $10-40$ \um, and 20\as\ in the wavelength range of $40-100$ \um.

\begin{figure}[t]
\plotone{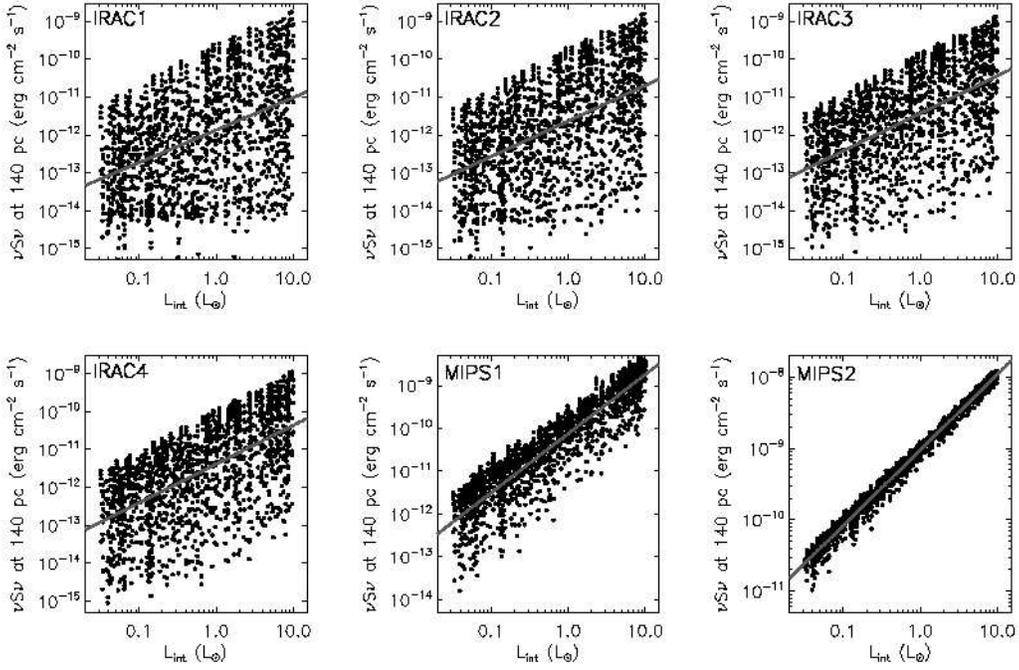}
\caption{\label{fig_fluxlintmodel}Same as Figure \ref{fig_fluxlintobs}, except for the 1460 model SEDs calculated from the grid of 2-D radiative transfer models rather than the 11 embedded protostars listed in Table \ref{lintlspitzertab}.}
\end{figure}

Following this description, a grid of 292 models were run and 1460 SEDs were obtained (one at each inclination).  The results are shown in Figure \ref{fig_fluxlintmodel} and, qualitatively, appear to confirm the correlation between $F_{70}$ and \lint.  The parameters derived from linear least-squares fits in log-log space for $F_{\nu}$ vs. \lint\ for the models are given in Table \ref{tab_fluxlintfits}.  It is clear from these models that there is in fact a tight correlation between $F_{70}$ and \lint\ (most of the scatter seen in the MIPS2 panel of Figure \ref{fig_fluxlintmodel} is due to inclination effects).  In fact, not only do the observations and models both show this correlation, but they also agree remarkably well on the quantitative relationship (within 3\% for the slope and within 1\% for the y-intercept) considering that, to date, only 11 sources have been modeled by us and are thus available to examine this from an observational standpoint.  Since $F_{70}$ is an observable quantity, the correlation we find between $F_{70}$ and \lint\ can be used to estimate \lint\ for sources which either lack sufficient data to constrain radiative transfer models or have such data but have not yet been modeled.  Using the results of the linear least-square fit to the modeled $F_{70}$ vs. \lint\ in log-log space, we derive this estimate to be
\begin{equation}\label{eq_lint}
\lint\ = 3.3 \times 10^8 \, F_{70} \, ^{0.94} \, \lsun \quad ,
\end{equation}
where $F_{70}$ is in cgs units (erg cm$^{-2}$ s$^{-1}$) and is normalized to 140 pc.

\begin{figure}[t]
\plotone{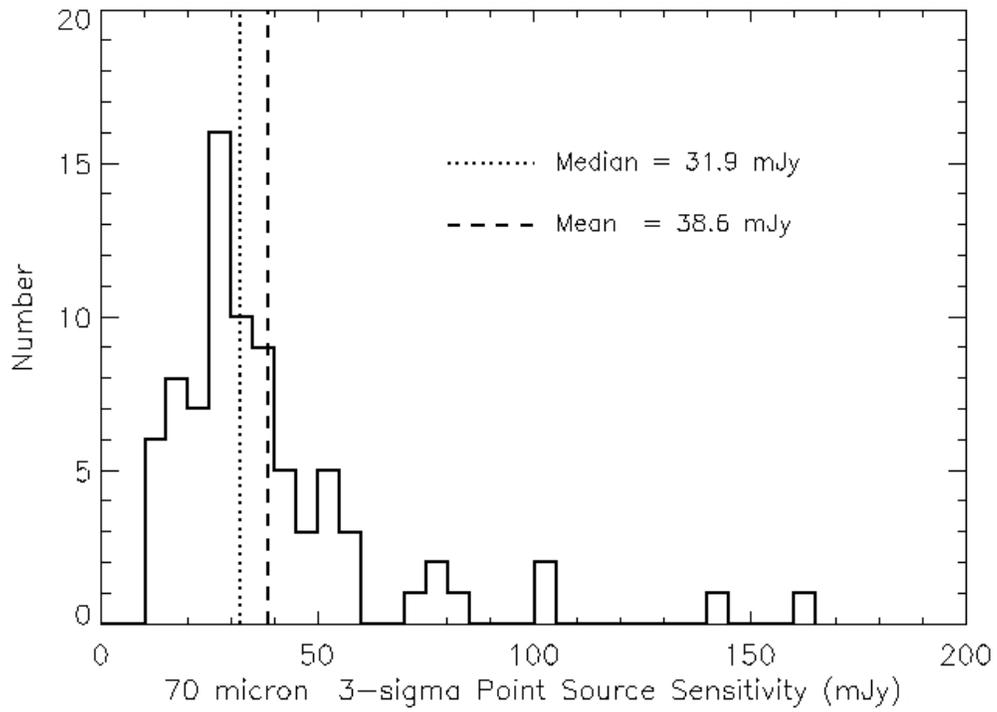}
\caption{\label{fig_mips2upperlimit}Histogram of the 70 \um\ 3$\sigma$ point source sensitivity for the 82 regions with dense cores observed by c2d.  The distribution has a mean and median of 38.6 mJy and 31.9 mJy, respectively.}
\end{figure}

Since a detection at 70 \um\ is crucial for constraining radiative transfer model determinations of \lint\ and gives an immediate, direct estimate of \lint, we require each source to be detected at 70 \um\ unless it is not observed at this wavelength or is located too close to another source detected at 70 \um\ to be resolved into a separate source (see \S \ref{need70}).  Thus, with these two exceptions, we effectively require candidates to be above the detection threshold at 70 \um.  This requirement sets the fundamental limit for our luminosity sensitivity.  Figure \ref{fig_mips2upperlimit} plots a histogram showing the 70 \um\ 3$\sigma$ point source sensitivity for each of the 82 regions with dense cores observed by c2d.  This sensitivity was derived by first calculating the standard deviation of the background intensity for each region, $\sigma_{\rm sky}$, in the image units of MJy sr$^{-1}$.  Approximating the 70 \um\ \emph{Spitzer} PSF as a Gaussian with $\theta_{\rm FWHM}$ equal to the diffraction limited resolution of $\sim 18$\as, this background intensity is then converted into units of mJy beam $^{-1}$ as follows:
\begin{equation}
\sigma_{\rm sky} \, \rm{(mJy} \, \rm{beam)}^{-1} = 1 \times 10^9  \, \left (\frac{\pi \, \theta_{\rm FWHM}^2}{4 \, ln 2}\right ) \, \sigma_{\rm sky} \, \rm{(MJy} \, \rm{sr)}^{-1}
\end{equation}
This gives the 3$\sigma$ point source sensitivity for each region, since the total flux density of a point source is equivalent to its flux density per beam.

\begin{figure}[t]
\plotone{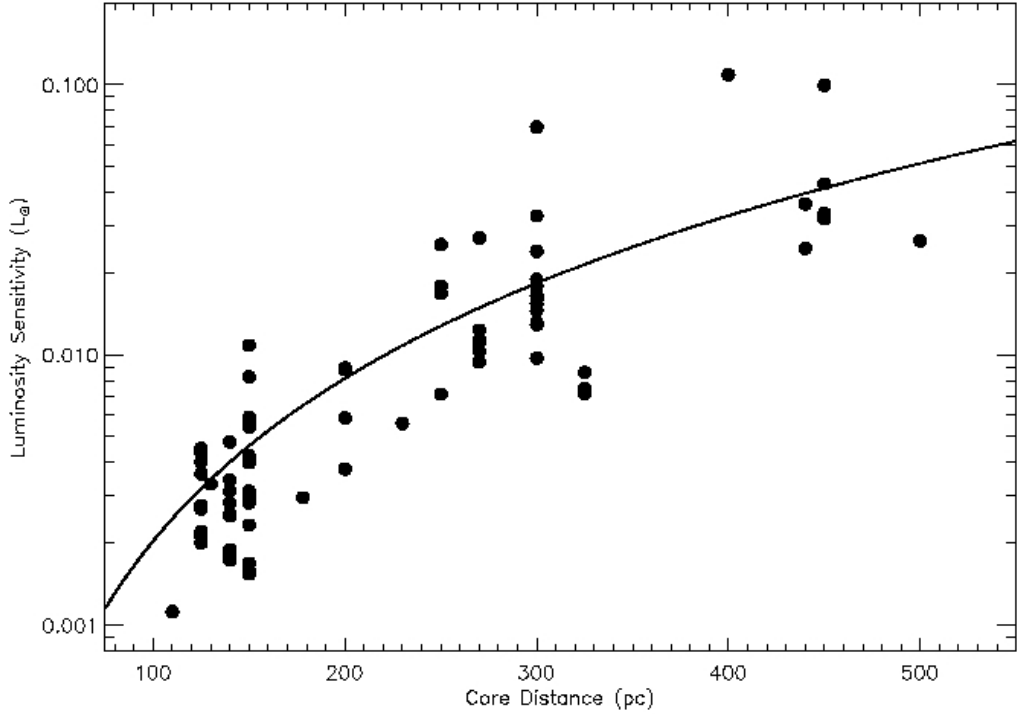}
\caption{\label{fig_lumlimit}Log of the luminosity sensitivity limit for embedded protostars in the 82 regions with dense cores observed by c2d ($L$) vs. the distance to each core ($d$), calculated by translating the 70 \um\ 3$\sigma$ point source sensitivity for each core into a luminosity sensitivity using the correlation found between $F_{70}$ and \lint\ and then scaling from 140 pc to the distance to the core.  The solid line shows the relation \lint\ $= 4 \times 10^{-3}$ $(d/140 \, \rm{pc})^2$ \lsun.}
\end{figure}

This 70 \um\ point source sensitivity can be directly translated into a luminosity sensitivity using Equation \ref{eq_lint}.  From this relationship, the mean 70 \um\ 3$\sigma$ point source sensitivity of 38.6 mJy translates into a luminosity sensitivity of $\sim 4 \times 10^{-3}$ \lsun\ at 140 pc.  Figure \ref{fig_lumlimit} plots the luminosity sensitivity limit for each of the 82 regions vs. the distance to the region, calculated by translating the 70 \um\ 3$\sigma$ point source sensitivity into a luminosity sensitivity using Equation \ref{eq_lint} and then scaling from 140 pc to the distance to the region.  The solid line shows the relation  \lint\ $=4 \times 10^{-3}$ $(d/140 \, \rm{pc})^2$ \lsun.  There is some scatter between the line and the actual data, caused by the variation in 70 \um\ point-source sensitivity from one region to the next, but this relation clearly provides an accurate estimate of the luminosity sensitivity.  On average, the c2d observations of regions with dense cores are sensitive to embedded protostars with internal luminosities \lint\ $\geq 4 \times 10^{-3}$ $(d/140 \, \rm{pc})^2$ \lsun.  This same argument holds true for the c2d observations of the clouds, as the 70 \um\ cloud maps reach similar depths as the cores.  This is a factor of 25 better than the sensitivity of IRAS quoted in \S \ref{intro}.

Applying these criteria results in the identification of 673 candidate embedded, low-luminosity protostars:  106 in the ensemble of dense core regions, 196 in Perseus, 112 in Chamaeleon II, 153 in Lupus, 57 in Ophiuchus, and 49 in Serpens.

\subsection{Relaxing the First Criterion}\label{relaxone}

The first of the seven criteria described above for selecting candidates from the \emph{Spitzer} c2d observations is a detection at 24 \um.  As discussed in \S \ref{criteria}, this is an important wavelength for determining whether or not a source features a rising (or flat) SED from $3.6-70$ \um.  However, it is possible that a source of interest has no detection at 24 \um.  For example, it might simply not be covered at 24 \um\ if it is located close to the edge of an observed region where the coverage between bands is not uniform, or it might be so deeply embedded that it is only detected at 70 \um.  The latter could be particularly interesting sources, either very young, deeply embedded protostars or perhaps even first hydrostatic cores, which are short-lived, hydrostatic objects in a phase between the initial collapse of a dense core and the formation of a Class 0 protostar predicted by theory but not yet found by observations (e.g., Boss \& Yorke 1995).

Thus, to make sure no sources are missed by imposing this first criteron, we search for any source not detected at 24 \um\ but detected at 70 \um.  We then apply the sixth criterion, \lir\ $\leq 0.5$ \lsun, to select the low luminosity sources that are the focus of this work.  This gives an additional 53 candidates:  13 in the ensemble of dense core regions, 7 in Perseus, 17 in Chamaeleon II, 10 in Lupus, 2 in Ophiuchus, and 4 in Serpens.  We add these to the candidates identified above for a total of 726 candidates.  However, only 4 of these 53 additional candidates survive the visual inspection described in \S \ref{visual}, and none of these surviving 4 are particularly strong candidates (see \S \ref{visual}).

\subsection{Visual Inspection}\label{visual}

The final step in the identification of candidates is visual inspection.  For each of the 726 sources identified above, we examined the images at the six wavelengths observed by \emph{Spitzer} and the source SED.  Throughout the course of this inspection, we found five reasons to reject candidates:  (1) The source is obviously a resolved galaxy; (2) The 70 \um\ detection is not real; (3) The source SED is not consistent with that of an embedded protostar; (4) The source is located in a region of nebulosity that produces a false infrared excess at the longer wavelengths; and (5) There is no real source.

The first reason is due to the fact that a source must meet certain requirements to be classified as a candidate YSO or candidate galaxy in the method of Harvey et al. (2007b).  Specifically a source must be detected in both epochs of observations (or in the single epoch if only one epoch was observed) with SNR $\geq$ 3 in all 4 IRAC bands and the 24 \um\ MIPS band.  Any source outside the area of overlap between all 5 bands, or simply not detected in any one of these bands, does not get evaluated by the method of Harvey et al. and thus can not be classified as a candidate YSO or galaxy.  This results in many galaxies making the candidate list.  Visual inspection removes those large and/or close enough to be resolved by IRAC; all others remain candidates at this stage.

The second reason relates to the fact that the exact sensitivity of the 70 \um\ c2d observations is a complicated function of the position within the scan map (clouds) or pointed observation (core regions), since the exact number of frames at a given position depends on the technical details of how \emph{Spitzer} executes observations.  In regions with fewer frames, often near the edges of maps, noise spikes are sometimes falsely identified as sources; we remove these from consideration.  Of the 53 candidates identified by searching for sources detected at 70 \um\ but not at 24 \um, 49 were removed for this reason, and the 4 that remain are all sources with questionable 70 \um\ detections that simply weren't obvious enough false detections to remove.

\begin{figure}[t]
\plotone{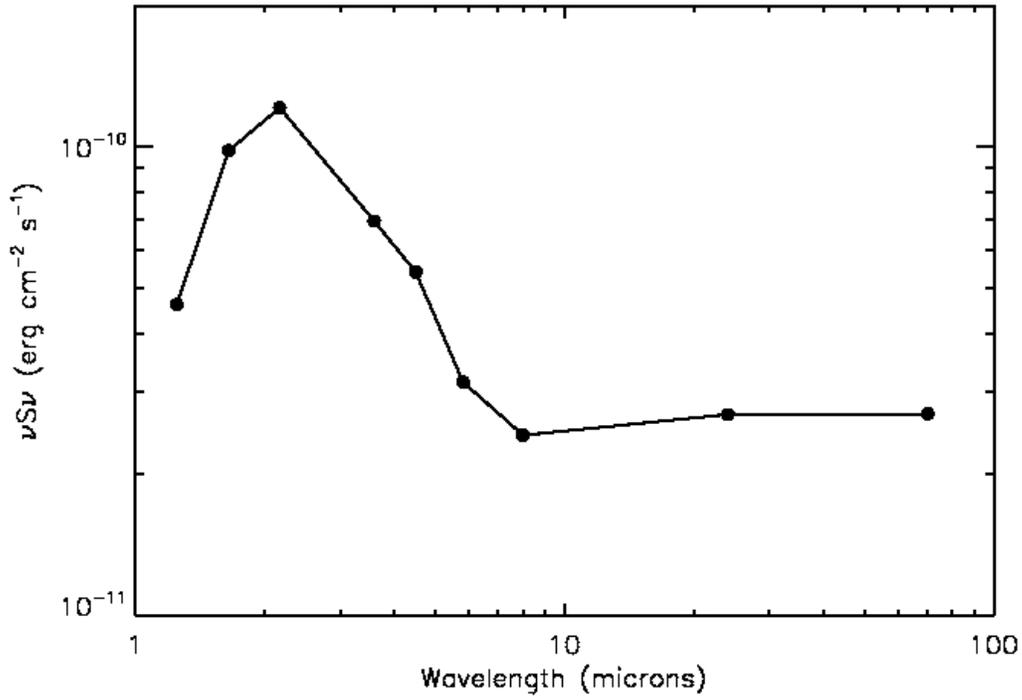}
\caption{\label{fig_sedinconsistent}SED of SSTc2d J032856.64+311835.6, a source in Perseus that is representative of candidates that pass all 7 selection criteria described in \S \ref{criteria} but feature SEDs inconsistent with being embedded sources.  This particular source is SSS 108, a previously known pre-main sequence object with a 2 \um\ excess indicative of a circumstellar disk.  Objects with these types of SEDs are rejected (see \S \ref{visual} for further information).}
\end{figure}

The third reason relates to the details of how rising sources were selected.  Figure \ref{fig_sedinconsistent} shows the SED of SSTc2d J032856.64$+$311835.6, a source in Perseus representative of those removed for this reason.  It is detected at all 6 \emph{Spitzer} wavelengths, features a rising flux between 8 and 24 \um, features a flat, but slightly rising, flux between 24 and 70 \um, has \lir\ $=0.37$ \lsun, and is classified as a candidate YSO by the classification method of Harvey et al. (2007b).  Thus, it fulfills all the criteria for selecting candidates discussed in \S \ref{criteria}.  However, it clearly emits more energy in the near-infrared ($1.25-2.17$ \um) than at 24 and 70 \um.  Comparison with 2-D models of sources (e.g., Whitney et al. 2003a; Whitney et al. 2003b; Robitaille et al. 2006) shows that, even at the most extreme pole-on inclinations, the effects of outflow cavities are unlikely to result in an observed SED more luminous in the near-infrared than at 24 and 70 \um.  A more likely explanation for objects featuring such SEDs is that they are more evolved YSOs surrounded by circumstellar disks but no longer embedded within dense cores.  Indeed, a search of the SIMBAD\footnote{Available at:  http://simbad.u-strasbg.fr/simbad/} database shows that this is SSS 108, which Aspin et al. (1994) and Aspin (2003) conclude is a pre-main sequence star with a small thermal excess at 2 \um\ (i.e., from a disk).  The nature of these objects with large infrared excesses at 24 and 70 \um\ compared to $3.6-8.0$ \um\ could potentially be very interesting; indeed, these SEDs appear similar to those of a sample of YSOs with disks featuring large inner holes presented by Brown et al. (2007).  Sources that meet our selection criteria but are more luminous in the near-infrared than at 24 and 70 \um\ may be interesting objects, but they are not relevant to this study and are thus removed from our sample.

The final two reasons for rejecting sources upon visual inspection both relate to the bandfilling process described in \S \ref{observations}.  The first reason is that stars are often detected only in the first two IRAC bands, due both to the decreasing \emph{Spitzer} sensitivity at longer wavelengths and the fact that \emph{Spitzer} observes the Rayleigh-Jeans portion of stellar SEDs.  Some stars detected only in these two bands are located, at least in projection, in regions of nebulosity.  For these objects, the bandfilling process can assign source fluxes in bands beyond IRAC2 that are contaminated by this extended nebulosity, creating false excesses and thus falsely rising SEDs.  These are obvious in that there is no real point-source in the longer wavelength images.  The second reason is that \emph{Spitzer} images of star-forming regions often exhibit copious amounts of nebulous, diffuse emission due to scattered light, shocked emisison from outflows, thermal emission from hot dust, PAH emission, etc.  In regions where this diffuse emission becomes clumpy, the source extraction may falsely detect a point-source at one wavelength that then gets band-filled at all other wavelengths.  Some of these fake sources may pass all 7 criteria for selection, but they are obvious upon visual inspection in that there is no point source present in the images.  Both types of sources are removed from consideration.

After removing sources for the reasons described above, we are left with 218 candidates: 49 in the ensemble of dense core regions, 70 in Perseus, 16 in Chamaeleon II, 42 in Lupus, 22 in Ophiuchus, and 19 in Serpens.  The source number from this work, group number indicating likelihood of being an embedded object (see \S \ref{confirmation}), \emph{Spitzer} source name, position, \lir, and distance to and name of the nearest source in the \emph{IRAS} point-source catalog are given for each candidate in Tables \ref{cores_candidates} $-$ \ref{serp_candidates}.  For the candidates in the dense cores, Table \ref{cores_candidates} also list the name of the core region in which each candidate is located and the assumed distance to this region; these distances have been derived by reddening or association with regions with known distances (T. Bourke, private communication).  The full list of distances to core regions will be presented by T. Huard et al. (2008, in preparation).  We emphasize here that these objects are only \emph{candidates}.  We have chosen to emphasize completeness over reliability, with the result being that many of these 218 candidates are in fact galaxies or more evolved YSOs no longer embedded in their dense cores.  Separating the wheat from the chaff is the focus of \S \ref{confirmation}.

\section{Proof of Concept}\label{proofofconcept}

In \S \ref{criteria}, we presented a set of criteria to select embedded, low-luminosity protostars from \emph{Spitzer} c2d observations.  We imposed certain restrictions on the shape of the $3.6-70$ \um\ SED in order to pick out the embedded objects (the fourth and fifth criteria), and justified their validity by comparing to both 1-D and 2-D models of embedded sources.  Here we further examine the ability of these criteria to identify embedded protostars.

\subsection{Comparison to Known Class 0 Objects}\label{proof_class0}

Class 0 objects were added to the original infrared spectral slope classification system of Lada (1987) by \andre\ et al. (1993) as objects younger and more deeply embedded than Class I objects and thus not detected in the infrared.  However, with the advent of sensitive infrared facilities such as \emph{Spitzer}, both Class 0 and I objects are detected in the infrared and are often indistinguishable from one another based on their infrared spectral slope $\alpha$ (e.g., Enoch 2007; Enoch et al. 2008a, in preparation; Kauffmann et al. 2008).  Robitaille et al. (2006) discussed the distinction between SED ``class'' and physical ``stage''.  Physically, the distinction between a Stage 0 and Stage I object is that a Stage 0 object still has more than half of its total mass in the circumstellar envelope (\andre\ et al. 1993).  Various evolutionary indicators can be used to separate embedded objects into Class 0 and Class I objects, which should correlate with the true physical stage (see \S \ref{evolstatus}).

Froebrich (2005) searched the literature and compiled a database of photometry of all known Class 0 objects.  This database is available online\footnote{Available at http://astro.kent.ac.uk/protostars/index.html} and is updated as new Class 0 objects are discovered and new photometry is published.  If we remove our sixth criterion, which selects only those embedded protostars with \lint\ $\la$ 1 \lsun, our criteria should select all of the objects in this database with available photometry between $3.6-70$ \um.

Since both \emph{IRAS} and the \emph{Infrared Space Observatory (ISO)} obtained photometry at 25 and 60 \um, there are many objects in this database with photometry at these two wavelengths but not at 24 and 70 \um\ (in other words, observed with \emph{IRAS} and/or {ISO}, but not with \emph{Spitzer}).  Thus, we first selected all sources in the database that had a detection at 60 or 70 \um; 80 of the 135 sources in the database meet this criterion.  We then imposed the same criteria as described in \S \ref{criteria} and \ref{relaxone}, omitting the sixth criterion as described above, repacing 24 and 70 \um\ with 25 and 60 \um, respectively, when necessary, and omitting the second, third, and seventh criteria because these require information not available from the database.

Of the 80 objects evaluated, 79 pass our criteria and are selected as embedded objects.  The one source that does not is IRAS 12500$-$7658, located in the Chamaeleon II molecular cloud.  According to the Class 0 database, this source has $F_{24} = 8.58\ee{-11} \pm 0.05\ee{-11}$ erg s$^{-1}$ cm$^{-2}$ and $F_{70} = 6.98\ee{-11} \pm 0.43\ee{-11}$ erg s$^{-1}$ cm$^{-2}$, and thus does not have a rising (or flat) flux between 24 or 70 \um, even when the uncertainties are taken into account.  This source was observed by c2d, and in fact these values are taken from Young et al. (2005), who presented MIPS data on the Chamaeleon II cloud using a preliminary version of the c2d pipeline.  In the final c2d source catalog for Chamaeleon II, which includes, among other things, improved source photometry and more accurate photometric uncertainties (Evans et al. 2007), this source has $F_{24} = 7.64\ee{-11} \pm 0.71\ee{-11}$ erg s$^{-1}$ cm$^{-2}$ and $F_{70} = 6.60\ee{-11} \pm 0.64\ee{-11}$ erg s$^{-1}$ cm$^{-2}$.  Using these new, more accurate values, IRAS 12500$-$7658 does indeed have a rising (or flat) SED between 24 or 70 \um\ when the uncertainties are taken into account.  Thus, all 80 sources are recovered; we identify all of the Class 0 objects.

\subsection{Comparison to Embedded Objects in Perseus, Serpens, and Ophiuchus}\label{proof_enoch}

The above exercise demonstrates that our criteria successfully identify the youngest, most heavily embedded protostars.  However, we are interested in the full population of embedded objects, not just the Class 0 objects.  To test our method for Class I objects, and to verify the above results, we compare our objects to those identified by Enoch (2007) and Enoch et al. (2008a, in preparation).  They searched for \emph{Spitzer} sources associated with dense cores identified by large-scale, uniform 1.1 mm Bolocam dust continuum emission surveys of Perseus, Serpens, and Ophiuchus.  They did not restrict themselves to the low-luminosity population, and since they had good-quality millimeter data available for all three clouds, they started by searching for sources associated with known dense cores.  Our work is complementary in that all sources identified here should be identified by Enoch et al., but we can extend our analysis to regions lacking such millimeter data.

\begin{figure}[t]
\plotone{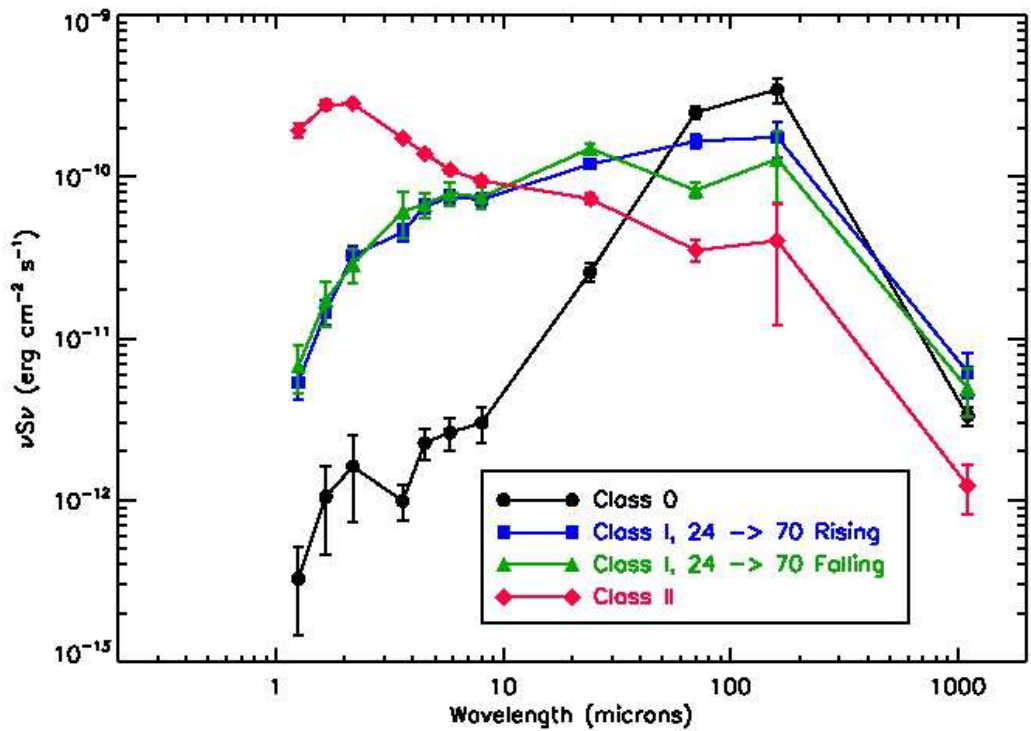}
\caption{\label{fig_enoch_avg_seds}Average SEDs, weighted by 1/\lbol, for the embedded protostars identified by Enoch (2007) and Enoch et al. (2008, in preparation) in Perseus, Serpens, and Ophiuchus based on a comparison between 1.1 mm Bolocam dust continuum emission maps and the \emph{Spitzer} c2d maps.  The average Class 0 SED is shown in black, the average SED for Class I sources with rising (or flat) fluxes from 24 to 70 \um\ is shown in blue, the average SED for Class I sources with decreasing fluxes from 24 to 70 \um\ is shown in green, and the average SED for the Class II objects identified by Enoch et al. is shown in red.}
\end{figure}

Comparing the results of both searches shows, in general, good agreement.  Enoch et al. separated their objects into Class 0 and Class I objects based on the bolometric temperature of each source (see \S \ref{lboltbol}).  After removing sources with \lir\ $\ge$ 0.5 \lsun\ from their sample, we identify all of their Class 0 sources.  This confirms the above result that we identify all Class 0 objects.  We also identify most of their Class I sources, but find a population of objects in their sample not identified as candidates in our work.  Inspection of this population of objects shows that they are not selected by our criteria because they do not have rising fluxes from 24 to 70 \um.  Out of the 37 Class I objects with \lir\ $\le$ 0.5 \lsun\ in the Enoch (2007) and Enoch et al. (2008a) sample, 15 (40\%) are not selected by our criteria for this reason.  These 15 objects range in luminosities from 0.03 $\leq$ \lir\ $\leq$ 0.4 \lsun.  Figure \ref{fig_enoch_avg_seds} shows the average SEDs, weighted by 1/\lbol\ so that the average is not dominated by the most luminous sources, for all sources identified by Enoch (2007) and Enoch et al. (2008a, in preparation), divided into four bins:  Class 0, Class I with rising (or flat) fluxes from 24 to 70 \um, Class I with decreasing fluxes from 24 to 70 \um, and Class II.

Following Whitney et al. (2003b), Enoch et al. (2008a, in preparation) use the bolometric temperature of each object to go beyond the simple Class 0/I division and divide their sources into Early Class 0, Late Class 0, Early Class I, and Late Class I objects.  They present the average SED for each of these four classes of objects; it is the Late Class I objects (300 K $\le$ \tbol\ $\le$ 650 K) that feature decreasing fluxes between 24 and 70 \um.  Thus, a possible conclusion is that our selection criteria miss the older Class I objects nearing the end of the embedded phase.

However, as described in \S \ref{criteria}, we developed our selection criteria based on comparison to models of protostars still embedded within their dense cores.  This includes both Class 0 and Class I protostars.  Two-dimensional models featuring outflow cavities predict rising or flat fluxes from 24 to 70 \um\ at even the most extreme pole-on inclinations, (e.g., Whitney et al. 2003a; Whitney et al. 2003b; Robitaille et al. 2006).  Indeed, Enoch et al. (2008a, in preparation) compare their average SEDs for Early and Late Class 0 and I objects to the two-dimensional models of each type of object presented by Whitney et al. (2003b) and note that their Late Class I objects differ from the models in that, among other things, they decrease in flux from 24 to 70 \um.

The nature of these objects is uncertain.  The overall similarities between the average SEDs of the Class I objects rising and falling between 24 and 70 \um\ (see Figure \ref{fig_enoch_avg_seds}) and the fact that these objects are associated with dense cores suggests they truly are embedded objects.  On the other hand, no published models of embedded protostars predict fluxes that decrease between these two wavelengths.  Robitaille et al. (2006) noted that the SED class and physical stage of an object do not always agree.  For example, a Stage II object located behind a large amount of material (such as a dense core) could feature a Class I SED.  It is possible that these objects with decreasing fluxes from 24 to 70 \um\ are not truly embedded protostars but heavily extincted Stage II objects located behind dense cores.  A future, detailed study of these objects is needed, but at present, their true physical stage remains unknown.  Our criteria will identify the majority of embedded, low-luminosity objects (100\% of the Class 0 objects; $\geq$ 60\% of the Class I objects).  Even if the objects with decreasing fluxes from 24 to 70 \um\ are in fact embedded objects, including them in our sample would not significantly change any of our results.

\section{Confirming the Candidates}\label{confirmation}

Many of the 218 candidates identified in \S \ref{id} are not truly embedded protostars.  The visual inspection described above removed a large number of false candidates, cutting our list down from 726 candidates to 218.  However, we only removed sources from consideration if it was obvious that they fell into one of the five types of false candidates discussed in \S \ref{visual}.  Candidates for which there was any ambiguity were kept, and at least some of these are expected to be false candidates.  Furthermore, as discussed above, a source is only evaluated by the classification method of Harvey et al. (2007b) for identifying candidate YSOs and candidate galaxies if it is detected in both epochs of observations (or in the single epoch if only one epoch was observed) with SNR $\geq$ 3 in all 4 IRAC bands and the 24 \um\ MIPS band.  Any source not covered by all 5 of these bands is ineligible for classification as either candidate galaxy or candidate YSO, and instead receives one of several general source types based on the shape of the SED (Evans et al. 2007).  Many galaxies thus slip into our candidate list because they pass criterion 7 of \S \ref{criteria}, whereas they would have been classified as candidate galaxies and rejected according to this criterion if they had been observed in all 5 bands.  Only those that are large and/or close enough to resolve are removed by the visual inspection described in \S \ref{visual}.  This is particularly relevant for the five large clouds, since the MIPS coverage is substantially larger than the IRAC coverage.

\begin{figure}[t]
\plotone{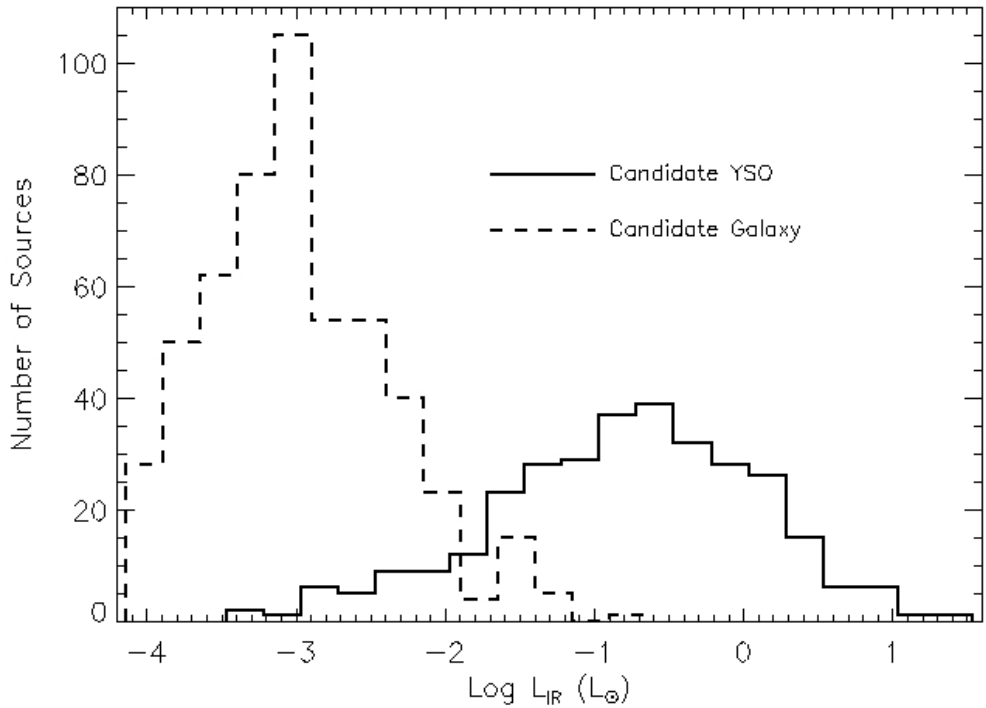}
\caption{\label{fig_lumhist}Distribution of luminosities for the 851 objects in the 82 regions with dense cores observed by c2d classified as either candidate YSOs or candidate galaxies by the classification method of Harvey et al. (2007b).  The solid line shows the distribution for the sources classified as candidate YSOs while the dashed line shows the same for the sources classified as candidate galaxies.  Of the 604 out of 851 sources with \lir\ $\leq$ 0.05 \lsun, 518 ($\sim$ 86\%) are classified as candidate galaxies.}
\end{figure}

To demonstrate the extent of extra-galactic contamination expected, we note that 148 of the 218 candidates have \lir\ $\leq$ 0.05 \lsun, which roughly corresponds to \lint\ $\la$ 0.1 \lsun\ using the relation between the two derived in \S \ref{criteria}.  This suggests a very large number of VeLLOs exist; however, out of these 148 sources, 114 are not considered by the Harvey et al. (2007b) classification method because they do not meet the requirement of SNR $\geq$ 3 detections in all 5 bands.  As Figure \ref{fig_lumhist} demonstrates, the majority of these low-luminosity sources are expected to be galaxies.  Out of 851 sources in the 82 regions with dense cores observed by c2d that are classified as either candidate YSOs or candidate galaxies, 604 have \lir\ $\leq$ 0.05 \lsun, and 518 of these 604 ($\sim$ 86\%) are classified as candidate galaxies.  Thus, as many as 98 of the 114 candidates not considered by the Harvey et al. (2007b) classification method because they do not meet the requirement of SNR $\geq$ 3 detections in all 5 bands may be extra-galactic in nature.  Clearly, further effort is required in order to determine which of the 218 candidates are in fact embedded, low-luminosity protostars.

\subsection{How to Prove a Source is Embedded}\label{prove}

We examined both the \emph{Spitzer} data and complementary data for each of the 218 candidates to search for evidence proving they are truly embedded objects.  We divide this evidence into two ``levels of certainty'':

\begin{enumerate}
\item Evidence exists proving that a source is associated, in projection on the sky, with a region of high volume density.
\item Evidence exists proving that this soure is actually embedded within the dense region (to remove the possibility of chance alignment).
\end{enumerate}
As discussed below, data on molecular line and dust continuum emission, molecular outflows, extinction, and infrared nebulosity are used to evaluate these conditions.  We note here that it is much easier to prove the first level than the second.

\subsubsection{Evaluating the First Level}\label{prove1}

The first level is evaluated by observing the region in a tracer of dense material.  Such tracers include millimeter dust continuum emission, which typically traces regions with volume densities of $n \ga 10^4$ cm$^{-3}$ (e.g., Enoch et al. 2007), inversion transition lines of NH$_3$, which also trace regions with volume densities of $n \ga 10^4$ cm$^{-3}$ (e.g., Ho \& Townes 1983; Benson \& Myers 1989), and rotational transition lines of \nthp, which trace similar material as the NH$_3$ lines (Caselli et al. 2002).  We thus looked for associations with millimeter dust continuum, NH$_3$, or \nthp\ sources using a number of different surveys of the regions covered by c2d in these tracers, some of which were specifically designed as complementary surveys to c2d; we list these surveys below.  In addition, we searched the SIMBAD database for other such observations besides those listed below.  In all cases, a candidate is said to be associated (in projection) with a millimeter dust continuum, NH$_3$, or \nthp\ source if it is located within one beam of the peak position of the source.

The surveys we used for the candidates in the 82 regions with dense cores are as follows:  An on-going 350 \um\ dust continuum survey of cores with the Submillimeter High Angular Resolution Camera II (SHARC-II) at the Caltech Submillimeter Observatory (CSO; Wu et al. 2007; M. Dunham et al. 2008, in preparation), a 450 and 850 \um\ dust continuum survey of 38 cores with the Submillimeter Common-User Bolometer Array (SCUBA) at the James Clerk Maxwell Telescope (JCMT; Young et al. 2006a), a 1.2 mm dust continuum survey of 37 cores with the Max Planck Millimeter Bolometer (MAMBO) at the IRAM 30 m telescope (Kauffmann et al. 2008), a 1.2 mm dust continuum survey of 151 southern cores with the SEST Imaging Bolometer Array (SIMBA) at the Swedish ESO Submillimeter Telescope (SEST; K. Brede et al. 2008, in preparation), an \nthp\ (1-0) survey of 38 cores observed with the Five College Radio Astronomy Observatory (FCRAO; C. De Vries et al. 2008, in preparation), an \nthp\ (1-0) survey of 59 cores also observed with the FCRAO (Caselli et al. 2002), and observations of the $(J,K)=(1,1)$ and $(2,2)$ lines of NH$_3$ for 264 cores compiled from the literature by Jijina et al. (1999).

The surveys used for Perseus are a 1.1 mm dust continuum survey of 7.5 deg$^2$ with Bolocam at the CSO (Enoch et al. 2006), an 850 \um\ dust continuum survey of 3.5 deg$^2$ with SCUBA at the JCMT (Kirk et al. 2006), a 450 and 850 \um\ dust continuum survey of 3.0 deg$^2$ with SCUBA at the JCMT (Hatchell et al. 2005; Hatchell et al. 2007a), and the SHARC-II survey described above, which includes several cores in Perseus.  For Chamaeleon II, we used a 1.3 mm dust continuum survey of 36 YSOs in Chamaeleon I and II with the $^3$He-cooled bolometer system at the SEST (Henning et al. 1993), and the SIMBA survey of 151 southern cores described above, which includes several cores in Chamaeleon II. For Lupus, a 1.2 mm dust continuum survey of $\sim$625 arcmin$^2$ with SIMBA at the SEST (Tachihara et al. 2007) was used.  For Ophiuchus, we used a 1.1 mm dust continuum survey of 10.8 deg$^2$ with Bolocam at the CSO (Young et al. 2006b), a 1.2 mm dust continuum survey of 1.0 deg$^2$ with SIMBA at the SEST (Stanke et al. 2006), and an 850 \um\ dust continuum survey of 700 arcmin$^2$ with SCUBA at the JCMT (Johnstone et al. 2000).  Finally, for Serpens we used a 1.1 mm dust continuum survey of 1.5 deg$^2$ with Bolocam at the CSO (Enoch et al. 2007), a 450 and 850 \um\ dust continuum map of 120 arcmin$^2$ with SCUBA at the JCMT (Davis et al. 1999), and a 1.2 mm dust continuum map of $\sim$150 arcmin$^2$ with MAMBO at the IRAM 30 m telescope (Djupvik et al. 2006).

Observations tracing dense material are not always available at the positions of the candidates.  This is especially true for candidates in the southern clouds and cores.  As an alternative, observations indicating high column density can be used to evaluate the first level.  A positive result does not guarantee this level is satisfied, as high column densities do not necessarily indicate high volume densities, but it does increase the likelihood that the source is an embedded object.

We utilize two indicators of high column density:  extinction maps and absorption against the mid-infrared background.  For the extinction maps, we used the maps created by the c2d team.  We give a brief description of them here; a more complete description can be found in Evans et al. (2007).  A line-of-sight extinction estimate is produced by the c2d pipeline for each source with a $1.25-24$ \um\ SED consistent with that of an extincted stellar photosphere, and these line-of-sight estimates were then convolved with uniformly spaced Gaussian beams.  Maps of different resolution were produced by using Gaussian beams with different FWHM.  We used the highest resolution maps available for each cloud (180\as\ for Perseus, 120\as\ for Chamaeleon, 120\as\ for Lupus I, 90\as\ for Lupus III, 90\as\ for Lupus IV, 240\as\ for Ophiuchus, and 90\as\ for Serpens); extinction maps are not yet available for the cores.  We did not attempt to derive quantitative estimates of the column density from the extinction maps; instead, we examined the maps at the positions of the candidates and identified any candidate located in a localized region of higher extinction as showing evidence for high column density.

\begin{figure}[t]
\plottwo{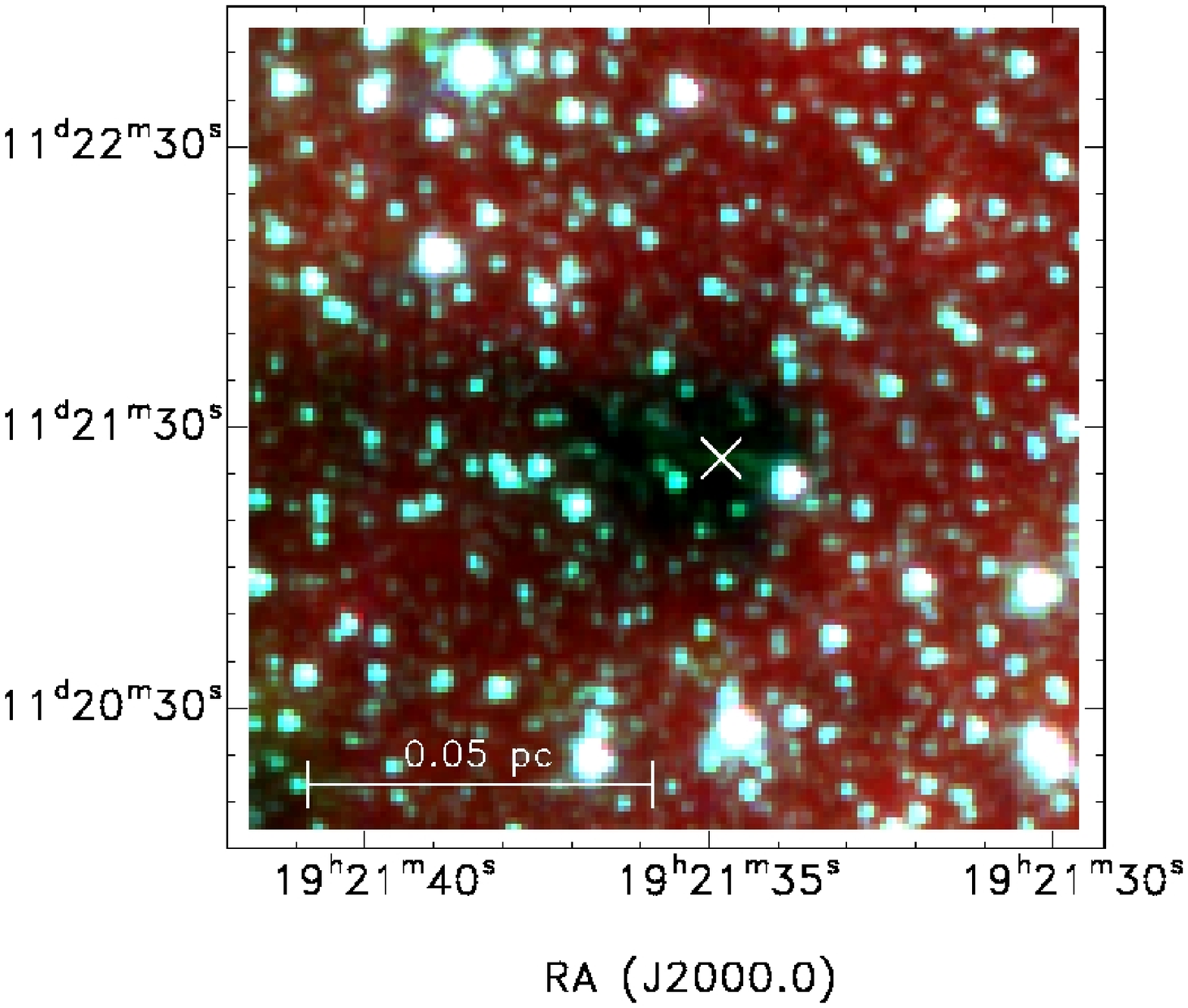}{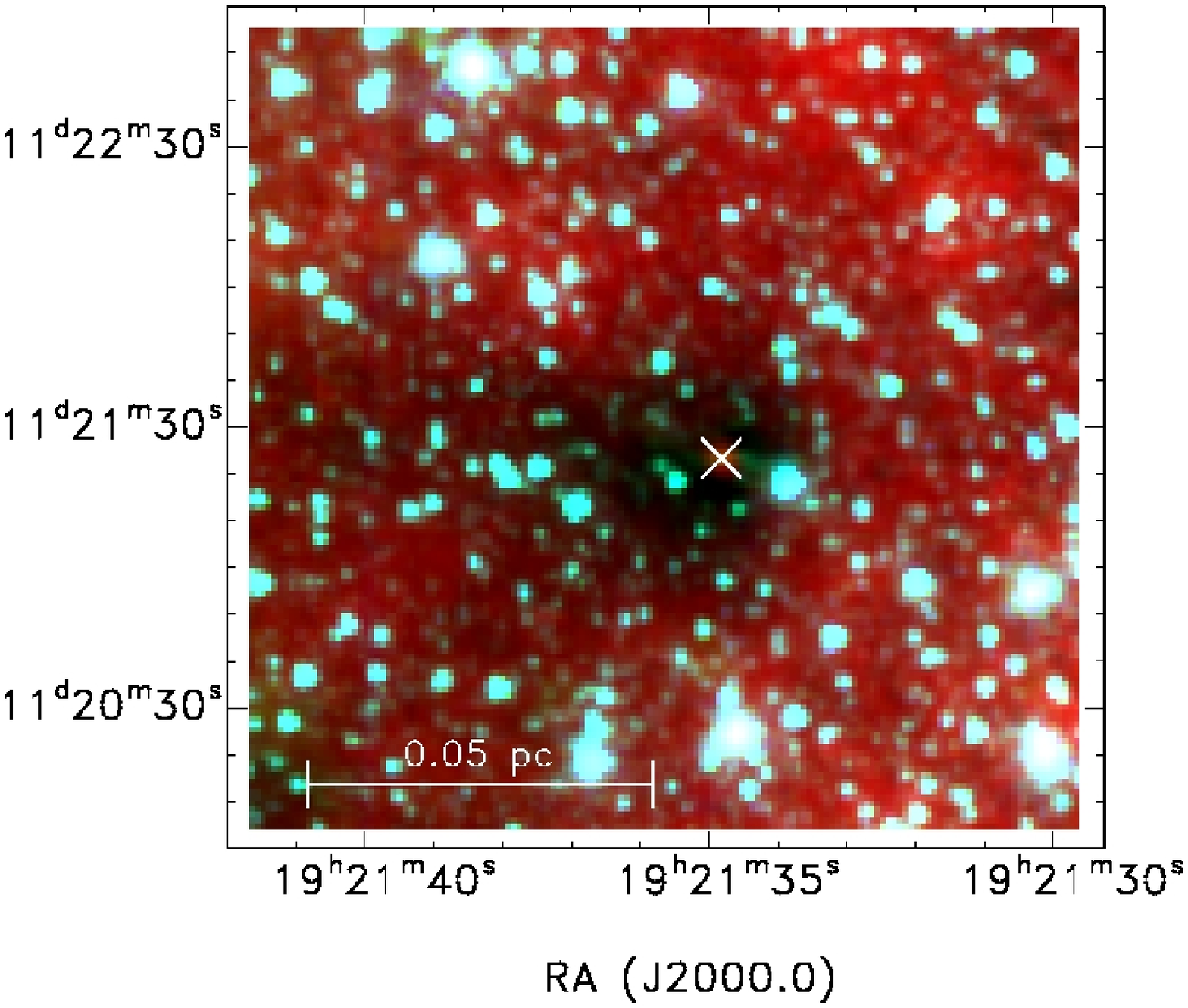}
\plottwo{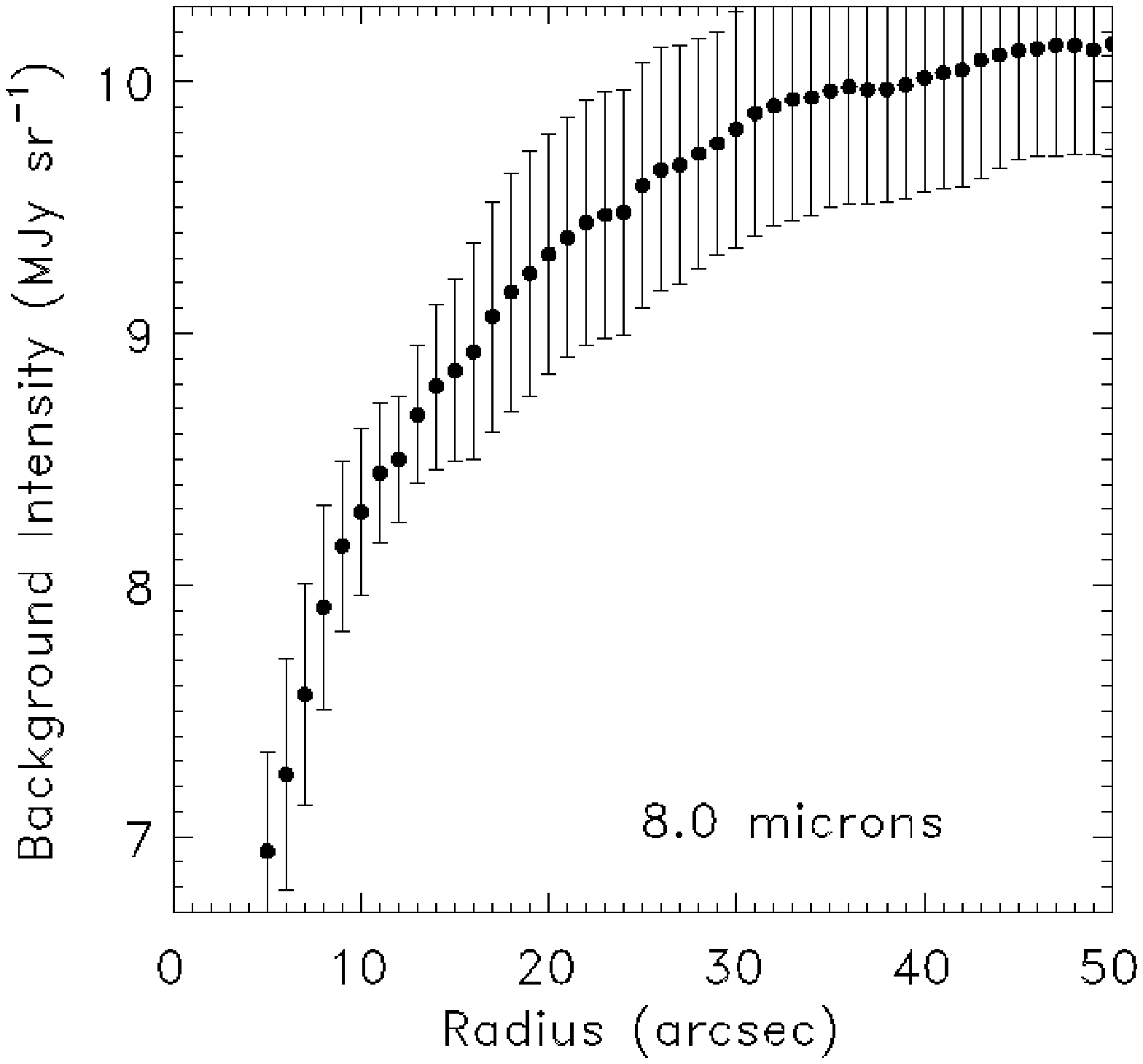}{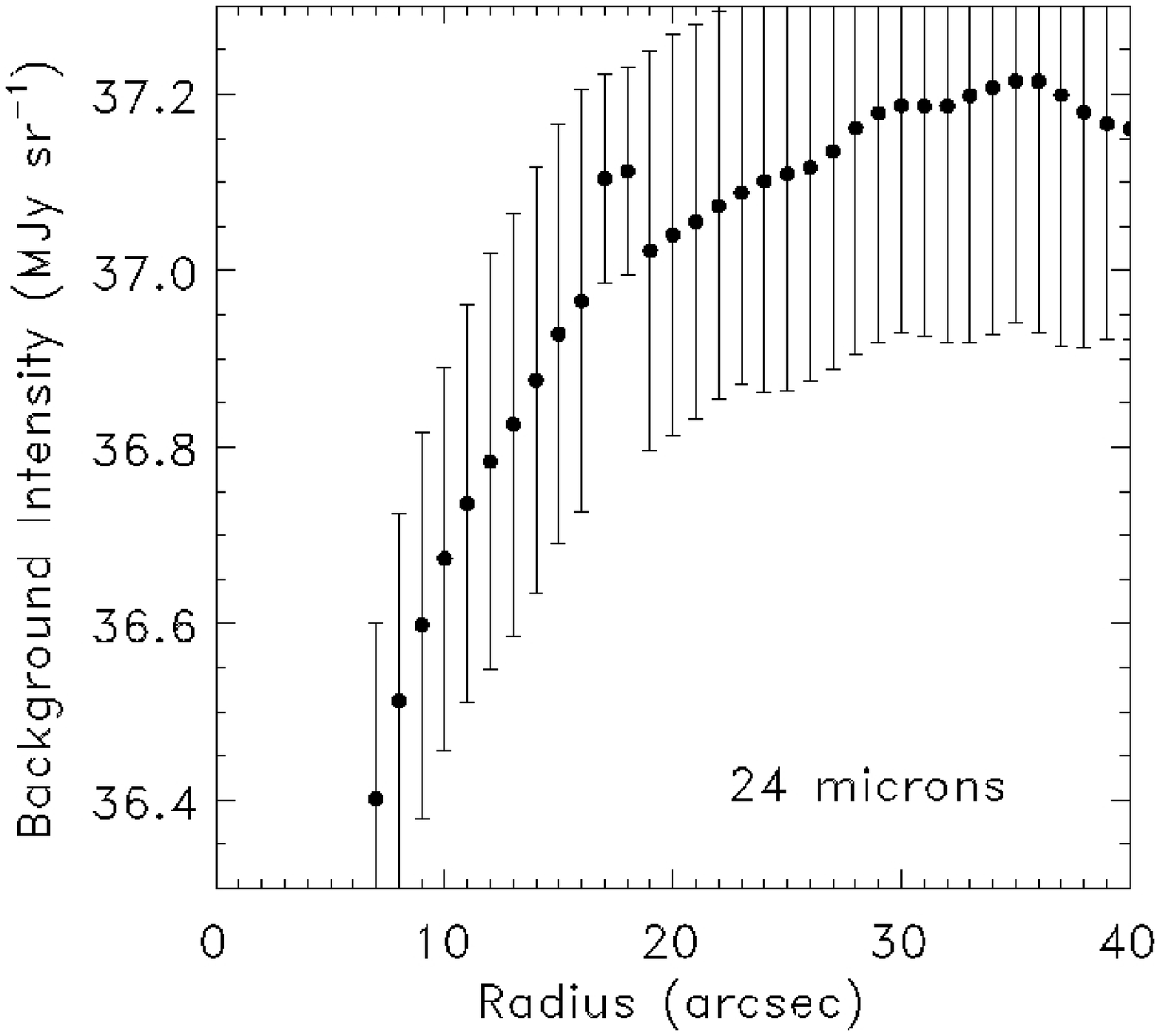}
\caption{\label{fig_darkcore}\emph{Top Left}:  Three-color image of L673-7 comprised of IRAC 1 (3.6 \um; \emph{blue}), IRAC 2 (4.5 \um; \emph{green}), and IRAC 4 (8.0 \um; \emph{red}).  The white cross marks the position of Source 031 from this work, which is not detected by IRAC; it is clearly seen to fall within a dark core at 8.0 \um.  \emph{Top Right}: Same as top left, except the red now shows the MIPS1 (24 \um) image.  A very red source is seen at the position of Source 031.  \emph{Bottom}:  Radial profiles of the background intensity at 8 and 24 \um, centered at the position of Source 031.  The background intensities are calculated in concentric annuli, each with a radius of 2\as.  The radius on the x-axis is the mid-point of the annulus (for example, at a radius of 10\as\, the background intensity was calculated in an annulus with its inner edge located 9\as\ from the source and its outer edge located 11\as\ from the source).  The calculations start at radii of 5 and 7\as\, respectively, for 8.0 and 24 \um.}
\end{figure}

For the absorption against the mid-infrared background, we looked for ``dark cores'' or ``shadows'' in the 8 and 24 \um\ images.  We examined both the images themselves and radial profiles of the background intensity.  Figure \ref{fig_darkcore} shows an example of a dark core at the position of the candidate in L673-7 (source 031).  While a quantitative analysis of the extinction profile and column density through the core is possible (e.g., Stutz et al. 2007), it is beyond the scope of this paper.  Here we simply identified any candidates that appeared to be located within such a dark core as showing evidence for high column density.  An important note is that, in both techniques, negative results do not rule out the possibility of high column density.  Very small, localized column density enhancements may not be seen in the extinction maps due to beam dilution, and variations in the strength of both the background and foreground emission will have a significant impact on the presence or absence of absorption against the mid-infrared background.  As a result, we consider cases where data are available to evaluate the existence of regions of high column density but return a negative result to be equivalent to cases where no such data are available.

\subsubsection{Evaluating the Second Level}\label{prove2}

The second level is primarily evaluated by searching the region around a source for a molecular outflow that is centered both spatially on the source and kinematically at the velocity of the dense core with which the source is associated.  Large-scale surveys of $^{12}$CO, the primary tracer of molecular outflows, with the necessary sensitivity, spatial resolution, and velocity resolution to detect molecular outflows are not as common as surveys of dust continuum.  In fact, the only such survey for the regions observed by c2d was a search for outflows in Perseus by Hatchell et al. (2007b), and even this was not an unbiased survey as they targeted known cores from a previous 850 \um\ dust continuum survey (Hatchell et al. 2005).  Thus, we searched the literature for each of the 218 candidates to find those with known molecular outflows.  Many candidates had no published $^{12}$CO observations, including several candidates in Perseus that were not part of the Hatchell et al. (2007b) survey.  Some of these were observed at the CSO as part of a search for molecular outflows, resulting in detections of outflows by us in L673-7 and L1251 (Sources 031, 044, and 045 from this work; M. Dunham et al. 2008, in preparation; J.-E. Lee et al. 2008, in preparation).  While a positive result will prove a source is embedded in a dense core, a negative result will not disprove this since the answer can depend on the available data.  As an example, L1014 (Source 038 from this work) showed no signs of driving an outflow from single-dish molecular line observations (Crapsi et al. 2005a), but a compact, low-velocity $^{12}$CO (2-1) outflow was discovered by Bourke et al. (2005) with the Submillimeter Array.  Thus, in evaluating whether each source shows evidence for being an embedded protostar, we consider cases where observations tracing molecular outflows exist but no outflow is detected to be equivalent to cases where no such observations exist.

We also use a 350 \um\ SHARC-II detection to satisfy the second level.  As with all submillimeter and millimeter observations of dust continuum emission, a detection with SHARC-II indicates high volume density.  However, unlike other observations of the dust continuum, Wu et al. (2007) concluded that, through a combination of temperature and instrumental effects, a SHARC-II 350 \um\ detection effectively always indicates that a core has a protostar embedded within it.

Many of the 218 candidates have not been observed either in $^{12}$CO or with SHARC-II at 350 \um.  Both a search for molecular outflows and an extension of the SHARC-II survey of dense cores presented by Wu et al. (2007) are on-going at the CSO (M. Dunham et al. 2008, in preparation), but in the meantime, this leaves us unable to evaluate the second level for many candidates.  Thus, we also examine the IRAC 2 (4.5 \um) image for each candidate to look for extended nebulosity or jets, suggesting the presence of molecular outflows and outflow cavities.  We specifically focus on 4.5 \um\ for two reasons:  the wavelength is short enough for scattered light off the edges of outflow cavities to often be visible, and the photometric band overlaps with emission lines of molecular hydrogen shocked by outflows (e.g., Noriega-Crespo et al. 2004).  Examples of this are shown for Sources 001 and 004 from this work in Dunham et al. (2006) and Bourke et al. (2006), respectively.  We do not separately identify ``extended nebulosity'' and ``jets''; instead we consider evidence for either to be the same thing.  While this is not strictly correct, it avoids the ambiguity sometimes inherent in assigning extended structure one label or the other.  Furthermore, both are likely indicators of outflow activity, the actual source property of relevance for this study.  A positive result does not guarantee that the second level is satisfied, but it does add to the likelihood that the source is an embedded object.  Once again, however, a negative result does not prove that the source is not embedded.  To again use L1014 (Source 038) as an example, it does not show any extended nebulosity or jet features in the 4.5 \um\ image (or any other \emph{Spitzer} image); only in deep near-infrared images is scattered light from an outflow cavity seen (Huard et al. 2006).  Thus, as above, we consider cases where 4.5 \um\ \emph{Spitzer} images are available but show no extended nebulosity and/or jets to be equivalent to cases where no such images are available.

\subsection{Assessing the Likelihood of Being Embedded}\label{groups}

Based on the results of evaluating the two levels of certainty described above, we divide the candidates into 6 groups of descending likelihood of being embedded protostars.  We list the groups in Table \ref{tab_groups} and describe them below:

\begin{enumerate}
\item Group 1 consists of candidates that are confirmed as embedded, low-luminosity protostars.  Existing observations confirm that sources in this group are embedded within regions of high volume density.
\item Group 2 consists of candidates that have a very high likelihood of being embedded, low-luminosity protostars.  Sources in this group are associated with regions of high volume density.  They show extended nebulosity and/or jets in 4.5 \um\ \emph{Spitzer} images, but they are not confirmed to be embedded within the high-density regions, either because no observations are available to evaluate this condition or such observations are available but give a negative result.
\item Group 3 consists of candidates that have a high likelihood of being embedded, low-luminosity protostars, although not as high as group 2 candidates.  Similar to group 2, group 3 candidates are associated with regions of high volume density but are not confirmed to be embedded within these regions.  Unlike group 2, however, group 3 candidates are not associated with extended nebulosity and/or jets at 4.5 \um, either because no such extended structure is detected or because 4.5 \um\ images are not available.
\item Group 4 consists of candidates that show evidence for being embedded, low-luminosity protostars but lack confirmation that either level of certainty is fulfilled.  Sources in this group have no available observations to evaluate whether or not they are associated with regions of high volume density and either no available observations or observations giving a negative result on whether or not they are embedded within regions of high volume density.  These sources either are associated with regions of high column density and show extended nebulosity and/or jets in 4.5 \um\ \emph{Spitzer} images (group 4a), are only associated with regions of high column density (group 4b), or only show extended nebulosity and/or jets in 4.5 \um\ \emph{Spitzer} images (group 4c).
\item Group 5 consists of candidates that might be embedded, low-luminosity protostars, but show no sign of being such objects in their limited available data.  Sources in this group have no available observations to evaluate whether or not they are associated with regions of high volume density and either no available observations or observations returning a negative result for proving they are embedded within dense regions, evaluating their association with regions of high column density, and identifying nebulosity or jets at 4.5 \um.
\item Group 6 consists of candidates that are most likely not embedded, low-luminosity protostars.  Observations show that sources in this group are not associated with regions of high volume density, with the important caveat that most dust continuum emission surveys used to search for regions of high volume density are only complete to cores with masses $\geq 0.1-1.0$ \msun\ (e.g., Enoch et al. 2007).

\end{enumerate}

The last column of Table \ref{tab_groups} lists the total number of the 218 candidates in each of the groups.  Table \ref{tab_divided_groups} lists, for all 218 candidates, sorted by group, the source number, cloud or core region in which the source is located, \lir, the status of 160 \um\ observations for this source, and the SIMBAD name of the source, where applicable.  If a candidate is not associated with an infrared or radio point source in SIMBAD, we give the name of the nearest extended dense core within a search radius of 1\am.  Note that this does not guarantee the candidate is embedded within this core.  

We find that 149 of the 218 candidates are in either group 5 or group 6 and thus show no evidence for being embedded, low-luminosity protostars.  A fraction of these 149 candidates are likely real embedded sources that simply lack complementary data of sufficient quality to prove this; these sources should be re-evaluated as additional surveys of these regions become available.  Most, however, are likely either galaxies or more evolved sources no longer embedded within dense cores, consistent with our earlier claim that as many as 98 of the candidates may be extra-galactic in nature (\S \ref{confirmation}).  This leaves 69/218 candidates that, at minimum, show some evidence for being embedded objects.  We are able to obtain complete SEDs for 50 of these 69 candidates; these SEDs are listed in Table \ref{tab_seds}.  We consider only these 50 sources in the rest of the paper.  Of the 19 remaining candidates, 16 are located close enough to other sources that they cannot be separated into individual sources at 70 \um; these are discussed below.  Two sources, 027 and 202, are not covered by the 70 \um\ c2d observations and have no obvious association with \emph{IRAS} 60 \um\ sources.  As 70 \um\ is an essential wavelength for this study (see below), we do not list these objects in Table \ref{tab_seds}.  The last remaining source is Source 132, which is identical to Source 017 as it was observed by both IRAC and MIPS in the map of the dense core DC303.8$-$14.2 and by MIPS in the map of Chamaeleon II.  We thus remove Source 132 from consideration to avoid duplication.

\subsection{Effects of Confused 70 \um\ Data}\label{need70}

Good quality 70 \um\ data are essential to this study.  Photometry at this wavelength is important for distinguishing between SEDs of embedded protostars and more evolved YSOs no longer embedded in dense cores (\S \ref{criteria}; \S \ref{proof_enoch}), and for deriving accurate protostellar internal luminosities (\S \ref{importance70}).  However, the MIPS 70 \um\ PSF is $\sim$ 18\as, compared to 6\as\ at 24 \um\ and 2\as\ at $3.6-8.0$ \um.  Embedded protostars spaced by more than 6\as\ but less than 18\as\ will be resolved into individual sources at 24 \um\ and thus identified by our selection criteria, but confused at 70 \um.  Various methods exist for dividing the total 70 \um\ flux among the confused sources, such as by the ratio of 24 \um\ fluxes of each source (e.g., Lee et al. 2006), but all such methods introduce large uncertainties.

Of the 69 candidates that show at least some evidence for being embedded objects, 16 are located too close to other objects to resolve into individual sources at 70 \um\footnote{The source numbers are 002, 006, 027, 048, 049, 055, 056, 073, 208, 210, 213, 214, 215, 108, 211, 212.}.  Many likely have \lint\ $\geq$ 1.0 \lsun\ and are thus not relevant to this study, since most have values of \lir, integrated only to 24 \um, greater than 0.1 \lsun.  However, a few might be embedded, low-luminosity protostars.  We do not list any of these 16 candidates in Table \ref{tab_seds} and do not consider them in the analysis discussed below.  Higher-resolution far-infrared data (e.g., Herschel) will be necessary to evaluate the nature of these objects.

\section{Discussion}\label{discussion}

\subsection{Estimating Internal Luminosities and the Luminosity Distribution}\label{lumint}

The internal luminosity is a key parameter in understanding the evolution of an object from dense core to star, but for low-luminosity, embedded protostars, it is not a directly observable quantity since a significant component of the bolometric luminosity will arise from external heating by the ISRF.  In \S \ref{id}, we presented two methods of estimating \lint\ based on observable quantities.  The first estimate is based on the result that an approximately constant ratio exists between \lint\ and \lir, where \lir\ is an observable quantity.  We call this estimate of the internal luminosity $L_{int}^{IR}$.  The second estimate is based on the tight correlation observed between the 70 \um\ flux, scaled to the value it would have if the source were located at 140 pc, and the internal luminosity.  This estimate, which we call $L_{int}^{70}$, is given by Equation \ref{eq_lint}.

\begin{figure}[t]
\plotone{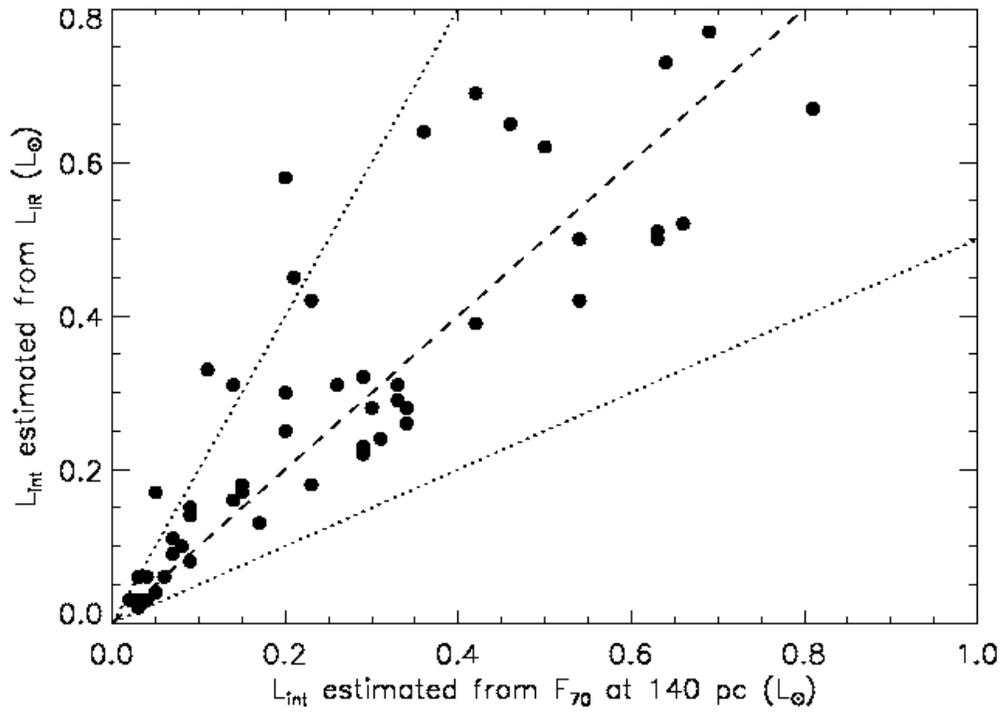}
\caption{\label{fig_lint_lint}$L_{int}^{IR}$, the internal luminosity estimated from the calculated value of \lir, vs. $L_{int}^{70}$, the internal luminosity estimated from the 70 \um\ flux scaled to its value at 140 pc, for the 50 objects listed in Table \ref{tab_seds}.  The dashed line shows the 1:1 line, while the dotted lines show the 2:1 and 1:2 lines.}
\end{figure}

Figure \ref{fig_lint_lint}, which plots $L_{int}^{IR}$ vs. $L_{int}^{70}$, demonstrates that the two estimates agree to within a factor of two of each other for nearly all of the sources.  We consider $L_{int}^{70}$ to be a more accurate estimate of \lint, both because nearly identical relationships between \lint\ and 70 \um\ flux normalized to 140 pc were derived separately from observations of protostars and a large grid of models and because $L_{int}^{IR}$ is based on a relationship between \lint\ and \lir\ that shows some variation from source to source.  Throughout the rest of this paper we use $L_{int}^{70}$ when we refer to internal luminosities.  Detailed radiative transfer modeling of each source is necessary to obtain more accurate internal luminosities.

\begin{figure}[t]
\plotone{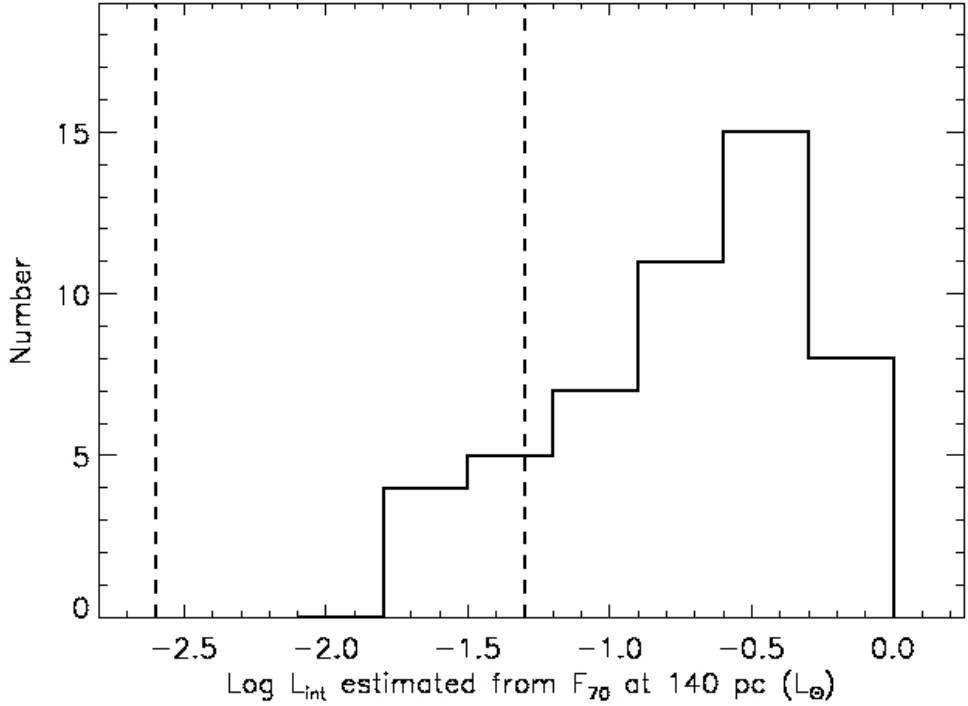}
\caption{\label{fig_lint_hist}Distribution of internal luminosities of the 50 embedded, low-luminosity protostars listed in Table \ref{tab_seds}.  The internal luminosity of each protostar was estimated from the observed 70 \um\ flux and the correlation between the two derived in \S \ref{id}.  The dashed lines shows our sensitivity limit to embedded protostars for the closest regions observed by c2d (\lint\ $= 2.5 \times 10^{-3}$ \lsun) and the most distant regions observed by c2d (\lint\ $= 5 \times 10^{-2}$ \lsun).}
\end{figure}

Figure \ref{fig_lint_hist} shows the distribution of Log(\lint) for the 50 objects listed in Table \ref{tab_seds}.  The decrease in the number of protostars near \lint\ $\sim 1$ \lsun\ is an artifact introduced by requiring that all sources have \lir\ $\leq 0.5$ \lsun, which is roughly equivalent, but not identical, to requiring that all sources have \lint\ $\leq 1.0$ \lsun\ given that the ratio of \lint$/$\lir\ has a weighted average of 1.7 but varies from source to source.  The full distribution of protostellar luminosities will be examined in a future paper (Evans et al. 2008, in preparation) that combines this work with other, complementary searches for embedded protostars in the c2d sample (e.g., Enoch 2007; Enoch et al. 2008a, in preparation; J\o rgensen et al. 2007; J\o rgensen et al. 2008, in preparation).  Here we only focus on the low end of the distribution.

\begin{figure}[t]
\plotone{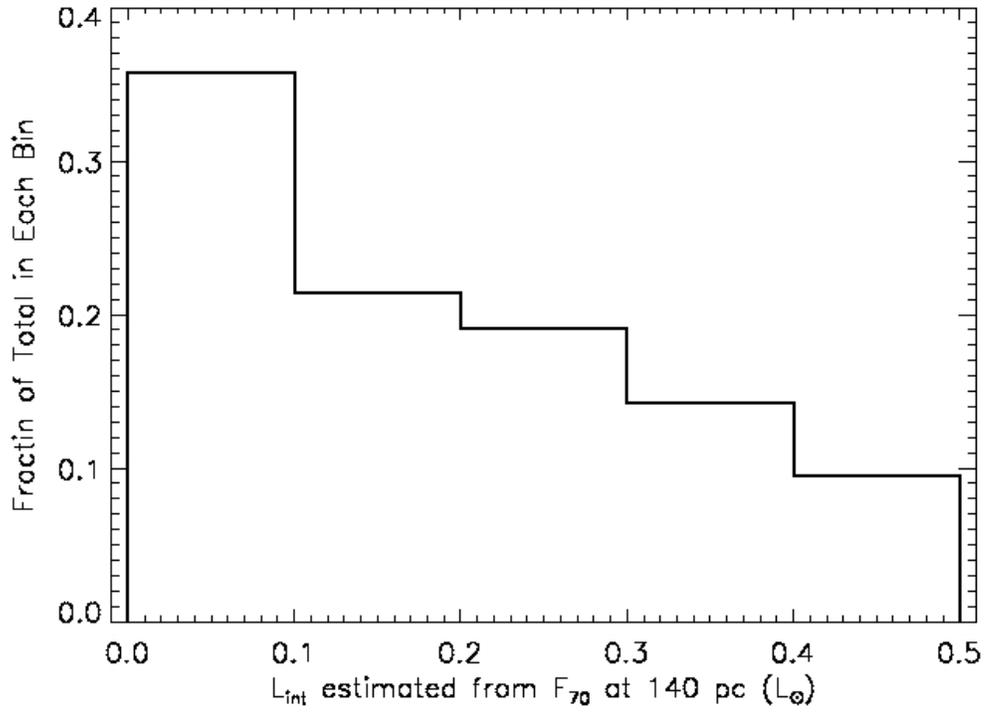}
\caption{\label{fig_lum_linnorm}Distribution of internal luminosities, in linear bins with a size of 0.1 \lsun, for all sources listed in Table \ref{tab_seds} with \lint\ $\leq$ 0.5 \lsun.}
\end{figure}

We find that 15/50 (30\%) of the objects identified here have \lint\ $\leq 0.1$ \lsun\ and are thus classified as VeLLOs.  Qualitatively speaking, this indicates that there are more embedded protostars at lower luminosities than at higher luminosities (if they were distributed evenly in luminosity, only 10\% would have \lint\ $\leq$ 0.1 \lsun).  To quantify this, Figure \ref{fig_lum_linnorm} shows the distribution of source internal luminosities, cut off at \lint\ $=$ 0.5 \lsun\ since our sample is likely incomplete near 1 \lsun.  A visual inspection of this figure suggests that the distribution is not uniform but increases with decreasing luminosity.  Applying a K-S test, we find that there is a 94.2\% probability that these sources are not drawn from a uniform luminosity distribution.  The implications of this increase in number of protostars with decreasing luminosity will be explored in \S \ref{vellos}.

Six VeLLOs have internal luminosities lower than our sensitivity limit for the most distant regions ($\sim$ $5 \times 10^{-2}$ \lsun), raising the question of how many such sources we miss in the more distant regions of our sample.  A discussion of correction for completeness due to the non-uniform distances to observed regions is presented in Appendix A.  Based on this discussion, we conclude that only $\sim 2$ sources with luminosities between our sensitivity limit for the closest regions (\lint\ $= 2.5 \times 10^{-3}$ \lsun) and the most distant regions (\lint\ $= 5 \times 10^{-2}$ \lsun) are missed.  Also based on Appendix A, we conclude that the intrinsic luminosity distribution does not continue to increase below \lint\ $= 0.1$ \lsun, although there are sources present all the way down to the sensitivity limit.  Finally, we note that we can not draw any conclusions about the possible presence of objects with extremely low luminosities below $\sim$ $10^{-3}$ \lsun.  Future infrared facilities with sensitivities much greater than that of \emph{Spitzer} will be required to search for such objects and determine the true lower limit to the intrinsic luminosity distribution.

The above conclusions are based on the analysis presented in Appendix A.  This analysis assumes that all sources with luminosities above our sensitivity limit of \lint\ $=4 \times 10^{-3}$ $(d/140 \, \rm{pc})^2$ \lsun\ are in fact identified.  To verify that this is indeed the case, we re-examine the criteria described in \S \ref{id}.  The only criterion that could potentially filter out sources with very low luminosities is the requirement that a source not be classified as a candidate galaxy in the classification method of Harvey et al. (2007b) unless $Log(P_{gal}) \leq -1.25$, since several of the steps in this classification method are based on the general fact that galaxies are faint (see Harvey et al. for details).

Using Perseus as an example, we re-applied our selection criteria, leaving out the step that filters out candidate galaxies with $Log(P_{gal}) \geq -1.25$.  This results in only an additional 36 sources, since most of the additional sources that would otherwise be selected are rejected because they are not detected at 70 \um.  We visually inspected these sources and rejected 14 based on either showing obvious galaxy morpologies or not being real sources (see \S \ref{visual} for details).  This left us with 22 sources.  The positions of all 22 sources were covered by at least one of the three submillimeter/millimeter dust continuum emission surveys of Perseus listed in \S \ref{prove1}, but none were associated with dense cores.  All 22 sources would thus be placed into group 6 in the terminology of \S \ref{groups}.  We conclude that we do indeed identify all sources above our sensitivity limit.

\subsection{Evolutionary Indicators}\label{lboltbol}

For the 50 objects with complete SEDs listed in Table \ref{tab_seds}, we calculate the following quantities:  the bolometric luminosity (\lbol), the ratio of bolometric to submillimeter luminosity (\lbolsmm), and the bolometric temperature (\tbol).  \lbol\ is calculated by intergrating over the full observed SED,
\begin{equation}\label{eq_lbol}
\lbol = 4\pi d^2 \int_0^{\infty} S_{\nu}d\nu \qquad ,
\end{equation}
while the submillimeter luminosity is calculated by integrating over the observed SED for $\lambda$ $\geq$ 350 \um,
\begin{equation}\label{eq_lsmm}
\lsmm = 4\pi d^2 \int_0^{\nu=c/350\, \mu m} S_{\nu}d\nu \qquad .
\end{equation}
The bolometric temperature is defined to be the temperature of a blackbody with the same flux-weighted mean frequency as the source (Myers \& Ladd 1993).  Following Myers \& Ladd, \tbol\ is calculated as
\begin{equation}\label{eq_tbol}
\tbol = 1.25 \times 10^{-11} \, \frac{\int_0^{\infty} \nu S_{\nu} d\nu}{\int_0^{\infty} S_{\nu} d\nu} \quad \rm{K} \qquad .
\end{equation}
We list \lbol, \lbolsmm, and \tbol\ for these 50 objects in the last column of Table \ref{tab_seds}.  The stated uncertainties reflect only those arising from measurement error.  Since we are integrating over finitely sampled SEDs, the values calculated for all three quantities will depend on how well each SED is sampled.  We discuss the errors introduced by incomplete sampling in Appendix B.

\subsubsection{Bolometric Luminosity-Temperature (BLT) Diagram}\label{blt}

\begin{figure}[t]
\plotone{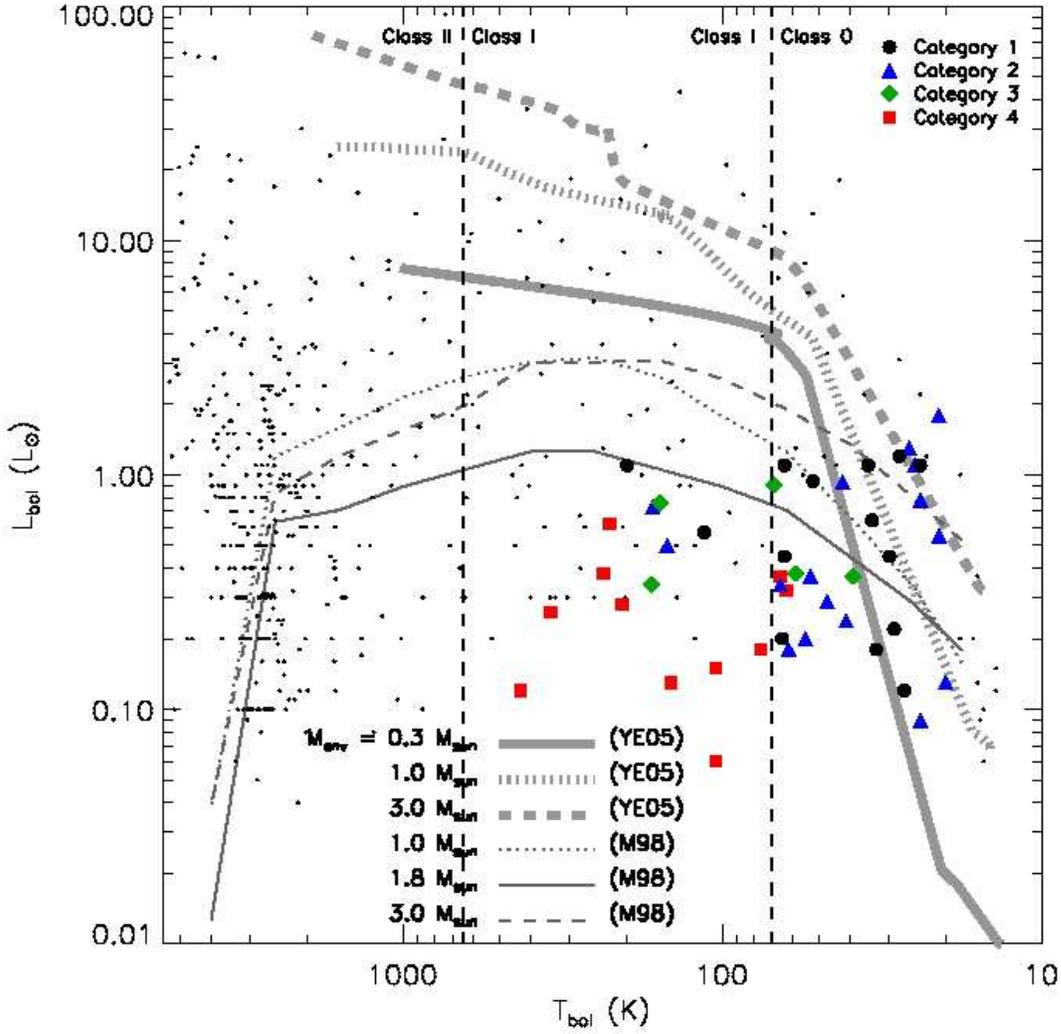}
\caption{\label{fig_blt}Figure 19 from Young \& Evans (2005), a \lbol\ vs. \tbol\ BLT, with the 50 objects listed in Table \ref{tab_seds} overlaid as large, colored points.  The meaning of the colors and symbols are described in the text.  The uncertainties in \lbol\ and \tbol, which are not shown on this plot, are dominated by the 20-60\% errors introduced by incomplete, finite sampling of the source SEDs (see Appendix 2).  The thick lines show the evolutionary tracks for the three models considered by Young \& Evans (2005), which differ in their initial envelope mass.  The thin lines show the evolutionary tracks for three models considered by Myers et al. (1998), which also differ in their initial envelope mass.  The small circles are observations from the literature compiled by Young \& Evans (2005).  The vertical dashed lines show the Class 0/I and Class I/II \tbol\ boundaries from Chen et al. (1995).}
\end{figure}

Figure \ref{fig_blt} places the 50 objects listed in Table \ref{tab_seds} on Figure 19 from Young \& Evans (2005), which plots \lbol\ vs. \tbol.  As discussed by Myers et al. (1998), who refer to this as a Bolometric Luminosity-Temperature (BLT) diagram, this is effectively a Hertzsprung-Russell diagram for protostars.  The lines show the evolutionary tracks followed by various models from Young \& Evans (2005) and Myers et al. (1998).  In general, models predict that protostars form at low values of \lbol\ and \tbol\ and move up and to the left (increasing \lbol, increasing \tbol) as the protostar grows.  Whether or not \lbol\ steadily increases or begins to decrease depends on the details of the mass accretion process.  Young \& Evans considered the collapse of a singular isothermal sphere in the ``standard model'' of star formation (Shu, Adams, \& Lizano 1987), which features a constant mass accretion rate over the entire protostellar phase.  Myers et al. (1998), on the other hand, assume that the accretion rate decreases exponentially with time and mass is lost from the system (from outflows, winds, etc.), explaining the different trends shown by their evolutionary tracks.

The colors/symbols of the points in Figure \ref{fig_blt} for the 50 objects from this work reflect how well the source SED is sampled.  All four colors/symbols have SEDs that are sampled at $1.25-70$ \um\ by 2MASS and \emph{Spitzer} and at millimeter wavelengths.  Black circles are also sampled at, at minimum, one wavelength in both the $100-200$ and $350-450$ \um\ wavelength ranges (category 1 SEDs in the notation of Appendix B).  Blue triangles are sampled between $350-450$ \um\ but not $100-200$ \um\ (category 2 SEDs).  Green diamonds are sampled between $100-200$ \um\ but not $350-450$ \um\ (category 3 SEDs).  Red squares are not sampled in either wavelength range (category 4 SEDs).  A detailed analysis of the errors introduced by the incomplete sampling in these various categories is presented in Appendix B; here we simply note that, depending on the category, the error in \lbol\ and \tbol\ is between approximately $20-60$\%.

There are many Class I sources (\tbol\ $\geq$ 70 K; Chen et al. [1995]) with \lbol\ below any evolutionary track by up to an order of magnitude.  The most extreme cases are all sources with category 4 SEDs, where \lbol\ is underestimated by $\sim$ 50\%, on average.  However, even accounting for this, many are still substantially below any of the evolutionary tracks.  A similar result was found by Enoch (2007) and Enoch et al. (2008a, in preparation), who suggest episodic accretion as a potential solution.  Another potential explanation for this discrepancy is that the sources with the lowest \lbol\ are simply those with envelope masses below those considered in the models.  However, this is ruled out by Enoch et al., who find good agreement between observations and models when comparing envelope mass vs. bolometric temperature despite the scatter seen in BLT diagrams, as well as several individual studies of VeLLOs that conclude at least some of the lowest-luminosity protostars are embedded in envelopes with masses similar to those of more luminous protostars (a few \msun; Young et al. 2004; Bourke et al. 2006; Dunham et al. 2006).

A similar population of Class 0 objects with luminosities below model predictions is not as apparent from Figure \ref{fig_blt}, and is also not seen by Enoch et al.  However, on average, external heating from the ISRF will add a larger contribution to \lbol\ for objects with more massive envelopes, thus such heating is likely to be more significant for Class 0 objects than Class I objects and can add up to several tenths of a solar luminosity to \lbol\ (e.g., Evans et al. 2001).  Since, as described in \S \ref{intro}, this distinction between \lbol\ and \lint\ is most relevant for very low-luminosity sources where external heating from the ISRF can dominate the total observed \lbol, a similar population of Class 0 objects with very low luminosities may be present but not readily apparent when examining a plot of \lbol\ vs. \tbol.  We return to this point in \S \ref{vellos}.

\subsubsection{Comparison of two Indicators of Evolutionary Status}\label{evolstatus}

\begin{figure}[t]
\plotone{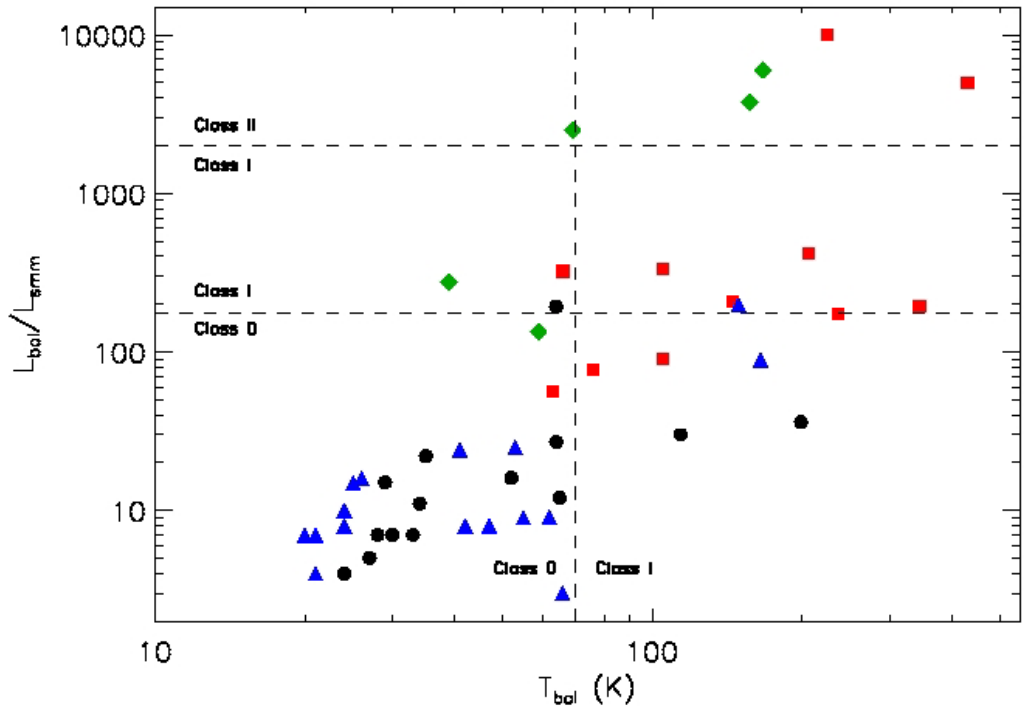}
\caption{\label{fig_lbs_tbol}\lbolsmm\ vs. \tbol\ for the 50 objects listed in Table \ref{tab_seds}.  The meaning of the colors/symbols are the same as for Figure \ref{fig_blt}.  The Class divisions in \tbol\ are from Chen et al. (1995), while the Class divisions in \lbolsmm\ are from Young \& Evans (2005).}
\end{figure}

Both \tbol\ and \lbolsmm\ are indicators of the evolutionary status of a protostar.  As the envelope mass decreases through a combination of accretion and dissipation processes, both quantities should increase.  Eventually, once the star has fully formed, \tbol\ will become equal to $T_{eff}$ and \lbolsmm\ will become very large.  We show a plot of \lbolsmm\ vs. \tbol\ for the objects considered here in Figure \ref{fig_lbs_tbol}.  The division between Class 0/I objects in \tbol\ is from Chen et al. (1995), while the division between Class 0/I and Class I/II objects in \lbolsmm\ is from Young \& Evans (2005).  The colors/symbols of the data points hold the same meaning as in Figure \ref{fig_blt}.

As stated in \S \ref{proof_class0}, the distinction between a Stage 0 and Stage I object is that a Stage 0 object still has more than half of the total mass of the system in the envelope, whereas less than half the total mass remains in the envelope of a Stage I object.  Both \tbol\ and \lbolsmm\ are used to separate embedded objects into Class 0 and Class I objects; in principle, these observational classes should correspond to Stage 0 and Stage I objects.  Young \& Evans (2005) concluded that \lbolsmm\ is a better measure of physical stage than \tbol, since their evolutionary models of different initial mass cores featured similar values of \lbolsmm\ but different values of \tbol\ when they reached the point at which half of the total mass had accreted onto the protostar.  They also found that embedded objects cross the Class 0/I boundary in \tbol\ ($\tbol=70$K; Chen et al. 1995) when they are still Stage 0 objects.  We see a similar trend in Figure \ref{fig_lbs_tbol} in that several objects are Class I objects according to \tbol\ but Class 0 objects according to \lbolsmm.  However, we caution that the details of where the observed SED is sampled affect the calculated value of \lbolsmm\ significantly more than the calculated value of \tbol; this explains the objects seen in Figure \ref{fig_lbs_tbol} with extremely high values of \lbolsmm\ (see Appendix B).  Great care must be taken to consider the errors introduced by incomplete sampling before using \lbolsmm\ as an evolutionary indicator for any particular object.

\subsection{Fraction of Starless Cores that Remain Starless}\label{starless}

Out of the 95 individual dense cores observed by c2d\footnote{Evans et al. (2007) list the 82 core regions surveyed by c2d.  Those regions that covered multiple cores are as follows (see also Evans et al. 2003 and references therein):  CB130-3 (CB130-1, CB130-2, and CB130-3), CG30-31 (CG30, CG31A, CG31B, CG31C), IRAM04191+1522 (IRAM04191+1522 and IRAS04191+1523), L1251 (L1251A, L1251C, and L1251E), L43 (L43-East and L43-RNO91), L673 (L673-SMM1 and L673-SMM2), and TMC1 (TMC1-1, TMC1-2, TMC1-1C-1, and TMC1-1C-2).}, 67 ($\sim$70\%) were classified as starless prior to being observed by \emph{Spitzer}.  The very first starless core observed, L1014, was found to actually harbor a very low-luminosity, embedded protostar (\lint\ $\sim$ 0.09 \lsun; Young et al. 2004; Bourke et al. 2005; Huard et al. 2005).  This naturally gives rise to the question of how many cores classified as starless prior to the launch of \emph{Spitzer} remain so after being observed with \emph{Spitzer}?  A substantial number of protostellar cores misclassified as starless cores would result in an incorrect estimate of the lifetime of the starless core phase\footnote{In actuality, this is an estimate of the lifetime of the portion of the starless core phase detectable by (sub)millimeter observations.  This is often referred to as the pre-stellar core phase (Kirk et al. 2005).}, as such estimates are derived by comparing the numbers of starless and protostellar cores  (e.g., Kirk et al. 2005; Enoch et al. 2008b).  Kirk et al. (2007) present the results of a search for embedded, low-luminosity protostars in \emph{Spitzer} observations of 22 starless cores previously studied using submillimeter data (Kirk et al. 2005); they find only one such source, L1521F-IRS (Bourke et al. 2006).  This suggests that $\sim$95\% of starless cores remain starless, leading them to conclude that the errors in starless core lifetime estimates introduced by previously undetected, low-luminosity protostars are smaller than the uncertainties in such estimates.

Our sample of 67 starless cores allows us to test these conclusions with a dataset approximately three times larger than that of Kirk et al. (2007).  Searching all 95 cores considered here, we have identified a total of 49 candidate embedded, low-luminosity protostars in 33 different cores (Table \ref{cores_candidates}).  Seventeen of these cores were classified as starless prior to being observed by \emph{Spitzer}.  If we restrict ourselves only to those sources in Groups $1-4$, those that, at minimum, show some additional evidence of being embedded protostars besides the shapes of their SEDs, we identify 29 candidates in 21 different cores, 9 of which were classified as starless prior to being observed by \emph{Spitzer}.  Thus, between 9 and 17 out of the 67 starless cores actually harbor embedded, low-luminosity protostars.  Approximately $75-85$\% of starless cores remain starless down to our luminosity sensitivity of \lint\ $\geq 4 \times 10^{-3}$ $(d/140 \, \rm{pc})^2$.  Most starless cores do indeed remain starless after being observed by \emph{Spitzer}, in agreement with Kirk et al. (2007), although the fraction is smaller than they found with their smaller sample.

\subsection{Non-Steady Mass Accretion}\label{vellos}

In this work, we have identified a sample of 50 embedded protostars with \lint\ $\leq$ 1.0 \lsun; 15 of these objects have \lint\ $\leq$ 0.1 \lsun\ and are thus classified as VeLLOs.  Assuming spherical mass accretion at the rate predicted by the standard model ($\dot{M}_{acc} \sim 2 \ee{-6}$ \msun\ yr$^{-1}$; Shu, Adams, \& Lizano 1987) onto an object with a typical protostellar radius of $R \sim 3$ \rsun, a protostar located on the stellar/substellar boundary ($M=0.08$ \msun) would have an accretion luminosity, $L_{acc} = \frac{GM\dot{M}_{acc}}{R}$, of $L \sim 1.6$ \lsun.  The objects identified here are difficult to understand in the context of the standard model of star formation.  VeLLOs, with luminosities more than an order of magnitude lower than the above calculation, are an extreme example of this problem.  If the objects identified here were observed edge-on through circumstellar disks their luminosities could possibly be underestimated, but, for at least some of them, this possibility can be eliminated (e.g., IRAM04191, Dunham et al. 2006).  Thus, they must either feature mass accretion rates far below that predicted by the standard model, masses far below the stellar/substellar boundary, or some combination of the two.  Several authors have invoked episodic accretion to explain the existence of these low-luminosity objects (e.g., Kenyon \& Hartmann 1995; Young \& Evans 2005; Enoch 2007; Enoch et al. 2008a, in preparation).

Strong evidence for non-steady mass accretion exists for the VeLLO IRAM 04191-IRS.  This object drives a well-defined, bipolar molecular outflow (\andre\ et al. 1999).  Taking the average mass accretion rate implied by the outflow and the dynamical time of this outflow ($\langle$$\dot{M}_{acc}$$\rangle$ $\sim 5 \ee{-6}$ \msun\ yr$^{-1}$ and $t_d \sim 10^4$ years), \andre\ et al. (1999) calculate that a protostellar mass of 0.05 \msun\ will have accreted over the lifetime of the outflow.  Accretion at this rate onto an object with a mass of $0.05$ \msun\ and a radius of 3 \rsun\ would give rise to an accretion luminosity of $L_{acc} \sim 2$ \lsun.  This is clearly inconsistent with the internal luminosity of \lint\ $\sim$ 0.08 \lsun\ derived by Dunham et al. (2006).  Dunham et al. suggested non-steady mass accretion to resolve this inconsistency:  the current mass accretion rate must be much lower than the average rate over the lifetime of the outflow.

\begin{figure}[t]
\plotone{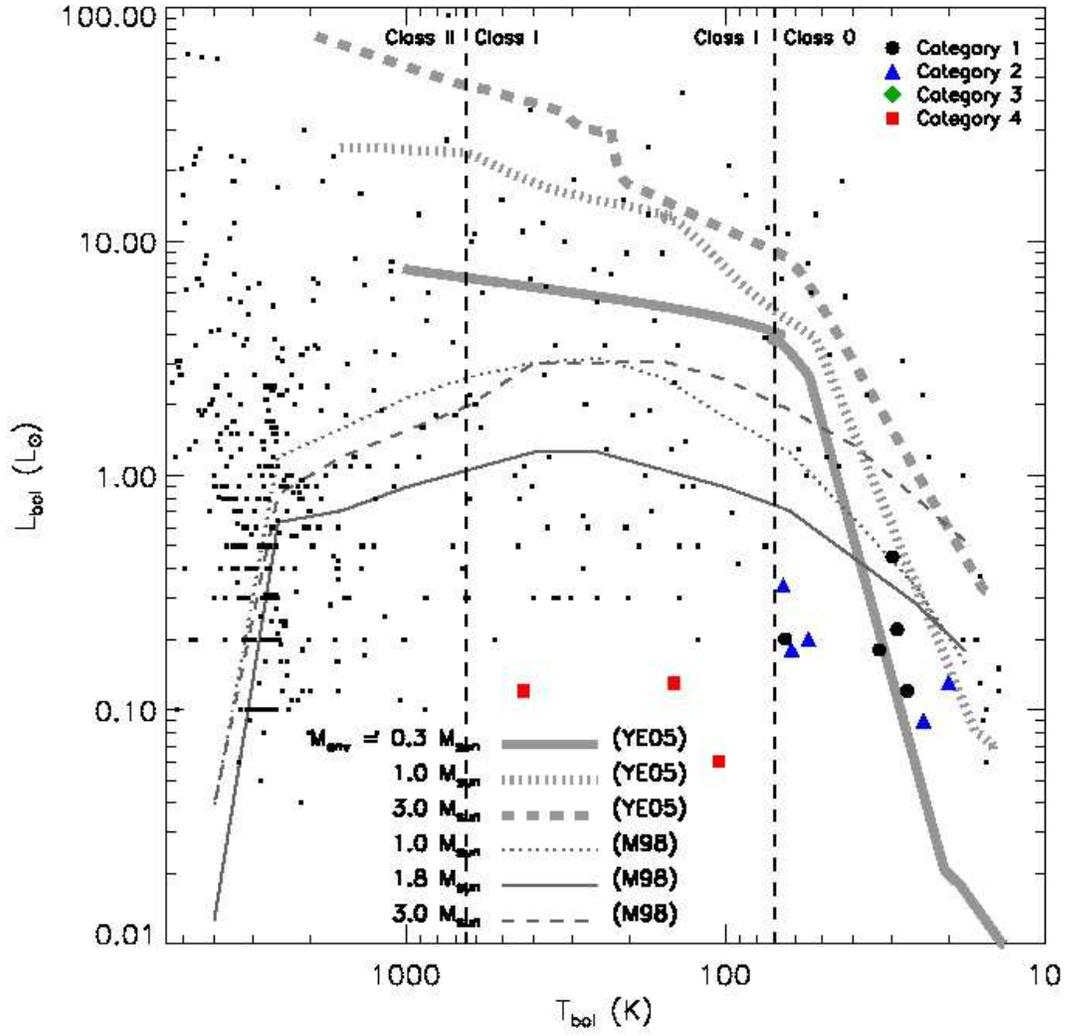}
\caption{\label{fig_blt_vello}Same as Figure \ref{fig_blt}, except only showing the 15 VeLLOs (objects with \lint\ $\leq$ 0.1 \lsun).}
\end{figure}

We show the location of the 15 VeLLOs in our sample on the BLT diagram in Figure \ref{fig_blt_vello}.  Most are located at \tbol\ $<$ 70 K and \lbol\ $>$ 0.1 \lsun\ and are thus Class 0 objects with very low internal luminosities but bolometric luminosities that generally appear consistent with evolutionary tracks.  If there is a population of Class 0 objects with luminosities lower than predicted by evolutionary tracks, as is the case for Class I objects (see \S \ref{blt}), the inconsistency between the luminosities of these Class 0 objects and the predictions of evolutionary tracks will likely be masked on a BLT diagram since, as described in \S \ref{blt}, the contribution to \lbol\ from external heating will be larger for younger objects with more massive envelopes.  In addition to \lbol\ and \tbol, considering time as a third dimension relevant to protostellar evolution can help illuminate whether or not these Class 0 objects with \lint\ $\leq$ 0.1 \lsun\ but \lbol\ $>$ 0.1 \lsun\ are truly as consistent with constant mass-accretion evolutionary tracks as they appear to be on a BLT diagram.

Assuming a constant rate of mass accretion, the protostellar mass $M$ will grow linearly with time.  If we assume the protostellar radius remains fixed, $L_{acc} \propto M$ and thus also grows linearly with time.  Adding in the assumption that star formation has been occuring continuously over a time much longer than the lifetime of the embedded phase, the number of protostars as a function of luminosity should be constant.  However, as discussed in \S \ref{lumint}, the intrinsic luminosity distribution is not uniform and instead increases with decreasing luminosity.  Even though many of the lowest luminosity objects appear consistent with the location of evolutionary tracks calculated assuming constant mass accretion, there are too many relative to higher luminosity sources to explain by constant mass accretion.

We have presented evidence for non-steady mass accretion in low-mass protostars, both from the above statistical argument and from a detailed study of one individual source (Dunham et al. 2006).  While we note here that non-steady mass accretion does not necessarily imply episodic mass accretion, these results, combined with other studies that find clear evidence for episodicity in outflow activity (e.g., HH211; Lee et al. 2007), strongly suggest that protostellar mass accretion is episodic in nature.  Future work is needed on both observational and theoretical fronts to better understand the implications of the objects identified in this study on the mass accretion process in low-mass star formation.  On the observational front, the detailed study of the outflow driven by IRAM04191 must be repeated for other sources known to drive outflows, and dedicated outflow searches are needed towards objects lacking such data.  Additionally, detailed chemical and physical studies of the objects identified in this work are needed to explore the nature of these sources beyond the simple evolutionary indicators calculated here.  Such studies are of particular relevance to understanding the evolution of starless cores and identifying those cores on the verge of collapse since VeLLOs have been discovered in cores that were classified as starless prior to \emph{Spitzer} observations but were not always those cores believed to be the most evolved and nearest to the onset of collapse (Crapsi et al. 2005b; Bourke et al. 2006).  Chemical studies can also be used to distinguish between low-luminosity protostars that have featured higher past luminosities (presumably through higher past mass accretion rates) and those that have always featured such low luminosities (e.g., Lee 2008).  Candidates identified in this study that show no evidence for being embedded in dense cores should be re-examined as additional data on each source becomes available (e.g., through large-scale surveys with Herschel, SCUBA-2, etc.) to ensure as complete a sample as possible.  Finally, the full sample of embedded, low-mass protostars must be assembled from this work and the various other searches for such objects outlined in \S \ref{intro}.

On the theoretical front, evolutionary models predicting the observational signatures of dense cores forming stars must be revisited.  These models must be able to explain both the existence of embedded protostars with luminosities much lower than predicted by current models and the very large scatter in \lbol\ observed at each value of \tbol\ in Figure \ref{fig_blt}.  Models that incorporate episodic mass accretion and the effects of source geometry may provide such explanations, but will require higher dimensions than the 1-D models considered by us to date.

\section{Conclusions}\label{conclusions}

We have conducted a search for all embedded protostars with \lint\ $\le$ 1.0 \lsun\ in the c2d dataset of nearby, low-mass star-forming regions.  We identify 218 candidates from the \emph{Spitzer} data alone; examining all available complementary data for each candidate results in a sample of 50 objects that show at least some evidence that they are indeed embedded within dense cores.  A summary of our major results is as follows:

\begin{itemize}
\item On average, the \emph{Spitzer} c2d data are sensitive to embedded protostars with \lint\ $\geq 4 \times 10^{-3}$ $(d/140 \, \rm{pc})^2$ \lsun, a factor of 25 better than the sensitivity of the \emph{Infrared Astronomical Satellite (IRAS)} to such objects.
\item The 70 \um\ flux and internal luminosity of a protostar are tightly correlated.  As the former is a directly observable quantity but the latter is not, this correlation gives a powerful method for estimating protostellar internal luminosities when detailed radiative transfer models for each source are lacking.
\item Of the 50 objects in our sample, 15 (30\%) have \lint\ $\leq$ 0.1 \lsun\ and are thus classified as VeLLOs.  The distribution of source luminosities is not uniform and instead increases with decreasing luminosity.  Accounting for incompleteness arising from non-uniform distances to the observed regions, we find sources down to the above sensitivity limit, indicating that the intrinsic luminosity distribution may extend to lower luminosities than probed by these observations.  Despite this, we are able to rule out a continued rise in the distribution below \lint\ $= 0.1$ \lsun.
\item Between $75-85$\% of cores classified as starless prior to being observed by \emph{Spitzer} remain starless down to the above luminosity sensitivity; the remaining $15-25$\% harbor low-luminosity, embedded protostars.  This is in general agreement with Kirk et al. (2007), who examined archival \emph{Spitzer} data of 22 starless cores and found only one to be harboring a low-luminosity protostar.  However, with our larger sample size, we are able to better constrain the fraction of cores previously classified as starless that in fact harbor such objects.  We confirm that recent estimates of starless core lifetimes (e.g., Kirk et al. 2005; Enoch et al. 2008b) do not feature large errors introduced by previously undetected, low-luminosity protostars.
\item The observed luminosity distribution for embedded objects with \lint\ $\le$ 1.0 \lsun\ is inconsistent with the simplest picture of star formation wherein mass accretes from the core onto the protostar at a constant rate.  Combining this result with other studies that find clear indications of episodic outflow activity strongly suggests that protostellar mass accretion is episodic in nature.
\end{itemize}

We have outlined several avenues of future work that must be pursued now that relatively complete and unbiased samples of embedded, low-mass protostars are being compiled.  Only with such future studies can we begin to build a coherent picture of low-mass star formation consistent with the growing observational database provided by systematic, large-scale surveys of low-mass star forming regions.

\acknowledgements
The authors thank M. Enoch for providing the data used in Figure \ref{fig_enoch_avg_seds}, as well as for several helpful discussions that have improved the quality of this paper.  We express our gratitude to the anonymous referee for several comments that have improved the quality of this publication.  We thank Jes J\o rgensen and Paul Harvey for providing helpfuul comments, and we also thank Jes J\o rgensen for his IDL scripts to display three-color images.  This work is based primarily on observations obtained with the \emph{Spitzer Space Telescope}, operated by the Jet Propulsion Laboratory, California Institute of Technology.  We thank the Lorentz Center in Leiden for hosting several meetings that contributed to this paper.  This publication makes use of the Protostars Webpage hosted by the University of Kent, as well as data products from the Two Micron All Sky Survey, which is a joint project of the University of Massachusetts and the Infrared Processing and Analysis Center/California Institute of Technology, funded by the National Aeronautics and Space Administration and the National Science Foundation.  These data were provided by the NASA/IPAC Infrared Science Archive, which is operated by the Jet Propulsion Laboratory, California Institute of Technology, under contract with NASA.  This research has made use of NASA's Astrophysics Data System (ADS) Abstract Service and of the SIMBAD database, operated at CDS, Strasbourg, France.  Support for this work, part of the Spitzer Legacy Science Program, was provided by NASA through contracts 1224608 and 1288658 issued by the Jet Propulsion Laboratory, California Institute of Technology, under NASA contract 1407.  Support was also provided by NASA Origins grants NNG04GG24G, NNX07AJ72G, and NAG5-13050.

\appendix
\section{Internal Luminosity Completeness and the Intrinsic Luminosity Distribution}

The regions observed by c2d are located at distances ranging from $125-500$ pc.  Using the result from \S \ref{criteria} that our observations are sensitive to embedded protostars with \lint\ $\geq 4 \times 10^{-3}$ $(d/140 \, \rm{pc})^2$ \lsun, the effect of these non-uniform distances is that our sensitivity to embedded protostars varies between $2.5 \times 10^{-3}$ and $5 \times 10^{-2}$ \lsun.  Two important questions thus arise:  (1) How many objects are lost because of these non-uniform sensitivities (stated another way, how many additional objects would we expect to identify if all observed regions were located at the distance of the closest regions)?, and (2) What conclusions can be drawn about the intrinsic luminosity distribution?  The latter is of particular significance to future work aimed at explaining the observed distribution of sources in \lbol-\tbol\ space through evolutionary modeling.

\subsection{Internal Luminosity Completeness}

\begin{figure}[t]
\plotone{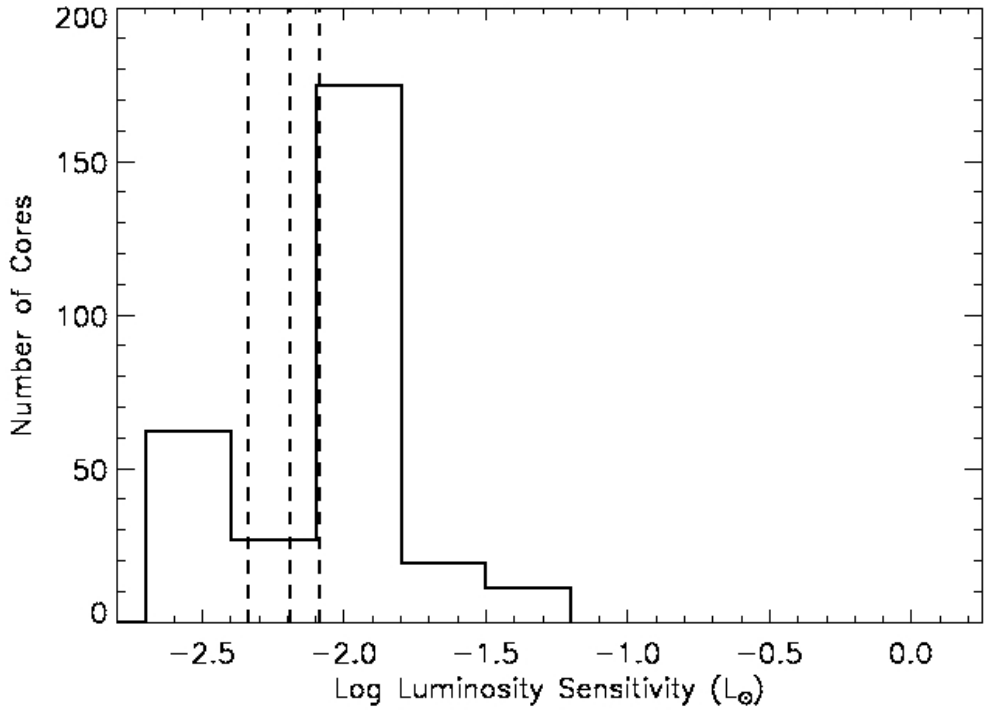}
\caption{\label{fig_lumsens}Distribution of the sensitivity to protostars embedded within the 300 dense cores in Perseus, Serpens, Ophiuchus, and the sample of 82 regions with dense cores targeted for observations.  The dashed lines show, from left to right, the sensitivity for cores in Lupus I and IV, Chamaeleon II, and Lupus III.}
\end{figure}

The regions observed consist of both targeted observations of 82 regions with 95 dense cores and unbiased surveys of 5 large, molecular clouds (\S \ref{observations}).  Thus, an analysis of completeness that uses the area observed at different distances is incorrect since the sample consists both of observations biased towards dense cores and observations with no such bias.  Instead, we start from the total number of dense cores observed at different distances since all embedded protostars are, by definition, within these cores.  The sample consists of 122 cores in Perseus (Enoch et al. 2006), 35 cores in Serpens (Enoch et al. 2007), and 48 cores in Ophiuchus (Young et al. 2006b), along with the targeted observations of 95 cores.  This yields a total of 300 cores; the distribution of the sensitivity to protostars embedded within these cores is shown in Figure \ref{fig_lumsens}.  We have no information on the number of dense cores in the Lupus and Chamaeleon II clouds; any such cores have comparable luminosity sensitivities to the rest of the sample (Figure \ref{fig_lumsens}).

To start with a simple analysis, we ask what percentage of the cores are located at distances close enough such that sources in each of the lowest three bins of Figure \ref{fig_lint_hist} could be detected.  Statistically, this should be equal to the percentage of sources in that bin we are able to detect.  All of the cores are located at distances such that sources in the bin -1.2 $\leq Log(\lint/\lsun) \leq$ -0.9 could be detected, thus we are complete to sources in this bin.  96\% of the cores are located at distances such that sources in the bin -1.5 $\leq Log(\lint/\lsun) \leq$ -1.2 could be detected; correcting for completeness raises the total number of sources in this bin from 5 to 5.2.  90\% of the cores are located at distances such that sources in the bin -1.8 $\leq Log(\lint/\lsun) \leq$ -1.5 could be detected; correcting for completeness raises the total number of sources in this bin from 4 to 4.4.  We conclude that we miss, at most, 1 source in the range -1.8 $\leq Log(\lint/\lsun) \leq$ -0.9; this is actually a strong upper limit since the percentage of cores with distances close enough to detect sources in these bins would increase if the dense cores in the Lupus and Chamaeleon II clouds were included.  However, this says nothing about the bins that are above our sensitivity limit for the nearest regions ($2.5 \times 10^{-3}$ \lsun) in which we detect no sources.  To examine the completeness in these bins, and confirm the above results, we turn to a Monte Carlo simulation.

For the Monte Carlo simulation, we create a sample of 10,000 embedded protostars.  We assign an internal luminosity to each protostar from a distribution specified at the start; in all cases, this distribution has a maximum of \lint\ $=$ 0.1 \lsun.  We then randomly place each of the 10,000 sources in one of the 300 dense cores.  If a source has \lint\ less than the sensitivity limit for the core in which it is placed, it is considered ``missed'' by our observations and rejected; otherwise, the source is kept.  We run the simulation 1000 times, average the results, and compare the observed distribution of \lint\ for the 15 VeLLOs with the average simulated observed distribution.

\begin{figure}[t]
\epsscale{0.75}
\plotone{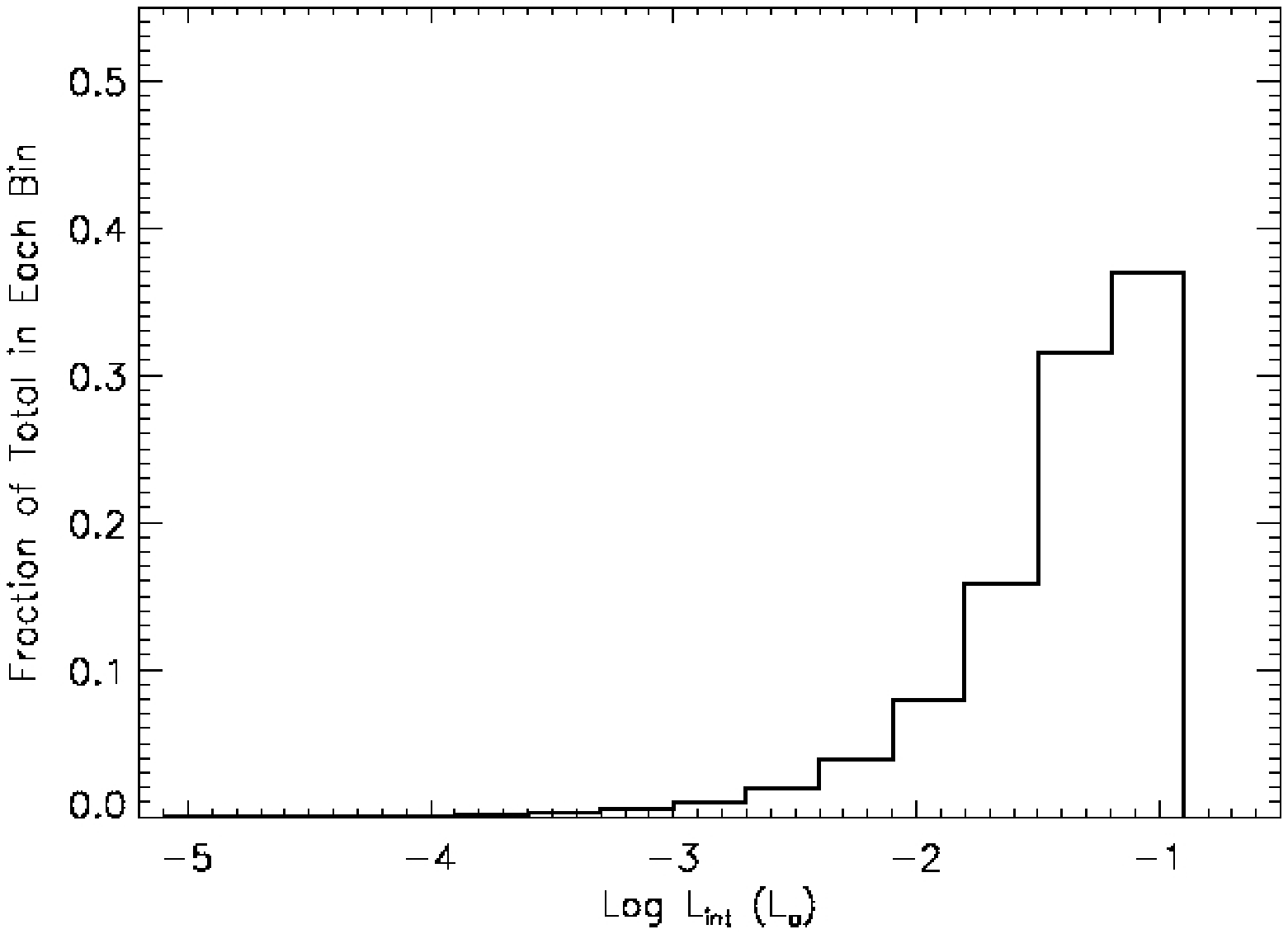}
\plotone{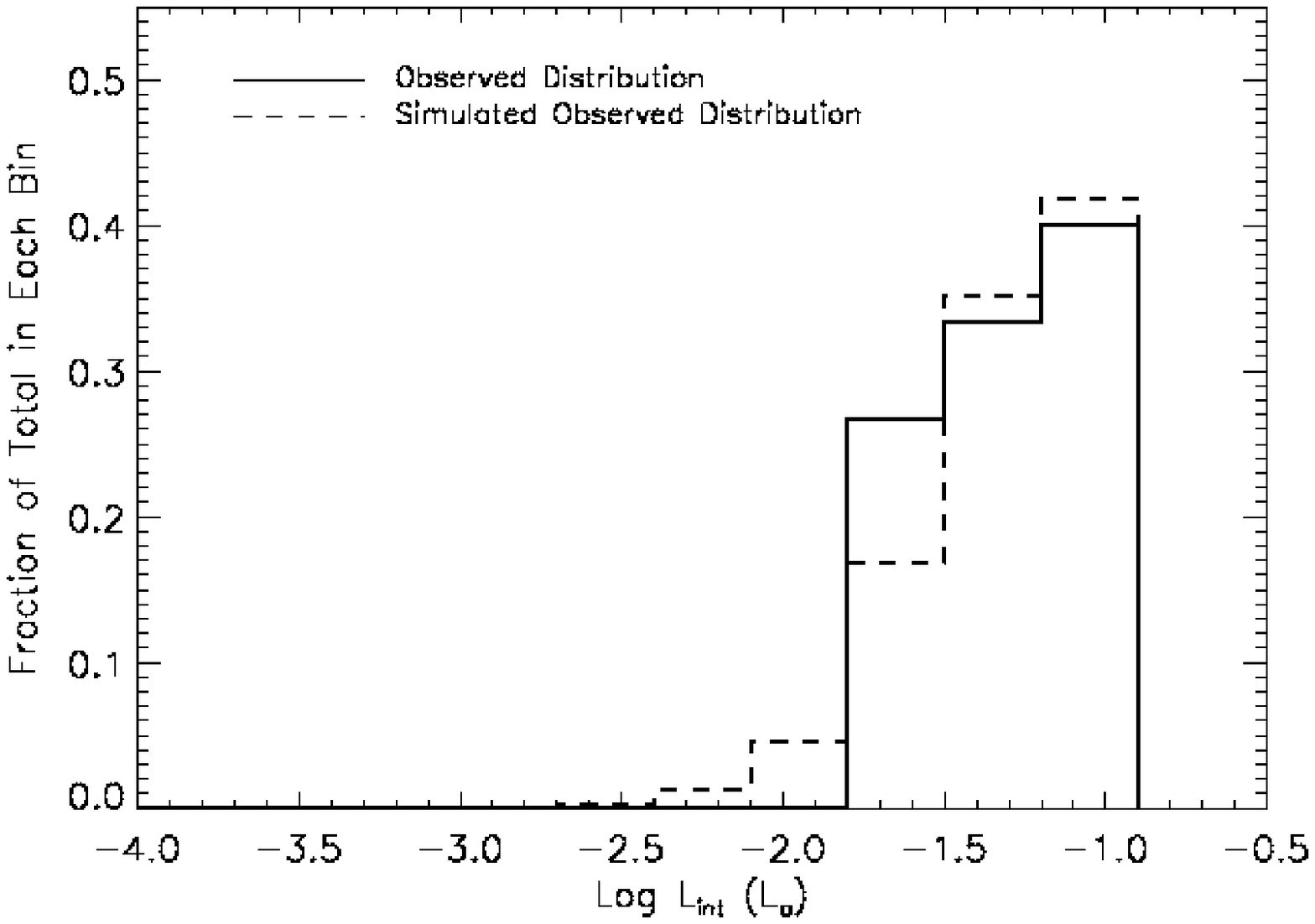}
\caption{\label{fig_montecarlo_run1}Monte Carlo simulation of the observation of 10,000 embedded protostars with internal luminosities distributed evenly between $10^{-5}$ and $10^{-1}$ \lsun\ and randomly placed in one of the 300 dense cores.  \emph{Top}: Normalized distribution of \lint\ for the 10,000 sources.  \emph{Bottom}:  Results of the simulation.  The solid line shows the observed distribution of \lint\ (normalized) for the 15 VeLLOs identified by this work, while the dashed line shows the simulated observed distribution of \lint\ (normalized) of the simulated sources.}
\end{figure}

Figure \ref{fig_montecarlo_run1} shows the results of a simulation where the 10,000 sources have internal luminosities evenly distributed (in linear space) between $10^{-5}$ and $10^{-1}$ \lsun.  The normalized observed distributions of \lint\ for the 15 VeLLOs identified by this work and for the simulated sources agree given the uncertainties due to small number statistics.  Furthermore, since we detect no sources with $Log(\lint) \leq -1.8$, we can set an upper limit of $\leq$ 1/15 sources (or $\leq$ 7\% of all sources detected) have such luminosities.  Out of all the sources detected in the Monte Carlo simulation, 6.1\% have such luminosities, consistent with the observed upper limit of 7\%.

Out of the 10,000 sources considered by this simulation, 1158 (12\%) have luminosities below the sensitivity limit of the core in which they are placed and are thus ``missed'' by our observations.  This means that the 15 VeLLOs identified are only 88\% of the total number of objects, indicating that approximately 2 sources are missed.  The exact details of the input luminosity distribution do not affect these results, as long as the simulation is consistent with the upper limit on the number of detected sources with $Log(\lint) \leq -1.8$ described above.  Input luminosity distributions that have a much larger fraction of sources at lower luminosities, such as a distribution flat in log space, would lead to an increase in the number of sources missed.  However, such distributions are inconsistent with the above upper limit (for example, a simulation with an input luminosity distribution flat in log space between $10^{-3}$ and $10^{-1}$ \lsun\ results in 24\% of the total sources detected having $Log(\lint) \leq -1.8$, clearly inconsistent with the observed upper limit of 7\%).  Thus, we conclude that $\sim 2$ sources are missed due to the non-uniform distances of the observed regions.

\subsection{Intrinsic Luminosity Distribution}

We now consider what constraints we can place on the low end of the intrinsic luminosity distribution.  Additional constraints will be provided by a future paper aimed at analyzing the shape of the full luminosity distribution, including higher luminosity sources (Evans et al. 2008, in preparation).  These constraints will serve as guides to future work aimed at explaining the full distribution of sources in \lbol-\tbol\ space.

\begin{figure}[t]
\plotone{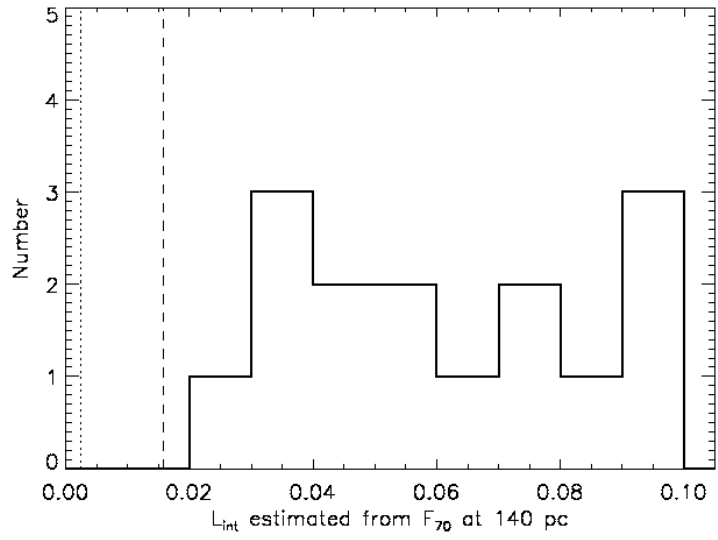}
\caption{\label{fig_lintvellos}Distribution of internal luminosities, in linear bins of 0.01 \lsun, for the 15 embedded protostars with \lint\ $\leq$ 0.1 \lsun.  The vertical dashed line shows \lint\ $= 10^{-1.8}$ \lsun\ $\approx 0.016$ \lsun.  The vertical dotted line shows \lint\ $= 0.0025$ \lsun.}
\end{figure}

As stated in \S \ref{lumint}, the intrinsic luminosity distribution for the objects identified here increases to lower luminosities.  To examine whether or not this increase continues to our sensitivity limit, Figure \ref{fig_lintvellos} shows the distribution of internal luminosities, in linear bins of 0.01 \lsun, for the 15 objects with \lint\ $\leq$ 0.1 \lsun.  Considering the small sample size, this distribution appears uniform down to \lint\ $= 0.02$ \lsun.  From the discussion above, approximately two sources are missing from this figure because of incompleteness arising from the non-uniform distances to the observed regions.  Additionally, a strong upper limit of 1 source is lost due to incompleteness for \lint\ $\geq 10^{-1.8}$ \lsun\ (denoted by the dashed line in Figure \ref{fig_lintvellos}), suggesting that the sources missing due to incompleteness likely fall in the lowest two bins.  Taking into account the small sample size, adding $\sim 2$ sources to these two bins results in a distribution that appears approximately uniform all the way down to the sensitivity limit (denoted by the dotted line in Figure \ref{fig_lintvellos}).  This suggests that the intrinsic luminosity distribution may extend to lower luminosities than probed by these observations.  However, even if all missed sources fall in the lowest bin of Figure \ref{fig_lintvellos}, this portion of the intrinsic luminosity distribution (\lint\ $\leq 0.1$ \lsun) would appear to be uniform rather than increasing to lower luminosities.  We thus conclude that the instrinsic luminosity distribution does \emph{not} continue to increase to our sensitivity limit, although K-S tests should be used to quantitatively confirm this result once a larger sample becomes available.

\begin{figure}[t]
\epsscale{0.75}
\plotone{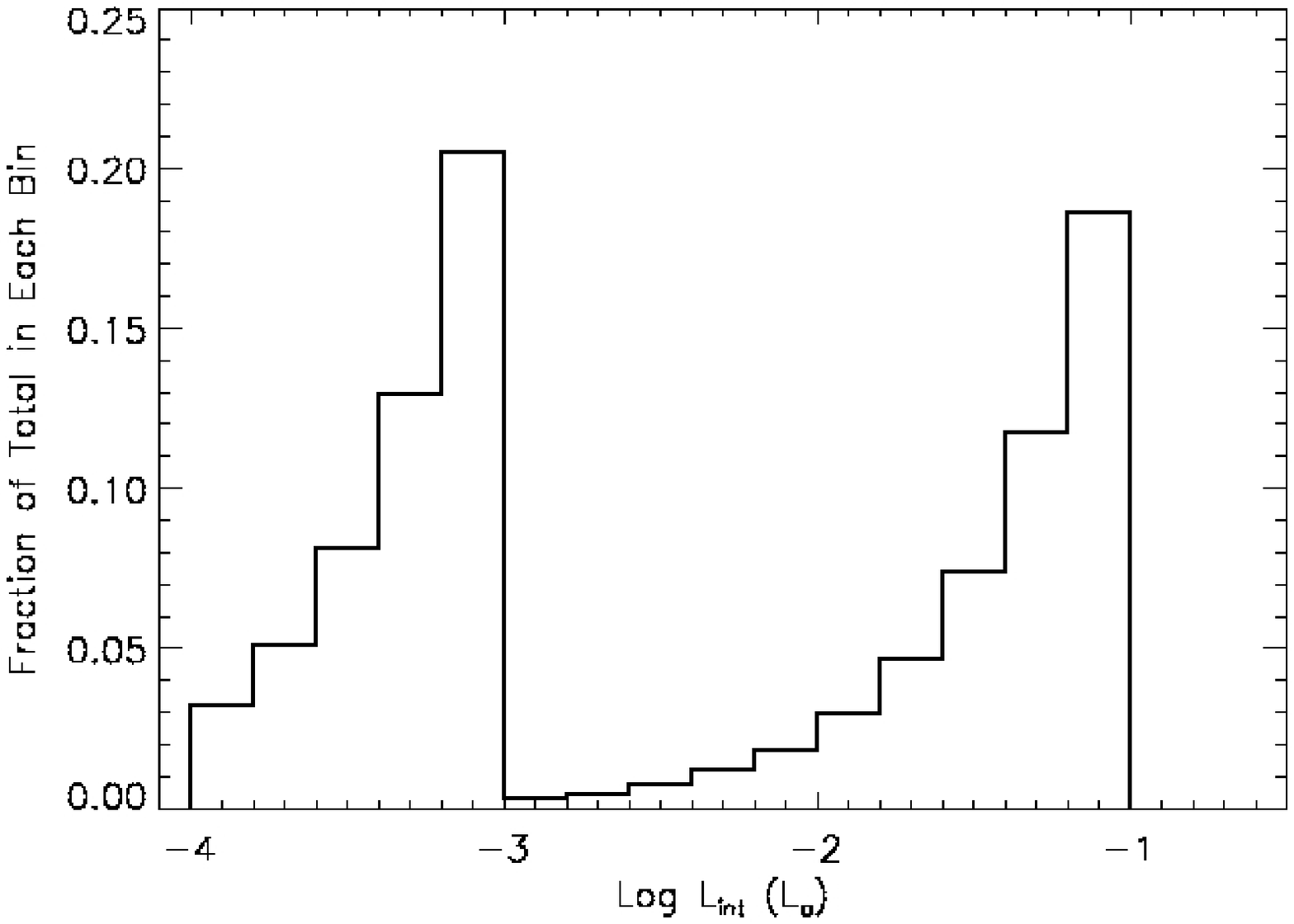}
\plotone{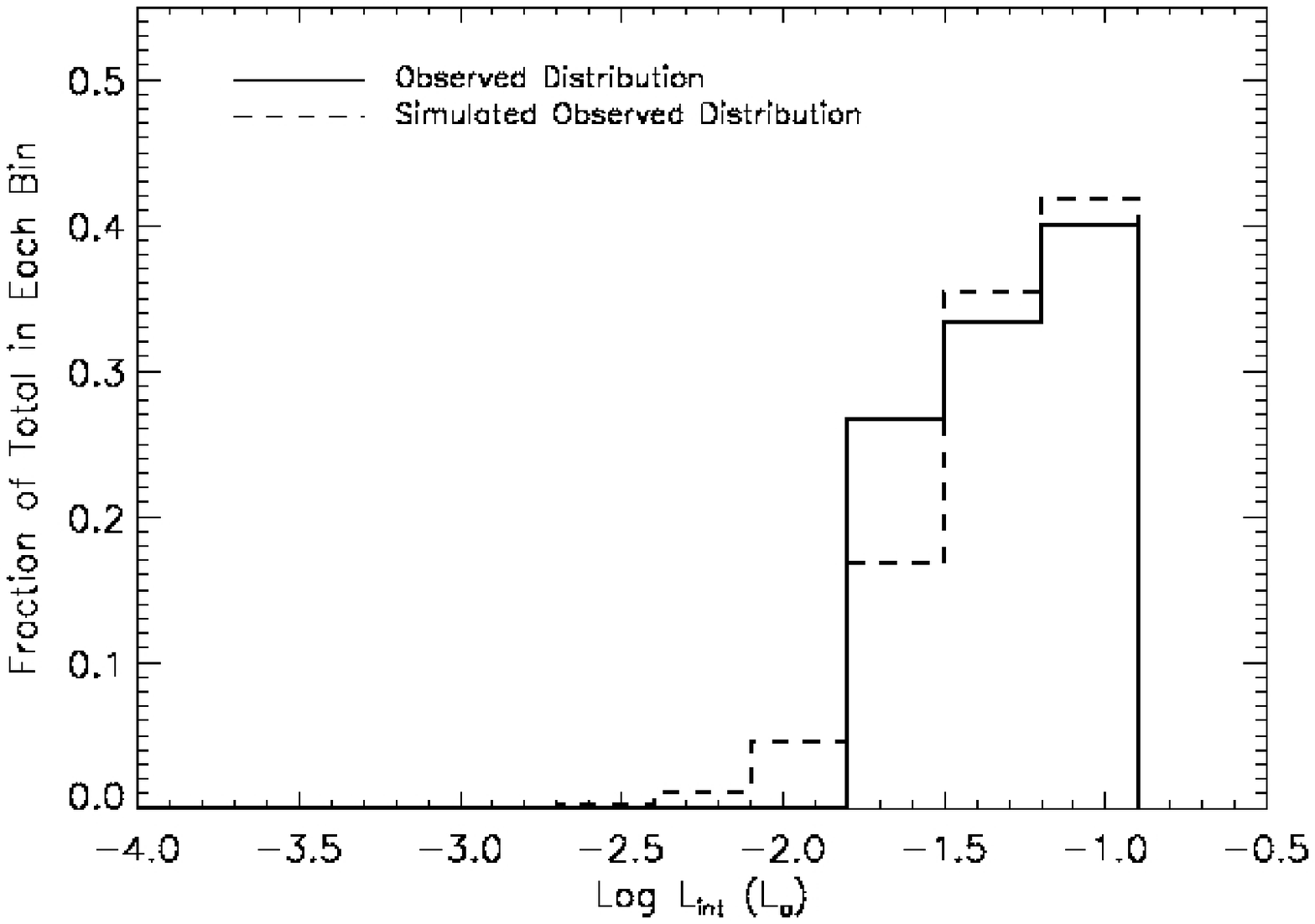}
\caption{\label{fig_montecarlo_run4}Same as Figure \ref{fig_montecarlo_run1}, except with the internal luminosities of 5,000 of the 10,000 sources evenly distributed between $10^{-4}$ and $10^{-3}$ \lsun\ and the internal luminosities of the other 5,000 sources evenly distributed between $10^{-3}$ and $10^{-1}$ \lsun.}
\end{figure}

This result provides an important constraint for future attempts at understanding these very low luminosity objects.  However, we caution that we can draw no conclusions about the possible presence of objects with luminosities below $\sim 10^{-3}$ \lsun.  To illustrate this with an example, Figure \ref{fig_montecarlo_run4} shows the results of a Monte Carlo simulation where half of the 10,000 sources have luminosities evenly distributed between $10^{-4}$ and $10^{-3}$ \lsun, while the other half have luminosities evenly distributed between $10^{-3}$ and $10^{-1}$ \lsun.  The simulated observed distribution matches the observed distribution quite well; the population of objects with \lint\ between $10^{-4}$ and $10^{-3}$ \lsun\ have no effect on the final result since none of them are detected.  A large population of embedded objects with such low luminosities would be very difficult to explain physically (see discussion of VeLLOs in \S \ref{vellos}), but we cannot rule out the presence of such objects based on these observations.

\section{Errors in Evolutionary Indicators From Incomplete, Finitely Sampled Spectral Energy Distributions}

The integrals defined in Equations \ref{eq_lbol} $-$ \ref{eq_tbol} are evaluated using the trapezoid rule to integrate over the finitely sampled source SEDs.  The contribution $\delta_i$ to the total integral from measurements $y_i$ and $y_{i+1}$, measured at $x_i$ and $x_{i+1}$, is
\begin{equation}\label{eq_trapezoid}
\delta_i = (x_{i+1} - x_i) \, \left (\frac{y_{i+1} + y_i}{2}\right ) \qquad .
\end{equation}
The total value of the integral is then the sum of each individual $\delta_i$.

Clearly, the result will depend on how well the source SED is sampled, especially near the peak.  Most of the material in the envelope of embedded objects is cold ($10-15$ K; e.g., Evans et al. 2001; Shirley et al. 2002).  The emission peaks from blackbodies with temperatures of 10 and 20 K are located at approximately 500 and 250 \um, respectively, overlapping with the $\sim50-300$ \um\ far-infrared regime that cannot be observed from the ground.  Incomplete sampling near the peak may thus introduce significant errors in the calculations of \lbol, \tbol, and \lbolsmm.

To quantify these errors, we consider the 13 group 1 candidates in Perseus.  Most of these sources have well-sampled SEDs, with 2MASS and \emph{Spitzer} detections from $1.25-70$ \um, \emph{Spitzer} detections at 160 \um\ and possibly \emph{IRAS} detections at 100 \um, SHARC-II detections at 350 \um, SCUBA detections at 450 and 850 \um, and Bolocam detections at 1.1 mm.  We construct four SEDs for each source:

\begin{enumerate}
\item We include all detections.  These ``category 1'' SEDs are those that are the most completely sampled.
\item We only include detections in the wavelength ranges $1.25 \leq \lambda \leq 70$ \um\ and $\lambda \geq 350$ \um.  These ``category 2'' SEDs are meant to simulate sources for which 2MASS and \emph{Spitzer} $1-70$ \um\ data, 350 or 450 \um\ submillimeter data, and millimeter wavelength data are available, but neither \emph{Spitzer} 160 \um\ nor \emph{IRAS} 100 \um\ data are available.
\item We only include detections in the wavelength ranges $1.25 \leq \lambda \leq 160$ \um\ and $\lambda \geq 850$ \um.  These ``category 3'' SEDs are meant to simulate sources for which 2MASS and \emph{Spitzer} $1.25-70$ \um\ data, \emph{Spitzer} 160 \um\ or \emph{IRAS} 100 \um\ data, and millimeter wavelength data are available, but no 350 or 450 \um\ submillimeter data are available.
\item We only include detections in the wavelength ranges $1.25 \leq \lambda \leq 70$ \um\ and $\lambda \geq 850$ \um.  These ``category 4'' SEDs are meant to simulate sources for which only 2MASS and \emph{Spitzer} $1.25-70$ \um\ and millimeter wavelength data are available.
\end{enumerate}

For each of the 13 sources, we calculate \lbol, \tbol, and \lbolsmm\ for each of the four versions of the SED.  Comparing the results will give an estimate of the errors introduced by incompletely sampling the source SED.

\subsection{Bolometric Luminosity}

\begin{figure}[t]
\plotone{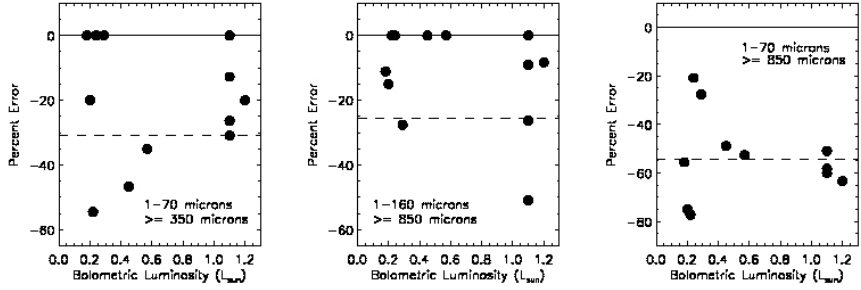}
\caption{\label{fig_errors_lbol}Percent errors between \lbol\ calculated from the category 2 (left; category 3, middle; category 4, right) SED and the category 1 SED, as defined in the text.  The dashed line shows the average value of this error for each category.}
\end{figure}

The three panels of figure \ref{fig_errors_lbol} shows the percent error for each source between \lbol\ calculated from the SEDs in categories $2-4$ and \lbol\ calculated from the SED in category 1.  For category $n$, where $n=2, 3, 4$, this percent error ($PE$) is defined as
\begin{equation}\label{eq_pe}
PE = 100 \times \frac{L_{bol,n} - L_{bol,1}}{L_{bol,1}}
\end{equation}

For all three cateogories, the calculated \lbol\ is an underestimate of the \lbol\ calculated from the more completely sampled Category 1 SEDs.  The average of the absolute values of $PE$, ignoring values of 0 which simply indicate that the complete SED for the source is not a Category 1 SED, are 31\%, 25\%, and 54\% for categories 2, 3, and 4, respectively.  As expected, including detections either between $100-160$ or $350-450$ \um\ significantly improves the accuracy of \lbol; adding detections in both ranges improves it further.

\subsection{Bolometric Temperature}

\begin{figure}[t]
\plotone{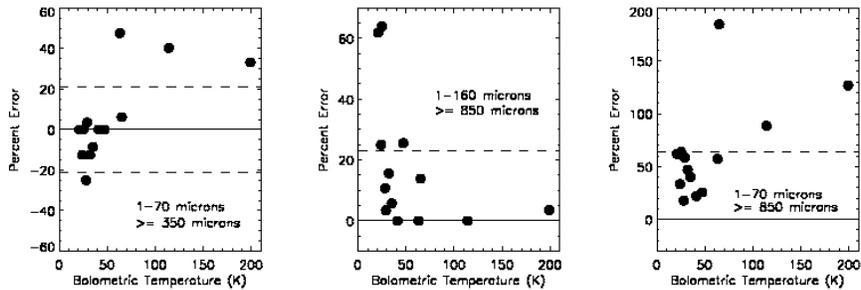}
\caption{\label{fig_errors_tbol}Same as Figure \ref{fig_errors_lbol}, except for \tbol\ rather than \lbol.}
\end{figure}

The results of the same analysis for \tbol\ as that performed above for \lbol\ are shown in Figure \ref{fig_errors_tbol}.  For category 3 and category 4 SEDs, the calculated \tbol\ is always higher than that calculated from a well-sampled SED.  This is expected since both categories leave out detections near the peak of the SED.  For category 2, however, the direction of the error in \tbol\ depends on the true value of \tbol.  Category 2 SEDs of the coldest sources (\tbol\ $\la$ 50 K) will lead to \emph{underestimates} of \tbol, whereas they will lead to overestimates for warmer sources (\tbol\ $\ga$ 50 K).

We calculate the percent error in \tbol\ in the same manner as above for \lbol.  The average of the absolute values of these percent errors are 21\%, 23\%, and 64\% for categories 2, 3, and 4, respectively.  As for \lbol, category 2 and 3 SEDs yield significantly more accurate values of \tbol\ than category 4 SEDs, but are not as accurate as Category 1 SEDs.

We note here that our results are in good agreement with Enoch (2007), who found that overall uncertainties for measured \lbol\ and \tbol\ values are approximately $20-50$\%, depending on whether or not 160 \um\ data are available.  We also note that distance uncertainties will add additional uncertainties to \lbol\ beyond those considered here.

\subsection{Bolometric to Submillimeter Luminosity}

We performed a similar anaysis for \lbolsmm\ as above for \lbol\ and \tbol.  The average values of the percent errors are 31\%, 1282\%, and 605\% for categories 2, 3, and 4, respectively, significantly larger than the errors in \lbol\ and \tbol.  The value of \lbolsmm\ calculated for category 4 SEDs overestimates the actual value; both \lbol\ and \lsmm\ are underestimated by not sampling the SED at all between 70 and 850 \um, but \lbol\ is underestimated less than \lsmm.  The same is true for category 3 SEDs, except the error is even larger because the calculation of \lbol\ is now less of an underestimate (see above), but the calculation of \lsmm\ remains the same.  On the other hand, \lbolsmm\ calculated from category 2 SEDs \emph{underestimates} the actual value, since \lbol\ is underestimated but \lsmm\ is accurately calculated.  The errors here are more than an order of magnitude smaller than Categories 3 and 4.  Clearly, great care must be taken to obtain the most well-sampled SEDs before attempting to draw conclusions from the value of \lbolsmm\ for a given source.

\clearpage
\begin{deluxetable}{lcrrr}
\tabletypesize{\scriptsize}
\tablewidth{0pt}
\tablecaption{\label{lintlspitzertab}Embedded Protostars Observed by \emph{Spitzer} with Published Models}
\tablehead{
		    & \colhead{\lir}	& \colhead{\lint}   & \colhead{\lbol}   &   \\
\colhead{Protostar} & \colhead{(\lsun)}	& \colhead{(\lsun)} & \colhead{(\lsun)} & \colhead{Reference\tablenotemark{a}}}
\startdata
CB130-3-IRS         & 0.05              & 0.15 $\pm$ 0.05   & 0.96              & 1 \\
IRAM 04191+1522	    & 0.02		& 0.08 $\pm$ 0.04   & 0.28              & 2 \\
IRAS 04191+1523	    & 0.25		& 0.4 $\pm$ 0.1	    & 0.64              & 3 \\
IRAS 18148$-$0440   & 8.19		& 13.0 $\pm$ 0.5    & 12.9              & 4 \\
IRAS 19156$+$1906   & 1.13		& 2.6 $\pm$ 0.5	    & 3.4               & 4 \\
L1014-IRS	    & 0.06		& 0.09 $\pm$ 0.03   & 0.3               & 5 \\
L1148-IRS	    & 0.08		& 0.1 $\pm$ 0.05    & 0.15              & 6 \\
L1221-IRS1	    & 1.12		& 2.5 $\pm$ 0.25    & 2.7               & 7 \\
L1221-IRS3	    & 0.36		& 1.0 $\pm$ 0.15    & 1.9               & 7 \\
L1521F-IRS	    & 0.01		& 0.05 $\pm$ 0.02   & 0.36              & 8 \\
L673-7-IRS	    & 0.02		& 0.04 $\pm$ 0.02   & 0.13              & 3 \\
\enddata                                                                            \\
\tablenotetext{a}{Reference for the radiative transfer model and determination of \lint\ and \lbol: \\
(1) H.J. Kim et al. (2008, in preparation); (2) Dunham et al. (2006); (3) M. Dunham et al. (2008, in preparation); (4) Shirley et al. (2002); (5) Young et al. (2004); (6) Kauffmann et al. (2005); (7) C. Young et al. (2008, in preparation); (8) Bourke et al. (2006.).}
\end{deluxetable}
\begin{deluxetable}{lcccccc}
\tabletypesize{\scriptsize}
\tablewidth{0pt}
\tablecaption{\label{tab_fluxlintfits}Linear Least-Squares Fits to Plots of Log F$_{\nu}$ vs. Log \lint.}
\tablehead{
\multicolumn{1}{l}{Wavelength} & \multicolumn{3}{c}{\textsc{Observations}} &\multicolumn{3}{c}{\textsc{Models}} \\
\colhead{(\um)} & \colhead{$m$} & \colhead{$b$} & \colhead{$\chi_r^2$} & \colhead{$m$} & \colhead{$b$} & \colhead{$\chi_r^2$}}
\startdata
3.6 & $0.57 \pm 0.42$ & $-11.07 \pm 0.32$ & 296 & $0.86 \pm 0.05$ & $-11.87 \pm 0.03$ & 3309 \\
4.5 & $0.88 \pm 0.26$ & $-10.54 \pm 0.20$ & 148 & $0.94 \pm 0.04$ & $-11.63 \pm 0.03$ & 2846 \\
5.8 & $0.97 \pm 0.29$ & $-10.54 \pm 0.22$ & 179 & $0.99 \pm 0.04$ & $-11.43 \pm 0.03$ & 2305 \\
8.0 & $0.83 \pm 0.27$ & $-10.72 \pm 0.22$ & 160 & $1.02 \pm 0.04$ & $-11.38 \pm 0.03$ & 2044 \\
24  & $0.87 \pm 0.20$ & $-10.05 \pm 0.17$ &  85 & $1.38 \pm 0.02$ & $-10.12 \pm 0.01$ &  433 \\
70  & $1.06 \pm 0.06$ &  $-9.26 \pm 0.06$ &   3 & $1.06 \pm 0.01$ &  $-9.02 \pm 0.01$ &   15 \\
\enddata \\
\end{deluxetable}
\begin{landscape}
\begin{deluxetable}{lcllcccccc}
\tabletypesize{\scriptsize}
\tablewidth{0pt}
\tablecaption{\label{cores_candidates}Candidate Low-Luminosity Embedded Objects in the Cores}
\tablehead{
\colhead{Source} & \colhead{} & \colhead{Spitzer} & \colhead{c2d} & \colhead{Dist.} & \colhead{RA} & \colhead{Dec} & \colhead{\lir \tablenotemark{d}} & \colhead{D$_{IRAS}$} & \colhead{Nearest IRAS} \\
\colhead{Number} & \colhead{Group\tablenotemark{a}} & \colhead{Source Name\tablenotemark{b}} & \colhead{Core Region\tablenotemark{c}} & \colhead{(pc)} & \colhead{(J2000)} & \colhead{(J2000)} & \colhead{(\lsun)} & \colhead{(\as)} & Source}
\startdata
001 & 1  & J042156.88$+$152946.0 & IRAM04191+1522 & $140 \pm 10$  & 04 21 56.88 & $+$15 29 46.0       & 0.023  &     43.6 & IRAS 04191$+$1523 \\
002 & 1  & J042200.07$+$153024.8 & IRAM04191+1522 & $140 \pm 10$  & 04 22 00.07 & $+$15 30 24.8 & \emph{0.033} &     17.4 & IRAS 04191$+$1523 \\
003 & 1  & J042200.41$+$153021.2 & IRAM04191+1522 & $140 \pm 10$  & 04 22 00.41 & $+$15 30 21.2       & 0.250  &     20.8 & IRAS 04191$+$1523 \\
004 & 1  & J042838.90$+$265135.6 &         L1521F & $140 \pm 10$  & 04 28 38.90 & $+$26 51 35.6       & 0.015  &    380.9 & IRAS 04260$+$2642 \\
005 & 1  & J044112.65$+$254635.4 &           TMC1 & $140 \pm 10$  & 04 41 12.65 & $+$25 46 35.4       & 0.383  &     15.7 & IRAS 04381$+$2540 \\
006 & 1  & J054429.26$+$090856.8 &           B35A & $400 \pm 40$  & 05 44 29.26 & $+$09 08 56.8 & \emph{0.141} &      8.1 & IRAS 05417$+$0907 \\
007 & 6  & J054443.94$+$090307.2 &           B35A & $400 \pm 40$  & 05 44 43.94 & $+$09 03 07.2 & \emph{0.006} &    405.4 & IRAS 05417$+$0907 \\
008 & 5  & J071357.60$-$482711.9 &          BHR22 & $450 \pm 50$  & 07 13 57.60 & $-$48 27 11.9       & 0.006  &    170.7 & IRAS 07128$-$4820 \\
\enddata\\
\tablenotetext{a}{Group indicating likelihood of being an embedded protostar (see \S \ref{groups}).}
\tablenotetext{b}{All source names are preceded by the prefix ``SSTc2d ''.}
\tablenotetext{c}{Name of the core region from Evans et al. (2007) in which this source is located.}
\tablenotetext{d}{Integrated luminosity using all available detections between 1.25 (2MASS J-band) and 70 \um.  Entries in italics denote that the calculation only extends to 24 \um\ due to no flux information available at 70 \um.}
\end{deluxetable}
\end{landscape}
\begin{deluxetable}{lclccccc}
\tabletypesize{\scriptsize}
\tablewidth{0pt}
\tablecaption{\label{per_candidates}Candidate Low-Luminosity Embedded Objects in Perseus}
\tablehead{
\colhead{Source} & \colhead{} & \colhead{Spitzer} & \colhead{RA} & \colhead{Dec} & \colhead{\lir \tablenotemark{c}} & \colhead{D$_{IRAS}$} & \colhead{Nearest IRAS} \\
\colhead{Number} & \colhead{Group\tablenotemark{a}} & \colhead{Source Name\tablenotemark{b}} & \colhead{(J2000)} & \colhead{(J2000)} & \colhead{(\lsun)} & \colhead{(\as)} & Source}
\startdata
050 & 5  & J032353.74$+$310244.5 & 03 23 53.74 & $+$31 02 44.5       & 0.010  &    440.4 & IRAS 03203$+$3048 \\
051 & 5  & J032430.79$+$310651.8 & 03 24 30.79 & $+$31 06 51.8       & 0.017  &    419.3 & IRAS 03219$+$3056 \\
052 & 5  & J032433.14$+$311433.0 & 03 24 33.14 & $+$31 14 33.0       & 0.008  &    407.6 & IRAS 03219$+$3103 \\
053 & 5  & J032503.14$+$310322.7 & 03 25 03.14 & $+$31 03 22.7       & 0.015  &    178.7 & IRAS 03222$+$3053 \\
054 & 5  & J032515.26$+$312511.3 & 03 25 15.26 & $+$31 25 11.3       & 0.008  &    689.4 & IRAS 03219$+$3103 \\
055 & 1  & J032536.22$+$304515.8 & 03 25 36.22 & $+$30 45 15.8 & \emph{0.061} &      4.3 & IRAS 03225$+$3034 \\
056 & 1  & J032539.12$+$304358.1 & 03 25 39.12 & $+$30 43 58.1 & \emph{0.432} &     90.4 & IRAS 03225$+$3034 \\
057 & 6  & J032621.65$+$305726.3 & 03 26 21.65 & $+$30 57 26.3       & 0.006  &    281.5 & IRAS 03232$+$3051 \\
\enddata\\
\tablenotetext{a}{Group indicating likelihood of being an embedded protostar (see \S \ref{groups}).}
\tablenotetext{b}{All source names are preceded by the prefix ``SSTc2d ''.}
\tablenotetext{c}{Integrated luminosity using all available detections between 1.25 (2MASS J-band) and 70 \um.  Entries in italics denote that the calculation only extends to 24 \um\ due to no flux information available at 70 \um.}
\end{deluxetable}

\begin{deluxetable}{lclccccc}
\tabletypesize{\scriptsize}
\tablewidth{0pt}
\tablecaption{\label{cham_candidates}Candidate Low-Luminosity Embedded Objects in Chamaeleon II}
\tablehead{
\colhead{Source} & \colhead{} & \colhead{Spitzer} & \colhead{RA} & \colhead{Dec} & \colhead{\lir \tablenotemark{c}} & \colhead{D$_{IRAS}$} & \colhead{Nearest IRAS} \\
\colhead{Number} & \colhead{Group\tablenotemark{a}} & \colhead{Source Name\tablenotemark{b}} & \colhead{(J2000)} & \colhead{(J2000)} & \colhead{(\lsun)} & \colhead{(\as)} & Source}
\startdata
120 & 5  & J124657.72$-$771916.0 & 12 46 57.72 & $-$77 19 16.0       & 0.004  &    372.4 & IRAS 12416$-$7703 \\
121 & 5  & J124816.42$-$781511.5 & 12 48 16.42 & $-$78 15 11.5       & 0.012  &    896.9 & IRAS 12440$-$7813 \\
122 & 5  & J124940.03$-$775657.1 & 12 49 40.03 & $-$77 56 57.1       & 0.006  &    897.8 & IRAS 12504$-$7745 \\
123 & 5  & J125254.34$-$780913.7 & 12 52 54.34 & $-$78 09 13.7       & 0.010  &    483.2 & IRAS 12504$-$7745 \\
124 & 3  & J125342.86$-$771511.5 & 12 53 42.86 & $-$77 15 11.5       & 0.184  &      5.7 & IRAS 12500$-$7658 \\
125 & 5  & J125507.66$-$765755.4 & 12 55 07.66 & $-$76 57 55.4       & 0.006  &    182.1 & IRAS 12522$-$7640 \\
126 & 5  & J125701.58$-$764834.9 & 12 57 01.58 & $-$76 48 34.9       & 0.050  &     10.5 & IRAS 12533$-$7632 \\
127 & 5  & J130100.05$-$770453.0 & 13 01 00.05 & $-$77 04 53.0       & 0.003  &    347.6 & IRAS 12571$-$7654 \\
\enddata\\
\tablenotetext{a}{Group indicating likelihood of being an embedded protostar (see \S \ref{groups}).}
\tablenotetext{b}{All source names are preceded by the prefix ``SSTc2d ''.}
\tablenotetext{c}{Integrated luminosity using all available detections between 1.25 (2MASS J-band) and 70 \um.  Entries in italics denote that the calculation only extends to 24 \um\ due to no flux information available at 70 \um.}
\end{deluxetable}

\begin{deluxetable}{lcllccccc}
\tabletypesize{\scriptsize}
\tablewidth{0pt}
\tablecaption{\label{lup_candidates}Candidate Low-Luminosity Embedded Objects in Lupus}
\tablehead{
\colhead{Source} & \colhead{} & \colhead{Spitzer} & & \colhead{RA} & \colhead{Dec} & \colhead{\lir \tablenotemark{d}} & \colhead{D$_{IRAS}$} & \colhead{Nearest IRAS} \\
\colhead{Number} & \colhead{Group\tablenotemark{a}} & \colhead{Source Name\tablenotemark{b}} & \colhead{Region\tablenotemark{c}} & \colhead{(J2000)} & \colhead{(J2000)} & \colhead{(\lsun)} & \colhead{(\as)} & Source}
\startdata
136 & 5  & J153706.36$-$334102.4 &   Lupus I & 15 37 06.36 & $-$33 41 02.4       & 0.004  &    682.1 & IRAS 15334$-$3340 \\
137 & 5  & J153751.12$-$335806.6 &   Lupus I & 15 37 51.12 & $-$33 58 06.6       & 0.004  &    680.6 & IRAS 15337$-$3350 \\
138 & 5  & J153800.05$-$340524.4 &   Lupus I & 15 38 00.05 & $-$34 05 24.4       & 0.004  &    709.8 & IRAS 15349$-$3407 \\
139 & 5  & J153826.95$-$325628.7 &   Lupus I & 15 38 26.95 & $-$32 56 28.7       & 0.003  &    178.5 & IRAS 15350$-$3246 \\
140 & 5  & J153833.84$-$334030.0 &   Lupus I & 15 38 33.84 & $-$33 40 30.0       & 0.004  &    511.7 & IRAS 15349$-$3337 \\
141 & 5  & J153834.66$-$344819.4 &   Lupus I & 15 38 34.66 & $-$34 48 19.4       & 0.003  &    486.1 & IRAS 15358$-$3444 \\
142 & 5  & J153914.59$-$341105.3 &   Lupus I & 15 39 14.59 & $-$34 11 05.3       & 0.003  &    532.3 & IRAS 15367$-$3405 \\
143 & 5  & J153953.90$-$334821.6 &   Lupus I & 15 39 53.90 & $-$33 48 21.6       & 0.005  &    186.7 & IRAS 15364$-$3339 \\
\enddata\\
\tablenotetext{a}{Group indicating likelihood of being an embedded protostar (see \S \ref{groups}).}
\tablenotetext{b}{All source names are preceded by the prefix ``SSTc2d ''.}
\tablenotetext{c}{Name of the Lupus region (I, III, or IV) in which this source is located.}
\tablenotetext{d}{Integrated luminosity using all available detections between 1.25 (2MASS J-band) and 70 \um.  Entries in italics denote that the calculation only extends to 24 \um\ due to no flux information available at 70 \um.}
\end{deluxetable}
\clearpage

\begin{deluxetable}{lclccccc}
\tabletypesize{\scriptsize}
\tablewidth{0pt}
\tablecaption{\label{oph_candidates}Candidate Low-Luminosity Embedded Objects in Ophiuchus}
\tablehead{
\colhead{Source} & \colhead{} & \colhead{Spitzer} & \colhead{RA} & \colhead{Dec} & \colhead{\lir \tablenotemark{c}} & \colhead{D$_{IRAS}$} & \colhead{Nearest IRAS} \\
\colhead{Number} & \colhead{Group\tablenotemark{a}} & \colhead{Source Name\tablenotemark{b}} & \colhead{(J2000)} & \colhead{(J2000)} & \colhead{(\lsun)} & \colhead{(\as)} & Source}
\startdata
178 & 5  & J162135.57$-$224351.6 & 16 21 35.57 & $-$22 43 51.6       & 0.004  &   1467.3 & IRAS 16200$-$2251 \\
179 & 5  & J162145.12$-$234231.7 & 16 21 45.12 & $-$23 42 31.7       & 0.033  &    246.9 & IRAS 16187$-$2339 \\
180 & 5  & J162458.99$-$252119.1 & 16 24 58.99 & $-$25 21 19.1       & 0.012  &      6.6 & IRAS 16219$-$2514 \\
181 & 3  & J162648.48$-$242838.6 & 16 26 48.48 & $-$24 28 38.6       & 0.103  &    366.6 & IRAS 16235$-$2416 \\
182 & 1  & J162705.23$-$243629.5 & 16 27 05.23 & $-$24 36 29.5       & 0.107  &     55.0 & IRAS 16240$-$2430 \\
183 & 3  & J162715.89$-$243843.1 & 16 27 15.89 & $-$24 38 43.1       & 0.440  &    163.5 & IRAS 16240$-$2430 \\
184 & 3  & J162717.57$-$242856.3 & 16 27 17.57 & $-$24 28 56.3       & 0.374  &     10.4 & IRAS 16242$-$2422 \\
185 & 1  & J162821.60$-$243623.4 & 16 28 21.60 & $-$24 36 23.4       & 0.046  &     11.6 & IRAS 16253$-$2429 \\
\enddata\\
\tablenotetext{a}{Group indicating likelihood of being an embedded protostar (see \S \ref{groups}).}
\tablenotetext{b}{All source names are preceded by the prefix ``SSTc2d ''.}
\tablenotetext{c}{Integrated luminosity using all available detections between 1.25 (2MASS J-band) and 70 \um.  Entries in italics denote that the calculation only extends to 24 \um\ due to no flux information available at 70 \um.}
\end{deluxetable}

\begin{deluxetable}{lclccccc}
\tabletypesize{\scriptsize}
\tablewidth{0pt}
\tablecaption{\label{serp_candidates}Candidate Low-Luminosity Embedded Objects in Serpens}
\tablehead{
\colhead{Source} & \colhead{} & \colhead{Spitzer} & \colhead{RA} & \colhead{Dec} & \colhead{\lir \tablenotemark{c}} & \colhead{D$_{IRAS}$} & \colhead{Nearest IRAS} \\
\colhead{Number} & \colhead{Group\tablenotemark{a}} & \colhead{Source Name\tablenotemark{b}} & colhead{(J2000)} & \colhead{(J2000)} & \colhead{(\lsun)} & \colhead{(\as)} & Source}
\startdata
200 & 6  & J182841.86$-$000321.2 & 18 28 41.86 & $-$00 03 21.2       & 0.153  &    240.0 & IRAS 18261$-$0009 \\
201 & 6  & J182844.02$+$005338.0 & 18 28 44.02 & $+$00 53 38.0 & \emph{0.146} &     99.1 & IRAS 18262$+$0050 \\
202 & 3  & J182844.78$+$005125.9 & 18 28 44.78 & $+$00 51 25.9 & \emph{0.016} &     40.9 & IRAS 18262$+$0050 \\
203 & 6  & J182844.95$+$005203.4 & 18 28 44.95 & $+$00 52 03.4 & \emph{0.209} &     14.2 & IRAS 18262$+$0050 \\
204 & 6  & J182854.91$+$001832.8 & 18 28 54.91 & $+$00 18 32.8       & 0.033  &    321.8 & IRAS 18267$+$0016 \\
205 & 6  & J182902.11$+$003120.6 & 18 29 02.11 & $+$00 31 20.6       & 0.032  &     70.9 & IRAS 18265$+$0028 \\
206 & 6  & J182902.83$+$003009.7 & 18 29 02.83 & $+$00 30 09.7       & 0.134  &     52.0 & IRAS 18265$+$0028 \\
207 & 6  & J182902.95$+$003008.3 & 18 29 02.95 & $+$00 30 08.3 & \emph{0.035} &     51.3 & IRAS 18265$+$0028 \\
\enddata\\
\tablenotetext{a}{Group indicating likelihood of being an embedded protostar (see \S \ref{groups}).}
\tablenotetext{b}{All source names are preceded by the prefix ``SSTc2d ''.}
\tablenotetext{c}{Integrated luminosity using all available detections between 1.25 (2MASS J-band) and 70 \um.  Entries in italics denote that the calculation only extends to 24 \um\ due to no flux information available at 70 \um.}
\end{deluxetable}

\begin{deluxetable}{lccccc}
\tabletypesize{\scriptsize}
\tablewidth{0pt}
\tablecaption{\label{tab_groups}Groups Showing Likelihood Each Candidate is an Embedded Protostar}
\tablehead{
 & \colhead{Associated with} & \colhead{Embedded in} & \colhead{Associated with} & \colhead{} & \colhead{} \\
\colhead{} & \colhead{Region of High} & \colhead{Region of High} & \colhead{Region of High} & \colhead{4.5 \um} & \colhead{Number} \\
\colhead{Group} & \colhead{Volume Density\tablenotemark{a}} & \colhead{Volume Density\tablenotemark{a}} & \colhead{Column Density\tablenotemark{a}} & \colhead{Jets/Nebulosity\tablenotemark{a}} & \colhead{of Candidates\tablenotemark{b}}}
\startdata
1  & Yes     & Yes        & N/A        & N/A        & 40 \\
2  & Yes     & No/Unknown & N/A        & Yes        & 8 \\
3  & Yes     & No/Unknown & N/A        & No/Unknown & 19 \\
4a & Unknown & No/Unknown & Yes        & Yes        & 1 \\
4b & Unknown & No/Unknown & Yes        & No/Unknown & 0 \\
4c & Unknown & No/Unknown & No/Unknown & Yes        & 1 \\
5  & Unknown & No/Unknown & No/Unknown & No/Unknown & 95 \\
6  & No      & N/A        & N/A        & N/A        & 54 \\
\enddata\\
\tablenotetext{a}{``Yes'' indicates data exists and satisfies this condition.  ``No'' indicates data exists and shows that this condition is not satisfied.  ``Unknown'' indicates no data exists to evaluate this condition.  As discussed in the text, ``No'' and ``Unknown'' are considered equivalent for all but ``Associated with Region of High Volume Density'' since negative results do not prove a source is not an embedded protostar.}
\tablenotetext{b}{Total number of the 218 candidates in this group.}
\end{deluxetable}
\begin{deluxetable}{lclccr}
\tabletypesize{\scriptsize}
\tablewidth{0pt}
\tablecaption{\label{tab_divided_groups}Candidates Divided Into Groups}
\tablehead{
                & \colhead{Source} &                      & \colhead{\lir\tablenotemark{a}} & & \\
\colhead{Group} & \colhead{Number} & \colhead{Core/Cloud} & \colhead{(\lsun)} & \colhead{MIPS3\tablenotemark{b}} & \colhead{SIMBAD}}
\startdata
1  & 001 & IRAM04191$+$1522 &       0.023  & N/A & NAME IRAM 04191$-$IRS         \\
1  & 002 & IRAM04191$+$1522 & \emph{0.033} & N/A & IRAS F04191$+$1523            \\
1  & 003 & IRAM04191$+$1522 &       0.250  & N/A & IRAS F04191$+$1523            \\
1  & 004 & L1521F           &       0.015  & N/A & NAME LDN 1521F$-$IRS          \\
1  & 005 & TMC1             &       0.383  & N/A & IRAS F04381$+$2540            \\
1  & 006 & B35A             & \emph{0.141} & N/A & Barnard 35A                   \\
1  & 017 & DC303.8$-$14.2   &       0.387  & N/A & BHR 86                        \\
1  & 021 & CB68             &       0.299  & N/A & [CB88] 68                     \\
\enddata                                                                                  \\
\tablenotetext{a}{Integrated luminosity using all available detections between 1.25 and 70 \um.  Italics denote that the calculation only extends to 24 \um\ due to no flux information available at 70 \um.}
\tablenotetext{b}{Indicates whether each candidate is associated with a 160 \um\ source.  Y:  Yes; N:  No; U:  Unknown - located off the edge of the 160 \um\ map; C/S: Unknown - located in a region that is confused and/or saturated at 160 \um; N/A: Not applicable for the dense cores.}
\tablenotetext{c}{Source 132 is the same as Source 017.}
\end{deluxetable}
\begin{deluxetable}{lcccccr}
\tabletypesize{\scriptsize}
\tablewidth{0pt}
\tablecaption{\label{tab_seds}SEDs of Groups 1-3}
\tablehead{
\colhead{Source} & \colhead{Wavelength} & \colhead{Flux Density} & \colhead{$\sigma$} & \colhead{Aperture\tablenotemark{a}} & & \colhead{Evolutionary}    \\
\colhead{Number} & \colhead{(\um)}      & \colhead{(mJy)}        & \colhead{(mJy)}    & \colhead{(\as)}    & \colhead{Reference\tablenotemark{b}} & \colhead{Indicators}}
\startdata
001 & 3.6   & 0.22   & 0.02   & ...  & 1 & $L_{int}^{IR}$ = 0.04 \lsun \\
    & 4.5   & 0.78   & 0.06   & ...  & 1 & $L_{int}^{70}$ = 0.05 \lsun \\
    & 5.8   & 0.94   & 0.08   & ...  & 1 & \lbol\ = 0.12 $\pm$ 0.02 \lsun \\
    & 8.0   & 0.63   & 0.06   & ...  & 1 & \lbolsmm\ = 5 $\pm$ 1 \\
    & 24    & 16.4   & 1.73   & ...  & 1 & \tbol\ = 27 $\pm$ 3 K \\
    & 60    & 500    & 100    & 60   & 2 &  \\
    & 70    & 875    & 120    & ...  & 1 &  \\
    & 90    & 800    & 160    & 60   & 2 &  \\
\enddata                                   \\
\tablenotetext{a}{Diameter of the aperture used for photometry at $\lambda > 100$ \um, where the sources are likely to be extended, when specified.  A ``T'' indicates the photometry is an attempt to derive the total source flux density rather than the flux density in a fixed aperture; see individual references for descriptions of the methods used to derive this quantity.}
\tablenotetext{b}{\textsc{References.}$-$  (1) \emph{Spitzer} observations from this work; (2) \andre\ et al. (1999); (3) Wu et al. (2007); (4) 2MASS; (5) \emph{IRAS}; (6) Young et al. (2006a); (7) Motte \& \andre\ (2001); (8) Bourke et al. (2006); (9) M. Dunham et al. (2008), in preparation; (10) Kirk et al. (2005); (11) Kauffmann et al. (2008); (12) Young et al. (2003); (13) Lehtinen et al. (2005); (14) K. Brede et al. (2008), in preparation; (15) Henning et al. (1993); (16) H.J. Kim et al. (2008, in preparation); (17) Young et al. (2004); (18) Hatchell et al. (2005); (19) Hatchell et al. (2007a); (20) Kirk et al. (2006); (21) Enoch et al. (2006); (22) Clark (1991); (23) Hodapp et al. (2005); (24) Tachihara et al. (2007); (25) Johnstone et al. (2000); (26) Stanke et al. (2006); (27) Young et al. (2006b); (28) Kun (1998); (29) \andre\ \& Montmerle (1994).}
\end{deluxetable}

\clearpage

\clearpage

\end{document}